\documentclass[reqno,12pt,a4paper,dvips]{article}

\usepackage{amsmath,amssymb,amsthm,times,mathptmx,array,color,macros}
\usepackage[dvips]{graphicx}
\usepackage{tikz}
\usepackage{titlesec}
\usepackage{setspace}

\titlespacing*{\section}{0pt}{3ex plus 1ex minus 0.4ex}{1ex plus 0.3ex}
\titlespacing*{\subsection}{0pt}{3ex plus 1ex minus 0.4ex}{1ex plus 0.3ex}
\titlespacing*{\subsubsection}{0pt}{3ex plus 1ex minus 0.4ex}{1ex plus 0.3ex}

\numberwithin{equation}{section}

\newcommand{\al}{\alpha}

\newcommand{\de}{\delta}
\newcommand{\De}{\Delta}
\newcommand{\ep}{\epsilon}
\newcommand{\la}{\lambda}

\newcommand{\si}{\sigma}

\newcommand{\vf}{\varphi}

\newcommand{\BL}{\bar{L}}

\newcommand{\bu}{\bar{u}}

\newcommand{\bw}{\bar{w}}

\newcommand{\bz}{\bar{z}}

\newcommand{\CB}{{\mathcal B}}

\newcommand{\BFE}{{\rm\mathbf E}}
\newcommand{\BFF}{{\rm\mathbf F}}

\newcommand{\CD}{{\mathcal D}}

\newcommand{\CH}{{\mathcal H}}

\newcommand{\CI}{{\mathcal I}}

\newcommand{\CL}{{\mathcal L}}
\newcommand{\CM}{{\mathcal M}}
\newcommand{\CN}{{\mathcal N}}
\newcommand{\CO}{{\mathcal O}}
\newcommand{\CP}{{\mathcal P}}

\newcommand{\CR}{{\mathcal R}}

\newcommand{\CU}{{\mathcal U}}
\newcommand{\CV}{{\mathcal V}}

\newcommand{\SE}{{\mathsf E}}
\newcommand{\SF}{{\mathsf F}}

\newcommand{\SH}{{\mathsf H}}

\newcommand{\SJ}{{\mathsf J}}

\newcommand{\sk}{{\mathsf k}}

\newcommand{\SM}{{\mathsf M}}

\newcommand{\SO}{{\mathsf O}}

\newcommand{\SQ}{{\mathsf Q}}
\newcommand{\SR}{{\mathsf R}}

\newcommand{\ST}{{\mathsf T}}
\newcommand{\SU}{{\mathsf U}}
\newcommand{\SV}{{\mathsf V}}
\newcommand{\SW}{{\mathsf W}}
\newcommand{\SX}{{\mathsf X}}
\newcommand{\SY}{{\mathsf Y}}
\newcommand{\SZ}{{\mathsf Z}}

\newcommand{\fg}{{\mathfrak g}}

\newcommand{\sa}{{\mathsf a}}

\newcommand{\sd}{{\mathsf d}}
\newcommand{\se}{{\mathsf e}}
\newcommand{\sg}{{\mathsf g}}
\renewcommand{\sf}{{\mathsf f}}
\newcommand{\sh}{{\mathsf h}}

\newcommand{\sq}{{\mathsf q}}
\newcommand{\spp}{{\mathsf p}}

\newcommand{\mst}{{\mathsf t}}
\newcommand{\su}{{\mathsf u}}
\newcommand{\sv}{{\mathsf v}}

\newcommand{\sw}{{\mathsf w}}

\newcommand{\sx}{{\mathsf x}}
\newcommand{\sy}{{\mathsf y}}
\newcommand{\sz}{{\mathsf z}}

\newcommand{\ot}{\otimes}
\newcommand{\fsl}{{\mathfrak s}{\mathfrak l}}
\newcommand{\pa}{\partial}
\newcommand{\ra}{\rightarrow}
\newcommand{\rf}[1]{\eqref{#1}}

\newcommand{\BR}{{\mathbb R}}

\newcommand{\BC}{{\mathbb C}}

\newcommand{\fr}[2]{{\textstyle\frac{#1}{#2}}}

\newcommand{\sst}{\scriptscriptstyle}

\newcommand{\alg}[1]{\mathfrak{#1}}

\newcommand{\func}[2]{#1 \left( #2 \right)}

\newcommand{\brac}[1]{\left( #1 \right)}
\newcommand{\sqbrac}[1]{\left[ #1 \right]}
\newcommand{\set}[1]{\left\{ #1 \right\}}
\newcommand{\st}{\mspace{5mu} : \mspace{5mu}}

\newcommand{\abs}[1]{\left| #1 \right|}

\newcommand{\ZZ}{\mathbb{Z}}

\newcommand{\CC}{\mathbb{C}}

\newcommand{\dd}{d}
\newcommand{\ii}{i}

\newcommand{\eps}{\varepsilon}

\newcommand{\wun}{\mathbf{1}}

\newcommand{\affine}[1]{\widehat{#1}}

\newcommand{\comm}[2]{\bigl[ #1 , #2 \bigr]}
\newcommand{\acomm}[2]{\bigl\{ #1 , #2 \bigr\}}

\newcommand{\normord}[1]{{} : #1 : {}} 

\newcommand{\SC}[2]{\func{\SV_{#1}}{#2}}
\newcommand{\VOp}[2]{\func{\SV_{#1}}{#2}}

\newcommand{\SLA}[2]{\alg{#1} \left( #2 \right)}
\newcommand{\AKMA}[2]{\widehat{\alg{#1}} \left( #2 \right)}
\newcommand{\SLSA}[3]{\alg{#1} \left( #2 \middle\vert #3 \right)}
\newcommand{\AKMSA}[3]{\widehat{\alg{#1}} \left( #2 \middle\vert #3 \right)}
\newcommand{\QEA}[2]{\mathcal{U}_{#1} \bigl( #2 \bigr)}

\newcommand{\qnum}[2]{\sqbrac{#1}_{#2}}

\newcommand{\eqnref}[1]{Equation~\eqref{#1}}
\newcommand{\eqnDref}[2]{Equations~\eqref{#1} and \eqref{#2}}

\newcommand{\secref}[1]{Section~\ref{#1}}
\newcommand{\secDref}[2]{Sections~\ref{#1} and \ref{#2}}
\newcommand{\appref}[1]{Appendix~\ref{#1}}

\DeclareMathOperator{\id}{id}

\newcommand{\cfts}{conformal field theories}

\newcommand{\qea}{quantum affine algebra}

\newcommand{\gsos}{screening charges}

\theoremstyle{plain}

\theoremstyle{remark}

\begin{document}

\thispagestyle{empty}

\begin{center}
{\LARGE \bf Integrability of a family of quantum field theories \\[2mm]
related to sigma models}

\vspace{1cm}

\begin{minipage}{0.45\textwidth}
\textcolor{white}{${}^1$ }{\large David Ridout$^{1,2}$} \\[2mm]
${}^1$ {\it Department of Theoretical Physics, \\
\phantom{${}^1$} Australian National University, \\
\phantom{${}^1$} Canberra, ACT 0200, Australia.}
\end{minipage}
\begin{minipage}{0.35\textwidth}
\textcolor{white}{${}^2$ }{\large J\"org Teschner$^2$} \\[2mm]
${}^2$ {\it Theory Group, DESY, \\
\phantom{${}^2$} Notkestra\ss{}e 85, \\
\phantom{${}^2$} D-22603, Hamburg, Germany.}
\end{minipage}

\vspace{6mm}

\today

\vspace{6mm}

\begin{minipage}{0.9\textwidth}
\begin{center}
{\bf Abstract}
\end{center}
\singlespacing
A method is introduced for constructing lattice discretizations of large classes of integrable quantum field theories. The method proceeds in two steps: The quantum algebraic structure underlying the integrability of the model is determined from the algebra of the interaction terms in the light-cone representation. The representation theory of the relevant quantum algebra is then used to construct the basic ingredients of the quantum inverse scattering method, the lattice Lax matrices and R-matrices. This method is illustrated with four examples: The Sinh-Gordon model, the affine $\fsl(3)$ Toda model, a model called the fermionic $\fsl(2|1)$ Toda theory, and the $N=2$ supersymmetric Sine-Gordon model. These models are all related to sigma models in various ways. The $N=2$ supersymmetric Sine-Gordon model, in particular, describes the Pohlmeyer reduction of string theory on $AdS_2\times S^2$, and is dual to a supersymmetric non-linear sigma model with a sausage-shaped target space.
\end{minipage}
\end{center}

\vspace{2ex}
\onehalfspacing

\section{Introduction}

\subsection{Motivation}

There is a growing family of quantum field theories that are
known or expected to be integrable at the quantum level.
If this is the case, then one may learn much about certain non-perturbative phenomena in these quantum field theories. One gains,
in particular, full control over interesting topics such as
non-perturbative dualities, giving deep insight into the nature and the
relevance of these in quantum field theory.
A particularly striking example is the conjectured
duality between the $N=4$ super Yang-Mills theory and
string theory on $AdS_5$ in the limit where the rank of the gauge group
is large.  There is considerable evidence
for the integrability of both theories and for their
equivalence as quantum theories, see \cite{AdS} for a review.

However, a proof of the integrability of these theories
has so far remained elusive. More generally, despite a lot of important
progress in the field of integrable models, there are
only a few quantum theories for which quantum integrability
has been fully established. In most cases, one needs
to regularize ultraviolet divergences. Integrability is hard,
if not impossible, to control in this process unless
the regularized theory is itself integrable. One of the
most successful regularization schemes uses integrable
lattice regularizations for which a certain supply of known
techniques is available.

Among the integrable lattice regularizations, the
spin-chain models seem to be the most popular. A spin-chain is
defined by choosing a collection of representations of a Lie algebra
(or some deformation thereof). These representations are then
associated with certain sites of a given lattice.
However, it is often not clear at the beginning if a
given spin chain will correspond to the quantum field theory that one
is trying to regularize. The proper definition of the continuum limit
may be intricate and important characteristics of the
theory may depend heavily on how exactly this limit is taken.

Another class of lattice regularizations exists which appears to
capture more of the features of the quantum field theory that the
lattice model is supposed to regularize.
We will call a lattice-discretization \emph{tailor-made} if
\begin{itemize}
\item the local degrees of freedom of the lattice model are in direct
relation to the field variables of the corresponding continuum quantum field
theory, and
\item the quantum algebraic structure underlying the integrability is the
same in the continuum models and the corresponding discretization.
\end{itemize}
Formulating these requirements more precisely is one of
our aims in this paper.
The first of these two features is, in particular, realized when
the variables of the lattice model can be identified with averages
of the continuum field variables over small regions of
space and/or time. The second is
naturally much more subtle. Another of our aims in what follows
is to explain in some detail how this can be precisely realized
for a certain family of examples.

From a practical point of view, it seems to be preferable to
use a tailor-made lattice-discretization when possible.
One then has very good reason to expect that
the continuum limit will be the quantum field theory which one is
interested in. It can also facilitate the
solution of the theories considerably --- important consequences
of the integrable structure are already under full control
in the discretized version, and these features remain essentially
unchanged when the continuum limit is taken. This remark applies
in particular to the functional relations obeyed by the generating
functions for the eigenvalues of the conserved quantities
(such functional relations are collectively known as T-, Q- or Y-systems).

\subsection{Aims}

To reiterate, our main aim in this paper is to present a method for
constructing tailor-made lattice regularizations that
appears to be applicable to a large class of models.
We illustrate this method with several physically relevant examples.
Very roughly, the method proceeds in two steps:
\begin{itemize}
\item First, we identify the algebraic structure underlying the
integrability of the model in question. This follows from the
algebra generated by the chiral halves of the interaction
terms. The consideration of these chiral halves is physically
well-motivated in the light-cone
representation, as we will explain in Section \ref{CFTint}.
The relevant algebraic structures for our examples turn out
to be quantum affine (super)algebras.
\item The second step then consists of constructing the
basic building blocks of the lattice
regularization from the representation
theory of the algebraic structure identified above.
Practically, this means computing Lax matrices
$L_n^{\pm}(\la)$ on the lattice using our knowledge of the
relevant quantum affine (super)algebra.
In doing this, it is crucial in our approach to
use a discrete light-cone representation for the
two-dimensional lattice. The monodromy matrices
may then be constructed in the form
\begin{equation}
\SM(\la) = L_N^-(\la)L_N^+(\la)\cdots L_1^-(\la)L_1^+(\la).
\end{equation}
The Lax matrices
$L_n^{\pm}(\la)$
represent parallel transport along the light-cone
directions in a two-dimensional discrete space-time. Our construction
will be similar, but not equivalent, to the previous constructions of this
type described in \cite{FV92,BBR}.
\end{itemize}

The four examples which we will consider in the following have
been chosen for their physical interest and because they appear
to be prototypical in the sense that they exhibit
a certain variety of different qualitative features.
These models are the Sinh-Gordon model, the
$\fsl(3)$ affine Toda theory, a model that we
call the fermionic $\fsl(2|1)$ affine Toda theory,
and the $N=2$ supersymmetric generalization of the
Sine-Gordon model.

The last two models are of particular interest.
They seem to be the first models containing
a mixture of fermions and bosons for which a
lattice regularization has been constructed. Moreover, the
$N=2$ supersymmetric generalization of the
Sine-Gordon model appears in the Pohlmeyer-reduction
of string theory on $AdS_2\times S^2$ \cite{GT}. Proving that
this theory is integrable supports the hope that
Pohlmeyer-reductions of string theories on anti-de Sitter spaces
can be consistently quantized.

We mention that all of the models under investigation share one important
feature: The presence of a non-compact boson $\phi_1$ with
exponential interactions $e^{\ep b\phi_1}$, $\ep=\pm 1,\pm 2$.
This feature is shared by all non-linear sigma models
with anti-de Sitter spaces as targets.
As we will explain in more detail, the presence of such
exponential interactions produces subtle divergences in the ultraviolet.
The proper treatment of these divergences produces
non-perturbative counterterms which dominate the
deep-quantum behavior of the theories, leading to
interesting duality phenomena \cite{T11}. 
In the case of the $N=2$ Sine-Gordon
model, one finds a dual description in terms of a
non-linear sigma model with a sausage-shaped target \cite{F96,HK}.
This means that the corresponding lattice model constructed
in this paper is simultaneously an integrable lattice regularization
for the $N=2$ supersymmetric sausage sigma model.

\subsection{Structure of this paper}

The structure of this paper is as follows.  Section \ref{models} first introduces the models of interest via their Lagrangian descriptions and discusses some of their basic features.  In Section \ref{classicalint}, the integrability of these models is discussed at the classical level. Zero curvature representations are given for the classical equations of motion, making the classical integrability of these models manifest.

Section \ref{CFTint} then reviews the known relations between quantum
affine algebras and the integrability of the bosonic affine Toda theories.
The algebra of the interaction terms in the light-cone representation
plays a crucial role. The fact that one can construct representations
of the nilpotent subalgebras of certain quantum affine algebras from these interaction terms leads, in certain cases, to direct constructions of quantum monodromy matrices.

Letting ourselves be guided by these examples, we continue in Section
\ref{supermodels} with the identification of the relevant
quantum algebraic structures underlying the fermionic $\fsl(2|1)$
affine Toda theory and the $N=2$ super Sine-Gordon model.
It turns out that we have to consider quantum affine
superalgebras in these cases.

In section \ref{lattice}, we reformulate the known lattice
discretization of the Sinh-Gordon model in way that serves as a paradigm for the construction to be presented for the other models.  We commence Section \ref{lattice-gen} by formulating a general recipe for the construction of integrable lattice discretizations that should be applicable to large classes
of integrable quantum field theories. This recipe is then
illustrated by working out the basic building blocks
(the Lax matrices) for the remaining three models
studied here.  The article concludes with a brief outlook and two appendices which discuss some technical points.

\section{The models of interest}\label{models}

We will be interested in the following
family of models which are related in various
ways, but also exhibit a certain variety of different qualitative features.
These models are of affine Toda type or
some generalization thereof. In the following, we will use the
anticipated relations with certain affine Lie (super)algebras as a classification tool.

\subsection{Lagrangian formulations}

Let us begin by listing the action functionals defining the models of
interest on the classical level.
\begin{itemize}

\item {\it The Sinh-Gordon model}.
This model is classically defined by the action
\begin{equation}\label{Ssl(2)}
S = \int d^2z\;
\Bigl( \frac{1}{4\pi}(\partial_\al\phi_1)^2+
\mu e^{-2b\phi_1}+\nu e^{2b\phi_1}\Bigr)
\end{equation}
and is formally related to the Sine-Gordon model by setting
$b= i\beta$.

\item {\it The $\SLA{sl}{3}$affine Toda theory}.
The action is
\begin{equation}\label{Ssl(3)}
S = \int d^2z\;
\Bigl(\;\frac{1}{4\pi}(\partial_\al\phi_1)^2+
\frac{1}{4\pi}(\partial_\al\phi_2)^2+\mu  e^{-b\phi_1} 2\cosh(\sqrt{3}b\phi_2)+\nu e^{2b\phi_1}\Bigr).
\end{equation}

\item {\it The fermionic $\SLSA{sl}{2}{1}$ affine Toda theory}.
Interesting new features arise when we consider models
containing fermions. As one of the simplest examples,
we shall consider the model defined classically by the action
\begin{multline}\label{dualcplxShG'}
S= \frac{1}{2\pi} \int d^2z \;
\biggl(\frac{1}{2} (\partial_\al\phi_1)^2
+  \bar\psi_{+}^{} \pa_-^{} \psi_{+}^{}
+  \bar\psi_{-}^{} \pa_+^{} \psi_{-}^{}
-\frac{b^2}{4}\psi_+^{}\bar\psi_+\psi_-\bar\psi_- \\
-2\pi {\mu}b {(\bar\psi_+\bar\psi_-+\psi_+\psi_-)} e^{-{b}\phi_1}
+8\pi^2\mu^2  e^{-2{b}\phi_1}+4\pi \nu e^{2b\phi_1}\bigg) .
\end{multline}
The reason for calling this model the fermionic $\SLSA{sl}{2}{1}$ affine Toda theory will be explained in \secref{classSL(2|1)}.



\item {\it The $N=2$ super Sine-Gordon model}.
We will also study a supersymmetric model, the $N=2$ super Sine-Gordon model.
The action is
\begin{multline} \label{N=2Liou}
S= \frac{1}{2\pi}\int d^2z \;
\biggl(\frac{1}{2} \bigl( (\partial_\al\phi_1)^2 +
(\partial_\al\phi_2)^2 \bigr)
+  \bar\psi_{+}^{} \pa_-^{} \psi_{+}^{}
+  \bar\psi_{-}^{} \pa_+^{} \psi_{-}^{} \biggr) \\
- b\int d^2z \;\Bigl(
\mu\bigl( \bar\psi_{+}^{} \bar\psi_{-}^{} e^{-\ii b\phi_2} +
\psi_{+}^{} \psi_{-}^{} e^{\ii b\phi_2} \bigr) e^{-b\phi_1}
+\nu
\bigl( \bar\psi_{+}^{}\bar\psi_{-}^{} e^{\ii b\phi_2} +
\psi_{+}^{}\psi_{-}^{} e^{-\ii b\phi_2} \bigr) e^{b\phi_1}\Bigr) \\
+4\pi \int d^2z \;
\bigl( \mu^2e^{2b\phi_1}+\nu^2e^{-2b\phi_1}-2\mu\nu\cos(2b\phi_2)\big).
\end{multline}
The $N=2$ supersymmetry can be made manifest using the superspace
formalism \cite{KU}.
\end{itemize}
An important parameter in each of the models that we are considering is the constant $b$ which appears in the exponential interaction terms. By a rescaling
of the fields, one may factor it out in front of the action,
leading one to identify $b^2$ with Planck's constant $\hbar$ as it
controls the strength of quantum fluctuations.
The above action functionals may therefore be used as the starting point for
constructing a perturbative expansion in the parameter $b$.
The method to be used is a variant of the background field method
in which one expands around a solution to the classical equations
of motion that follow from these functionals.

\subsection{Descriptions as perturbed free field theories}\label{altacts}

Another way to approach the definition of these models is to
quantize the field theories whose action $S_0$ is
obtained by setting $\mu=\nu=0$ in their respective action functionals.
One then tries to define the interaction terms as certain
composite operators constructed from the quantum fields
present in the theory defined by $S_0$,
leading to a prescription for the evaluation of the correlation
functions as formal series in powers of $\mu$ and $\nu$.
In the implementation of this procedure, one typically
encounters two types of difficulties:
\begin{itemize}
\item The treatment of ultraviolet divergences requires the renormalization of
both the composite fields appearing in the interaction terms
and the coupling constants.
\item The dependence of the
correlation functions on $\mu$ and $\nu$
involves non-perturbative behavior which {depends sensitively}
on the choice of infrared regularization.
\end{itemize}
In this section, we shall briefly discuss the first of these
issues for the interesting regime corresponding to $b=i\beta$,
$\beta\in\BR$.  The problem of constructing the interaction terms
turns out to be fairly tame in this case in the sense that
there exist formulations of the models in which
standard free field normal ordering suffices.
For real values of $b$, which is the case of our main interest, there arise
additional subtleties in the ultraviolet behavior of the theories
which will be discussed in Section \ref{realb}.

The description as perturbed free field theories is
absolutely straight-forward for the actions \rf{Ssl(2)}
and \rf{Ssl(3)}. Defining the exponential functions of the
fields $\phi_1$ and $\phi_2$ by standard free field normal
ordering will be sufficient. The situation is more
subtle in the remaining two cases.

\subsubsection{The fermionic $\SLSA{sl}{2}{1}$ affine Toda theory as a perturbed free field theory}\label{pertFFsl(2|1)}

Instead of \rf{dualcplxShG'}, let us consider the action
\begin{multline}\label{Ssl(2|1)-f}
S= \frac{1}{2\pi}\int d^2z \;
\biggl(\frac{1}{2} (\partial_\al\phi_1)^2
+  \bar\psi_{+}^{} \pa_-^{} \psi_{+}^{}
+  \bar\psi_{-}^{} \pa_+^{} \psi_{-}^{}
-\frac{b^2}{4}\psi_+^{}\bar\psi_+\psi_-\bar\psi_-  \\
-2\pi  {\mu}b (\bar\psi_+\bar\psi_-+\psi_+\psi_-) e^{-{b}\phi_1}
+2\pi \nu e^{2b\phi_1}\bigg) ,
\end{multline}
which differs only by dropping the term proportional to $\mu^2$.
Setting $\mu=\nu=0$ yields an action $S_0$ which describes
a free bosonic field $\phi_1$ and a decoupled massless Thirring
model. The terms proportional to $\mu$ and $\nu$ are considered
to be interactions coupling the bosonic and fermionic fields.

One should note, however, that the action \rf{Ssl(2|1)-f}
is not suitable for constructing the semiclassical expansion
in powers of $b$. In the limit $b\ra 0$, the products of the terms
proportional to $e^{-{b}\phi_1}$ generate
the finite additional contribution
$8\pi\mu^2 \int d^2z \; e^{-2{b}\phi_1}$ to the action.
Indeed, let us consider the following contribution at order $\mu^2$:
\begin{equation}
\mu^2 b^2\int d^2z_1 d^2z_2\;
\bar\psi_+(z_1)\bar\psi_-(\bz_1) \normord{e^{-b\phi_1(z_1,\bz_1)}}
\psi_+(z_2)\psi_-(\bz_2) \normord{e^{-b\phi_1(z_2,\bz_2)}}.
\end{equation}
Directly taking $b\ra 0$ would produce a
non-integrable singularity $\sim|z_1-z_2|^{-2}$ from the
fermion operator product expansion 
\begin{equation} \label{FermionOPE}
\psi_\ep(z) \bar\psi_{\ep'}(w) =
\frac{-2i \ep\delta_{\ep\ep'}}{z-w} + \ldots
\end{equation}
We need to introduce a cut-off $\epsilon$ and
split the integral into a contribution from $|z_1-z_2|<\epsilon$
and the rest. For small $\epsilon$, we get a good
approximation for the contributions from $|z_1-z_2|<\epsilon$
by using the operator product expansion:
\begin{equation}
\int d^2z \int_{|w|<\epsilon}d^2w  \;
\frac{4\mu^2 b^2}{|w|^{2+b^2}} \normord{e^{-2b\phi_1(z)}} =
-8\pi \frac{\mu^2}{\epsilon^{b^2}}\int d^2z \;
 \normord{e^{-2b\phi_1(z)}}.
\end{equation}
The term on the left has a finite limit for $b\ra 0$ which
is $\epsilon$-independent. It can be taken into account by
adding the term $8\pi\mu^2 \int d^2z  e^{-2{b}\phi_1}$ to \rf{Ssl(2|1)-f}.
The resulting action is exactly \rf{dualcplxShG'}.

In order to arrive at a description of this model as a perturbed
free field theory, it is useful to
apply the boson-fermion
correspondence to the model defined by \rf{Ssl(2|1)-f}.  This
yields the action
\begin{equation}\label{Ssl(2|1)}
S = \int d^2z\;
\bigg(\frac{1}{4\pi}(\partial_\al\phi_1)^2+\frac{1}{4\pi}(\partial_\al \phi_2)^2-
2\mu b  e^{-{b}\phi_1}\cos(\sqrt{\kappa}\phi_2)
+\nu e^{{2}{b}\phi_1}\bigg)  ,
\end{equation}
where the parameters $b$ and $\kappa$ in
\rf{Ssl(2|1)} are related by $b^2=\kappa-2$.
The action \rf{Ssl(2|1)} was the starting point for the investigation of this
model in \cite{F95}.


\subsubsection{The $N=2$ Sine-Gordon model as a perturbed
free field theory}

In the case of the $N=2$ Sine-Gordon model with action \rf{N=2Liou},
we may take $S_0$ to be defined by the terms in the first line
of \rf{N=2Liou}, treating the terms in the second line as perturbations
and considering the terms in the third line of \rf{N=2Liou}
as counterterms generated from the renormalization of the
perturbations in the limit $b\ra 0$. Bosonizing the fermions in the $N=2$ Sine-Gordon model, we obtain the action
\begin{multline}\label{N=2Liou'}
S = \frac{1}{4\pi}\int d^2z \;
\Bigl( (\partial_\al\phi_1)^2+(\partial_\al\phi_2)^2+(\partial_\al\phi_3)^2 \Bigr)
\\
-\mu b\int d^2z \; 2\cos\bigl(\sqrt{2}\phi_3+b\phi_2\bigr) e^{-b\phi_1}
-\nu b\int d^2z \; 2\cos\bigl(\sqrt{2}\phi_3-b\phi_2\bigr) e^{b\phi_1}.
\end{multline}
In this form, one easily recognizes the model as a special case of the
so-called SS-model introduced by Fateev \cite{F96}.

\subsection{The ultraviolet behavior of real exponential interactions}\label{realb}


Turning to the case of our main interest, $b\in\BR$, it is worth noting
that the exponential interactions now
lead to rather subtle ultraviolet behavior. As an illustration, let
us consider the simple example of Liouville theory:
\begin{equation}\label{Sliou}
S = \int \frac{d^2z}{\pi}\;\bigl(\partial_z\phi \partial_{\bz}\phi+\pi\mu e^{2{ b}\phi}\bigr) .
\end{equation}
Consider those $n$-th order terms in the perturbative expansion of this action which contain
\begin{equation}\label{nthorder}
\frac{(-\mu)^n}{n!}\int d^2u_1\cdots\int d^2u_n \;
e^{2{b}\phi(u_1,\bu_1)}\cdots e^{2{b}\phi(u_n,\bu_n)} .
\end{equation}
By using the operator product expansion
\[
e^{2{b}\phi(z,\bz)} e^{2{b}\phi(w,\bw)} \sim
|z-w|^{-4{{b}^2}}e^{4{b}\phi(w,\bw)} ,
\]
it is easy to see that there are singularities
produced by the possible ``clustering'' of integration variables. If $m$ of the integration variables are close to coinciding,
one may effectively represent the product of the $m$ fields
$e^{2{b}\phi(u_1,\bu_1)}\cdots e^{2{b}\phi(u_{m},\bu_{m})}$
by $e^{2m{b}\phi(u_{m},\bu_{m})}$. It follows that the integration over
$u_{m+1}$
encounters an effective singularity of the form $|u_m-u_{m+1}|^{-2mb^2}$.
As a function of $b^2$, one will therefore encounter poles in
perturbative computations when $b^2$ is rational.
Even if one excludes the rational values of $b^2$
from consideration, there will still be a
small denominator problem to surmount.  For taking $b^2$
irrational means that the summation over $n$ will produce terms
in which $nb^2$ comes arbitrarily close to the values where
\rf{nthorder} has poles.

It can be argued \cite{T11}
that the proper renormalization of these singularities
necessitates dual interactions which contain exponential operators
proportional to $e^{\pm b^{-1}\phi_1}$. At the moment, the
lattice regularization seems to be the most powerful approach to the 
quantization of these
theories as is illustrated by the results obtained for the Sinh-Gordon
model and for Liouville theory in \cite{ByT,T,ByT2}.

\subsection{Description as perturbed conformal field theories} \label{Conflim}

It is important to note that all of the models above share one salient
feature: They have interaction terms proportional to $e^{\ep b \phi_1}$, $\ep=\pm 1,\pm 2$, that become strong when
$\phi_1\ra\pm \infty$. If however, one sets $\nu=0$ in the above
actions, one obtains models in which all interactions vanish
for $\phi_1\ra\infty$. This is closely related to the appearance
of conformal invariance in the $\nu=0$ models.
The following table summarizes the resulting models
and their chiral algebras.
\begin{center}
\begin{tabular}{c|c|c}
Massive model & Limit $\nu=0$ & Chiral symmetry \\ \hline\hline
$\SLA{sl}{2}$ affine Toda & Liouville theory & Virasoro algebra \\ \hline
$\SLA{sl}{3}$ affine Toda & conformal Toda theory & $W_3$ algebra \\ \hline
$\SLSA{sl}{2}{1}$ affine Toda & Sine-Liouville theory & Parafermion algebra \\ \hline
$N=2$ super Sine Gordon & $N=2$ Liouville theory & $N=2$ superconformal algebra
\end{tabular}
\end{center}
All of these \cfts{} are non-rational. The key features, including the spectrum and the three-point functions, are known in the cases of Liouville theory, Sine-Liouville theory and $N=2$ Liouville theory.

\section{Classical integrability} \label{classicalint}

\subsection{The Sinh-Gordon model}\label{SG-KdV}

The classical Sinh-Gordon model is a dynamical system
whose degrees of freedom are described by
a field $\phi(x,t)$ defined on
$(x,t)\in S^1_R \times \BR$ (assuming periodic boundary conditions $\phi(x+R,t)=\phi(x,t)$).
The dynamics of this model
may be described in the Hamiltonian formalism in terms of
$\phi(x,t)$ and $\Pi(x,t) = \pa_t\phi(x,t)$, the Poisson brackets being
\begin{equation}\label{PB}
\{ \Pi(x,t) , \phi(x',t) \}  = 2\pi \de(x-x') .
\end{equation}
The time-evolution of an arbitrary observable $O(t)$ is then given as
\begin{equation}
\pa_tO(t) = \{ H , O(t) \} ,
\end{equation}
with the Hamiltonian $H$ being defined as
\begin{equation} \label{Hdef}
H = \int_0^{R}\frac{dx}{4\pi}\;\sqbrac{\Pi^2 + (\pa_x\phi)^2 + 8\pi\mu
\cosh(2b\phi)}.
\end{equation}

It is well known that the equation of motion for the
Sinh-Gordon model can be represented as the zero curvature condition
\begin{equation}\label{ZCC}
\bigl[ \pa_t-U_t(x,t;\lambda) , \pa_x-U_x(x,t;\lambda) \bigr] = 0 ,
\end{equation}
where the matrices $U_x(x,t;\lambda)$ and $U_t(x,t;\lambda)$ are given by
\begin{subequations} \label{LaxSG}
\begin{align}
U_x(x,t;\lambda) &=
\begin{pmatrix}
b\pa_t\phi/2 & m(\la e^{-b\phi} +\la^{-1}e^{+b\phi})\\
m(\la e^{+b\phi} +\la^{-1}e^{-b\phi}) & -b\pa_t\phi/2
\end{pmatrix}, \\
U_t(x,t;\la) &=\begin{pmatrix}
b\pa_x\phi/2 & m(\la e^{-b\phi} -\la^{-1}e^{+b\phi})\\
m(\la e^{+b\phi} -\la^{-1}e^{-b\phi}) & -b\pa_x\phi/2
\end{pmatrix},
\end{align}
\end{subequations}
and where $m$ is related to the coupling constant $\mu$ by $m^2=\pi b\mu$.
The constant $\la \in \CC$ is known as the spectral parameter.

The classical integrability of the Sinh-Gordon model
follows from the existence of sufficiently many conserved quantities. These conserved quantities
are generated from the trace of the monodromy matrix of the connection $\pa_x-U_x(x,t;\lambda)$:
\begin{equation}\label{Tdef}
T(\la) = {\rm tr}(M(\la)) ,\qquad M(\la) = \CP\exp\left(\int_0^{R}dx\;U_x(x,t;\lambda)\right) .
\end{equation}
The Poisson brackets for the elements of the matrix $M(\la)$ can be written in the form
\begin{equation}\label{FPBR}
\big\{  M(\la)  \overset{\otimes}{,}  M(\mu)  \big\}  =
\big[  R(\la/\mu)  ,  M(\la)\ot M(\mu)  \big] ,
\end{equation}
where $R(\la)$ is the matrix
\begin{equation}
R(\la) = \frac{\la+\la^{-1}}{\la-\la^{-1}}\frac{\SH\ot \SH}{2}+\frac{2}{\la-\la^{-1}}(\SE\ot \SF+\SF\ot \SE)
\end{equation}
with
\begin{equation}\label{efh}
\SE=
\begin{pmatrix}
0& 1 \\
0 & 0
\end{pmatrix}
,\qquad
\SH=
\begin{pmatrix} 1 & 0 \\
0 & {-1}
\end{pmatrix}
,\qquad
\SF=
\begin{pmatrix} 0& 0 \\
1 & 0
\end{pmatrix}
.
\end{equation}
The mutual Poisson commutativity $\bigl\{T(\la),T(\mu)\bigr\}=0$ follows easily from \rf{FPBR}.

\subsubsection{Light-cone representation}\label{lc}

Another useful representation of the zero curvature condition
\rf{ZCC} is obtained by passing to the light-cone coordinates
$x_\pm=t\pm x$ and the corresponding derivatives
$\pa_\pm=\frac{1}{2}(\pa_t\pm\pa_x)$. The zero curvature condition
\rf{ZCC} can now be written as
\begin{equation}\label{ZCClc}
\bigl[  \pa_+-U_+(\lambda)  ,  \pa_--U_-(\lambda)  \bigr]  =  0  ,
\end{equation}
where the matrices $U_+ = \tfrac{1}{2} \brac{U_t + U_x}$ and $U_- = \tfrac{1}{2} \brac{U_t - U_x}$ are given by
\begin{subequations}\label{LaxSG+-}\begin{align}
U_+(\la) & = +\frac{b}{2} \SH \pa_+\phi+m  \SE_1  e^{-b\phi}+
m \SE_0  e^{b\phi} ,\\
U_-(\la) & = -\frac{b}{2} \SH \pa_-\phi-m  \SF_1 e^{-b\phi}-
m \SF_0  e^{b\phi} .
\end{align}\end{subequations}
Here, we have used the notation
$\SE_1=\la \SE$, $\SE_0=\la \SF$, $\SF_1=\la^{-1}\SF$,
$\SF_0=\la^{-1}\SE$ which is motivated by the
relationship to the affine Lie algebra $\widehat{\fsl}(2)$
(this will be important for us later). Recall that
the affine Lie algebra $\widehat{\fsl}(2)$ has Chevalley
generators $E_i$, $H_i$, $F_i$, $i=0,1$. It is easy to
see that the identifications
\begin{equation}\label{evrepsl2}
\pi_{a,\la}(E_i)=\SE_i, \qquad
\pi_{a,\la}(F_i)=\SF_i, \qquad
\pi_{a,\la}(H_1)=-\pi_{a,\la}(H_0)=\SH ,
\end{equation}
define a representation of $\widehat{\fsl}(2)$ in which the
central element $H_0+H_1$ is represented by zero.

The zero curvature condition \rf{ZCC} implies that
\begin{equation}\label{Tmod}
M(\la)  =  \CP\exp\left(\int_0^{R}dx\;U_x(x,t;\lambda)\right)
 = \CP\exp\left(\int_{\mathcal{C}} ds\;
\frac{dx^{\al}}{ds} U_{\al}(\la)\right) ,
\end{equation}
for any contour $\mathcal{C}$ that can be deformed into
${\mathcal{C}}_0=\{(x,t) \st 0 \leqslant x \leqslant R \}$, preserving the start and end points.
We may, in particular,
choose the ``saw-blade'' contour ${\mathcal{C}}_N=\bigcup_{k=1}^N{\mathcal{C}}_k^+\cup
{\mathcal{C}}_k^-$, where ${\mathcal{C}}_k^\pm$ are the light-like segments
\begin{equation}
\begin{aligned}
{\mathcal{C}}_k^+&=
\big\{(k\De+u,t+u)\st 0 \leqslant u \leqslant \De/2\big\},\\
{\mathcal{C}}_k^- &=
\big\{(k\De+v,t+\De-v)\st \De/2 \leqslant v \leqslant \De\big\}
\end{aligned}
\qquad\text{($\De:=R/N$).}
\end{equation}
\begin{center}
\begin{tikzpicture}[auto,thick,scale=1.5]
\draw [dotted] (-2.75,0.5) to (-2,0.5);
\draw [->] (-2,0) to (-1,1);
\draw [->] (-1,1) to node [swap] {$\scriptstyle \mathcal{C}_{k-1}^-$} (0,0);
\draw [->] (0,0) to node {$\scriptstyle \mathcal{C}_k^+$} (1,1);
\draw [->] (1,1) to node [swap] {$\scriptstyle \mathcal{C}_k^-$} (2,0);
\draw [->] (2,0) to node {$\scriptstyle \mathcal{C}_{k+1}^+$} (3,1);
\draw [->] (3,1) to (4,0);
\draw [dotted] (4,0.5) to (4.75,0.5);
\end{tikzpicture}
\end{center}
This allows us to rewrite $M(\la)$ as
\begin{equation}\label{Tmodmod}
M(\la)  =  L_N^{-}(\la)L_N^+(\la)\cdots
L_1^{-}(\la)L_1^+(\la) ,
\end{equation}
where
\begin{equation}\label{lctransp}
L_k^{+}(\la) :=
\CP\exp\left(\int_{{\mathcal{C}}_k^{+}} dx_+\;U_{+}(\la)\right) ,\quad
L_k^{-}(\la) :=
\CP\exp\left(\int_{{\mathcal{C}}_k^{-}} dx_-\;U_{-}(\la)\right).
\end{equation}
This representation of the monodromy matrix $M(\la)$
will be a particularly useful starting point for the
quantization.

\subsubsection{Massless limit} \label{massless}

The Sinh-Gordon model is well known to be related to (m)KdV-theory.
This can be seen as follows. The massless limit $m\ra 0$ turns the Sinh-Gordon equation into the
equation for the massless free field, whose general solution is
\begin{equation}
\phi(x,t) = \phi_+(x_+)+\phi_-(x_-) .
\end{equation}
Interesting integrable structures can be preserved in the massless limit if
the limit
$m\ra 0 $ is combined with the limit $\la\ra \infty$ or $\la\ra 0$,
keeping
$\la_+=m\la$ or $\la_-=m\la^{- 1}$ fixed, respectively.

In order to discuss the limit $m\ra 0$, $\la\ra \infty$ with
$\la_+=m\la$ fixed, for example,
it will be useful to consider
the saw-blade contour ${\mathcal{C}}_N$ with $N=1$ which leads to
the factorization
\begin{equation}\label{M+factor}
M(\la;m) = N_-(\la;m) N_+(\la;m).
\end{equation}
In the limit under consideration, we see that $N_-(\la;m)$ becomes
a simple diagonal matrix while $N_+(\la;m)\ra N_+(\la_+)$, say, does not.
The main point to observe is that
\begin{equation}
T_+(\la_+):={\rm Tr}(N_+(\la_+))
\end{equation}
is a functional of $\phi_+(x_+)$ from which one may
obtain the conserved quantities of the
(m)KdV hierarchy in the asymptotic expansion for large $\la_+$.

In order to explain this statement in more detail,
let us first rewrite $T_+(\la_+)$ in a way that
makes manifest that it is a functional of the
left-moving part $\phi_+(x_+)$ only.
To this aim, let us use the gauge transformation
\begin{equation}\label{kdvgauge}
\pa_+- W_+(x_+)  :=  g^{-1}(x,t)
(\pa_+-U_+(\la_+)) g(x,t) ,
\end{equation}
with matrix  $g(x,t)$ chosen as $g(x,t):=
e^{{b}\SH(\phi_+(x_+)-\phi_-(x_-))/2}.$ The matrix $W_+(x_+)$
is found to be
\begin{equation}
W_+(x_+)\,=\,\la_+(e^{-2b\phi_+(x_+)}\SE+e^{+2b\phi_+(x_+)} \SF)\,.
\end{equation}
It will furthermore be convenient to consider
$M_+(\la_+):=(g(0,0))^{-1}N_+(\la_+)g(0,0)$.
It is then easy to show
that $M_+(\la_+)$ can be represented in terms of the
Lax connection $W_+(\la_+)$ as
{\begin{equation}
M_+(\la_+)\,=\,e^{\pi b\SH p_+}
\CP\exp\left(\int_{0}^R dx_+\;W_{+}(x_+)\right) ,
\end{equation}
where $p_+=(\phi_+(R)-\phi_+(0))/2\pi$.}
It now remains to observe that
the Hamiltonian functions  $H_n^+$
of the (m)KdV theory
are obtained from the asymptotic expansion of $\log(T_+(\la_+))$
as
\begin{equation}
\frac{1}{2\pi}\log(T_+(\la_+))  \sim  \la_++\sum_{n=1}^{\infty}  c_n^{}  H_{n}^+  \la_+^{1-2n}, \qquad \text{for $\la_+\ra\infty$.}
\end{equation}
The $c_n$ are normalization constants whose precise forms will not be needed in the following. With a proper 
choice of the $c_n$ we find, for example, 
that $H_1^+=\int_0^Rdx_+\,U(x_+)$ and
$H_2^+=\int_0^Rdx_+\,(U(x_+))^2$, where
\begin{equation}
U(x_+)\,=\,(\pa_+\phi_+(x_+))^2-\frac{1}{b}\pa_+^2\phi_+(x_+)\,.
\end{equation}
Let us also note that
the Poisson brackets following from \rf{PB} for $\phi_+$ are
\begin{equation}
\big\{\phi_+(u) ,\phi_+(v) \big\}_+  =  \frac{\pi}{2} {\rm sgn}_R(u-v) ,
\end{equation}
where ${\rm sgn}_R(u)$ is the sign function for $|u|<R/2$, continued to all real $u$ via ${\rm sgn}_R(u+R)={\rm sgn}_R(u)+1$.
The Hamiltonian functions $H_n^+$ will
then generate the (m)KdV-flows $W_+$ via
\begin{equation}
\pa_{t_n^+}W_+(t_1^+,t_2^+,\dots)  =  \big\{  H_n^+  ,  W_+(t_1^+,t_2^+,\dots)
 \big\}_+ ,
\end{equation}
where one should identify $x_+$ and $t_1^+$.

In the limit $m\ra 0$, $\la\ra 0$ with $\la_-=m\la^{-1}$ fixed,
{a similar development leads to
\begin{equation}
M_-(\la_-) =
\CP\exp\left(-\int_{0}^{-R} dx_-\;W_{-}(x_-)\right)
e^{\sb{\pi bp_-\SH}} ,
\end{equation}
with $W_-(x_-)=\la_-(e^{2b\phi_-(x_-)}\SE+e^{-2b\phi_-(x_-)} \SF)$ and $p_- = (\phi_-(0)-\phi_-(-R))/2\pi$.}
The matrix $M_-(\la_-)$ defines the integrable structure of the
right-moving part $\phi_-(x_-)$ in a way that is analogous to what
was described above. {Note that
\begin{equation}
2 \pi \brac{p_+ - p_-} = \phi_+(R)+\phi_-(-R)-\phi_+(0)-\phi_-(0) = \phi(R,0) - \phi(0,0) = 0,
\end{equation}
hence $p_+ = p_- \equiv p$.}

\subsection{Classical $\SLA{sl}{3}$ affine Toda theory} \label{ClassSL3Toda}

The classical equations of motion of the $\SLA{sl}{3}$-Toda theory are
\begin{subequations} \label{sl3EOMs}
\begin{align}
-\pa_+\pa_-\phi_1 &= 2\pi\nu e^{2b\phi_1}-
2\pi\mu e^{-b\phi_1}\cosh(\sqrt{3}b\phi_2) ,\\
-\pa_+\pa_-\phi_2 &= 2\pi\mu \sqrt{3} e^{-b\phi_1}\sinh(\sqrt{3}b\phi_2) .
\end{align}
\end{subequations}
In order to formulate the zero curvature representation of the equations,
let us introduce the Chevalley generators $E_i$, $H_i$, $F_i$, $i=0,1,2$, of the affine Lie algebra $\widehat{\fsl}(3)$. They satisfy in particular the relations
\begin{equation}
\comm{H_i}{E_j} = A_{ij}E_j ,\qquad
\comm{H_i}{F_j} = -A_{ij}F_j ,\qquad
\comm{E_i}{F_j} = \de_{ij}H_i,
\end{equation}
where $A$ is the Cartan matrix
\begin{equation}
A = \begin{pmatrix}
2 & -1 & -1 \\
-1 & 2 & -1 \\
-1 & -1 & 2
\end{pmatrix}
.
\end{equation}
Let $\widehat{\fsl}(3)_0$ be the loop algebra defined by setting the central element $H_0+H_1+H_2$ to zero.
We may then define the following $\widehat{\fsl}(3)_0$-valued fields:
\begin{subequations} \label{sl3Lax}
\begin{align}
\CU_+(\la) & = +\frac{b}{2}
\Big(H_1 \pa_+(\phi_1+\fr{1}{\sqrt{3}}\phi_2)
+H_2 \pa_+(\phi_1-\fr{1}{\sqrt{3}}\phi_2)\Big) \notag \\
&\qquad\qquad +m  \big(E_1  e^{-b(\phi_1+\sqrt{3}\phi_2)/2}+
E_2  e^{-b(\phi_1-\sqrt{3}\phi_2)/2}+
E_0  e^{b\phi_1}\big) , \\
\CU_-(\la) & = -\frac{b}{2}
\Big(H_1 \pa_-(\phi_1+\fr{1}{\sqrt{3}}\phi_2)
+H_2 \pa_-(\phi_1-\fr{1}{\sqrt{3}}\phi_2)\Big) \notag \\
&\qquad\qquad  -m  \big(F_1  e^{-b(\phi_1+\sqrt{3}\phi_2)/2}+
F_2  e^{-b(\phi_1-\sqrt{3}\phi_2)/2}+
F_0  e^{b\phi_1}\big) .
\end{align}
\end{subequations}
The zero curvature condition
\begin{equation}\label{ZCClc'}
\bigl[  \pa_+-\CU_+(\lambda)  ,  \pa_--\CU_-(\lambda)  \bigr]  =  0
\end{equation}
reproduces \rf{sl3EOMs} if $\mu=\nu=m^2/2\pi b$.
In order to get the corresponding Lax matrices, note that we could use any
representation of $\widehat{\fsl}(3)_0$. Of particular interest are the two fundamental representations realized on $\BC^3$. These may be defined by
\begin{subequations} \label{evrep3}
\begin{align}
\pi_{a,\la}(E_0) &= \la E_{31}, &
\pi_{a,\la}(H_0) &= E_{33}-E_{11}, &
\pi_{a,\la}(F_0) &= \la^{-1} E_{13}, \\
\pi_{a,\la}(E_1) &= \la E_{12}, &
\pi_{a,\la}(H_1) &= E_{11}-E_{22}, &
\pi_{a,\la}(F_1) &= \la^{-1} E_{21}, \\
\pi_{a,\la}(E_2) &= \la E_{23}, &
\pi_{a,\la}(H_2) &= E_{22}-E_{33}, &
\pi_{a,\la}(F_2) &= \la^{-1} E_{32}
\end{align}
\end{subequations}
and
\begin{subequations} \label{evrep3'}
\begin{align}
\pi_{a,\la}'(E_0) &= -\la E_{31}, &
\pi_{a,\la}'(H_0) &= E_{33}-E_{11}, &
\pi_{a,\la}'(F_0) &= -\la^{-1} E_{13}, \\
\pi_{a,\la}'(E_1) &= +\la E_{23}, &
\pi_{a,\la}'(H_1) &= E_{22}-E_{33}, &
\pi_{a,\la}'(F_1) &= +\la^{-1} E_{32}, \\
\pi_{a,\la}'(E_2) &= +\la E_{12}, &
\pi_{a,\la}'(H_2) &= E_{11}-E_{22}, &
\pi_{a,\la}'(F_2) &= +\la^{-1} E_{21},
\end{align}
\end{subequations}
respectively, where $E_{ij}$ denotes the matrix with $1$ in the $\brac{i,j}$-th entry and zero everywhere else. The resulting Lax matrices differ only by the permutation of some matrix elements and some signs.
We will, however, find interesting differences between the lattice versions of these Lax matrices when we consider discretizations in Section \ref{BoussinesqLat}.

\subsection{The fermionic $\fsl(2|1)$ affine Toda theory} \label{classSL(2|1)}

Turning our attention to the theory defined classically by
the action \rf{dualcplxShG'}, we observe an interesting feature:
The presence of fermions necessitates consideration of Lie superalgebras
for the formulation of a zero curvature condition. Let us consider the
affine Lie superalgebra $\widehat{\fsl}(2|1)$ with Cartan matrix
\begin{equation} \label{eqnCartanSL(2|1)}
A =
\begin{pmatrix}
0 & -1 & +1 \\
-1 & 2 & -1 \\
+1 & -1 & 0
\end{pmatrix}
.
\end{equation}
This superalgebra has
Chevalley generators $E_i$, $H_i$, $F_i$, $i=0,1,2$, with
$E_0$, $E_2$, $F_0$ and $F_2$ fermionic, all other generators being bosonic.
They satisfy in particular the relations
\begin{equation}\label{superS-C}
\comm{H_i}{E_j} = A_{ij}E_j ,\qquad
\comm{H_i}{F_j} = -A_{ij}F_j ,\qquad
E_i F_j -(-1)^{p_i p_j}F_jE_i = \de_{ij}H_j,
\end{equation}
in which $p_i \in \set{0,1}$ denotes the parity of $E_i$ and $F_i$.
The loop algebra $\widehat{\fsl}(2|1)_0^{}$ is again defined by setting $H_0+H_1+H_2=0$.

We now introduce a real bosonic field $\vf$ and two
complex fermionic fields $\chi_+$, $\chi_-$ (depending on both $x_+$ and $x_-$).  These fermions anticommute among themselves and also anticommute with the fermionic generators of $\AKMSA{sl}{2}{1}$.  With these, we construct
\begin{subequations}
\begin{align}
\CU_+(\la) &= -\bigl( H_1 \pa_+ \vf - 2(H_0-H_2) \chi_+ \bar\chi_+ \bigr) \notag \\
&\qquad + m \bigl(
E_1 e^{2\vf} + 2 E_2 \chi_+ e^{-\vf} + 2 E_0 \bar{\chi}_+ e^{-\vf}
\bigr) + m^2 e^{-2\vf} \acomm{E_0}{E_2}, \label{U+sl(2|1)} \\
\CU_-(\la) &= +\bigl( H_1 \pa_- \vf -2(H_0-H_2) \chi_- \bar\chi_- \bigr) \notag \\
&\qquad -m \bigl(
F_1 e^{2\vf} + 2 F_2 \chi_- e^{-\vf} + 2 F_0 \bar{\chi}_- e^{-\vf}
\bigr) + m^2 e^{-2\vf} \acomm{F_2}{F_0}. \label{U-sl(2|1)}
\end{align}
\end{subequations}
The zero curvature condition \rf{ZCClc'} then
yields the system of equations
\begin{subequations}
\begin{align} \label{bosesl(2|1)}
0&=\pa_+\pa_-\vf+\frac{m^2}{2}(e^{4\vf}-m^2e^{-4\vf})
-m^2(\chi_+\chi_-
+\bar\chi_+\bar\chi_-)e^{-2\vf} ,\\ \label{fermisl(2|1)b}
0&=\pa_+(\chi_-e^{-\vf})+(\pa_+\vf+
2\chi_+^{}\bar\chi_+^{})\chi_-^{}e^{-\vf}
- m^2\bar\chi_+e^{-3\vf} ,\\ \label{fermisl(2|1)c}
0&=\pa_-(\chi_+e^{-\vf})+(\pa_-\vf +
2\chi_-^{}\bar\chi_-^{})\chi_+^{}e^{-\vf}
+ m^2\bar\chi_-e^{-3\vf}.
\end{align}
\end{subequations}
These equations are equivalent to those following from
\rf{dualcplxShG'} once we identify
\begin{equation}
\phi=2b^{-1}\vf ,\qquad 2\pi b\nu=m^2 ,\qquad 2\pi \mu b=m^2 ,\qquad
{\psi_\pm=2\sqrt{2}b^{-1}\chi_\pm} .
\end{equation}
We take the fact that the equations of motion follow from a Lie-algebraically
defined Lax pair, similar to the ones used in the purely bosonic affine Toda
theories, as a justification for calling this theory the fermionic affine Toda theory associated with $\fsl(2|1)$.

The fundamental representation of $\widehat{\fsl}(2|1)_0$ is defined on
the vector superspace $\BC^{2|1}$ with two bosonic basis vectors $v_1$, $v_2$ and one fermionic basis vector $v_3$.  With respect to this basis, the elementary matrices $E_{13}$, $E_{23}$, $E_{31}$ and $E_{32}$
are fermionic (parity-reversing), whereas the rest of the $E_{ij}$ are bosonic.
The fundamental representation for $\widehat{\fsl}(2|1)_0^{}$ is then
\begin{subequations} \label{evrep(2|1)}
\begin{align}
\func{\pi_{a,\la}}{E_0} &= \la E_{23}, &
\func{\pi_{a,\la}}{H_0} &= E_{22}+E_{33}, &
\func{\pi_{a,\la}}{F_0} &= +\la^{-1} E_{32}, \\
\func{\pi_{a,\la}}{E_1} &= \la E_{12}, &
\func{\pi_{a,\la}}{H_1} &= E_{11}-E_{22}, &
\func{\pi_{a,\la}}{F_1} &= +\la^{-1} E_{21}, \\
\func{\pi_{a,\la}}{E_2} &= \la E_{31}, &
\func{\pi_{a,\la}}{H_2} &= -E_{11}-E_{33}, &
\func{\pi_{a,\la}}{F_2} &= -\la^{-1} E_{13}.
\end{align}
\end{subequations}
The second fundamental representation may be obtained from this by exchanging $J_0$ and $J_2$, $J = E,H,F$.
We may use the representation \eqref{evrep(2|1)} to get
conserved quantities from traces of the path-ordered integrals
of the Lax matrix $U_x:=\pi_{a,\la}(\CU_+-\CU_-)$.

\subsection{The $N=2$ Sine-Gordon model} \label{ClassN=2}

\subsubsection{Supersymmetry and equations of motion}

In order to write the equations of motion in a manifestly
supersymmetric way, let us introduce the complex superfield
$\Phi(x_+,x_-;\theta_+,\theta_{-})$,
which depends upon the additional Grassmann variables $\theta_+$, $\theta_{-}$,
together with the super-derivatives
\begin{equation}
\CD_+=\frac{\pa}{\pa \theta_+}+\theta_+\frac{\pa}{\pa x_+} ,\qquad
\CD_{-}=\frac{\pa}{\pa \theta_{-}}+\theta_{-}\frac{\pa}{\pa x_{-}} .
\end{equation}
With these definitions, we will write the classical equations of motion
for the $N=2$ Sine-Gordon model in the form
\begin{equation}\label{EOM:N=2}
\CD_-\CD_{+}\Phi  =  m^2\sinh(2\bar\Phi) ,\qquad
\CD_-\CD_{+}\bar\Phi  =  m^2\sinh(2\Phi) ,
\end{equation}
where $\bar\Phi$ denotes the complex conjugate superfield.
Written out in terms of component fields,
\begin{equation}
\Phi = \vf+\theta_+\chi_++\theta_-\chi_-+\al\theta_+\theta_- ,
\qquad
\bar\Phi = \bar\vf+\theta_+\bar\chi_++\theta_-\bar\chi_-+
\bar\al\theta_+\theta_- ,
\end{equation}
one finds the equations
\begin{subequations}\label{N=2compEOM}
\begin{align}
\pa_+\pa_-\vf &= 2m^2 \brac{2\bar\chi_+\bar\chi_-\sinh(2\bar\vf) - \bar\al\cosh(2\bar\vf)}, &
\pa_+^{}\chi_-^{} &= -2m^2 \bar\chi_+^{}\cosh(2\bar\vf) ,\\
\al &= m^2\sinh(2\bar\vf), &
\pa_-^{}\chi_+^{} &= +2m^2 \bar\chi_-^{}\cosh(2\bar\vf).
\end{align}
\end{subequations}
It is straight-forward to verify that these equations are equivalent to the
equations of motion of the $N=2$ Sine-Gordon theory if one identifies the
respective fields as
\begin{equation}\label{N=2ident}
\phi = 2b^{-1}\vf ,\qquad \psi_\pm = 2\sqrt{2}b^{-1}\chi_\pm ,\qquad
m^2 = 2\pi b \mu ,
\end{equation}
where $\phi$ is the complex combination $\phi_1+i\phi_2$.

\subsubsection{The super-Lax representation} \label{secSuperLax}

In order to construct a zero curvature representation for the
equation of motion \rf{EOM:N=2} of the  $N=2$ Sine-Gordon model,
let us consider the
affine Lie superalgebra $\widehat{\fsl}(2|2)$ with Cartan matrix
\begin{equation} \label{eqnCartanSL(2|2)}
A =
\begin{pmatrix}
0 & +1 & 0 & -1 \\
+1 & 0 & -1 & 0 \\
0 & -1 & 0 & +1 \\
-1 & 0 & +1 & 0 \\
\end{pmatrix}
.
\end{equation}
This superalgebra has fermionic Chevalley generators $E_i$, $F_i$,  $i=0,1,2,3$, and bosonic generators $H_i$ of the Cartan subalgebra,
satisfying the relations \rf{superS-C}. We will again restrict
attention to the loop algebra
$\widehat{\fsl}(2|2)_0^{}$ defined by setting
$H_0+H_1+H_2+H_3=0$.
With these definitions, let us consider the
super-zero curvature condition \cite{InKa1,InKa2}
\begin{equation}\label{S-ZCC}
\CD_+\CL_-(\la)+\CD_-\CL_+(\la)-\{ \CL_+(\la) , \CL_-(\la) \}
 = 0 ,
\end{equation}
where
\begin{subequations}
\begin{align}
\CL_+(\la) &:= -\frac{1}{2} \brac{H \CD_+ \Phi + \bar H \CD_+ \bar\Phi}
+ Z \Xi_+ + m \bigl(
E_1 e^{\Phi} + E_2 e^{\bar\Phi} - E_3 e^{-\Phi} - E_0 e^{-\bar\Phi} \bigr), \\
\CL_-(\la) &:= +\frac{1}{2} \brac{H \CD_- \Phi + \bar H \CD_- \bar\Phi}
- Z \Xi_- - m \bigl(
F_1 e^{\Phi} + F_2 e^{\bar\Phi} - F_3 e^{-\Phi} - F_0 e^{-\bar\Phi} \bigr).
\end{align}
\end{subequations}
Here, we have used the notation
$H:=H_2-H_0$, $\bar H:=H_1-H_3$ and $Z:=H_0+H_2$.
The zero curvature condition \rf{S-ZCC} implies, on top of
the equations of motion \rf{EOM:N=2}, the additional equation
\begin{equation}\label{AuxEOM}
\CD_-\Xi_+-\CD_+\Xi_- = m^2(\cosh(2\Phi)-\cosh(2\bar\Phi)).
\end{equation}
{This does not constrain the $\Xi_\pm$ uniquely.  Rather, it means that there is some freedom to choose the $\Xi_\pm$ to solve \rf{AuxEOM} without constraining $\Phi$ any further than \rf{EOM:N=2} does.  For later purposes, we note that the equation for the coefficient $\xi_\pm$ of $\theta_\pm$ in $\Xi_\pm$ is
\begin{equation}
\pa_+ \xi_- + \pa_- \xi_+ = 4m^2 \brac{\chi_+ \chi_- \cosh (2\vf) - \bar\chi_+ \bar\chi_- \cosh (2\bar\vf)} = \pa_+ (\chi_+ \bar\chi_+) + \pa_- (\chi_- \bar\chi_-),
\end{equation}
where we have also used \eqref{N=2compEOM}.  It follows that setting
\begin{equation} \label{XiChoice}
\xi_{\pm} = \chi_{\pm} \bar\chi_{\pm}
\end{equation}
is a particularly natural choice, consistent with the equations of motion \eqref{EOM:N=2}.}

In order to get the corresponding super-Lax matrices,
one may, for instance, evaluate the Lax matrices $\CL_{\pm}$ in the
fundamental representation $\pi_{a,\la}$
of $\widehat{\fsl}(2|2)_0$ which may be defined by
\begin{subequations}\label{sl(2|2)fund-a}
\begin{align}
\pi_{a,\la}(E_0) &= \la E_{41}, & \pi_{a,\la}(H_0) &= +E_{11}+E_{44}, & \pi_{a,\la}(F_0) &= +\la^{-1} E_{14}, \\
\pi_{a,\la}(E_1) &= \la E_{13}, & \pi_{a,\la}(H_1) &= -E_{11}-E_{33}, & \pi_{a,\la}(F_1) &= -\la^{-1} E_{31}, \\
\pi_{a,\la}(E_2) &= \la E_{32}, & \pi_{a,\la}(H_2) &= +E_{22}+E_{33}, & \pi_{a,\la}(F_2) &= +\la^{-1} E_{23}, \\
\pi_{a,\la}(E_3) &= \la E_{24}, & \pi_{a,\la}(H_3) &= -E_{22}-E_{44}, & \pi_{a,\la}(F_3) &= -\la^{-1} E_{42},
\label{sl(2|2)fund-d}\end{align}
\end{subequations}
There is of course another fundamental representation, but we will restrict our attention to this one in what follows.

\subsubsection{Ordinary Lax representation}

The zero curvature condition \rf{S-ZCC} implies a zero
curvature condition of the usual form \rf{ZCClc'},
where
\begin{equation}\label{LvsCL}
\CU_\pm(\la) = \CL_\pm^{(1)}(\la)-\big(\CL_\pm^{(0)}(\la)\big)^2,
\end{equation}
given the expansion $\CL_\pm(\la) = \CL_\pm^{(0)}(\la) + \theta_\pm\CL_\pm^{(1)}(\la) + \ldots$ \ Indeed, \rf{S-ZCC} guarantees the existence
of solutions to the equations
\begin{equation}\label{flatness}
(\CD_+-\CL_+(\la)) \Psi(\la) = 0 ,\qquad
(\CD_--\CL_-(\la)) \Psi(\la) = 0 .
\end{equation}
Expanding in $\theta_\pm$, one easily finds from \rf{flatness}
that the lowest component $\Psi^{(0)}(\la)$ of the superfield
$\Psi(\la)$ satisfies the equations
\begin{equation}\label{b-flatness}
(\pa_+-\CU_+(\la)) \Psi^{(0)}(\la) = 0 ,\qquad
(\pa_--\CU_-(\la)) \Psi^{(0)}(\la) = 0 ,
\end{equation}
with the $\CU_\pm(\la)$ defined in \rf{LvsCL}. The Lax matrices are
explicitly given by
\begin{subequations}\label{boseLaxsl(2|2)}
\begin{align}
\CU_+(\la) &= -\frac{1}{2}(H \pa_+\vf+\bar H \pa_+\bar\vf)
+Z \chi_+\bar\chi_+ \nonumber\\
&\quad +2m
\big(  E_1^{} \chi_+^{} e^{\vf}+E_2^{} \bar\chi_+^{} e^{\bar\vf}
+E_3^{} \chi_+^{} e^{-\vf}+E_0^{} \bar\chi_+^{} e^{-\bar\vf}\big) \nonumber \\
& \quad -m^2\big(\{E_1^{},E_2^{}\} e^{\vf+\bar\vf}-\{E_{2}^{},E_3^{}\} e^{\bar\vf-\vf}
+\{E_3^{},E_0^{}\} e^{-\vf-\bar\vf}-\{E_0^{},E_1^{}\}
e^{\vf-\bar\vf}\big) , \label{U+sl(2|2)} \\
\CU_-(\la) &= +\frac{1}{2}(H \pa_+\vf+\bar H \pa_+\bar\vf)-
Z \chi_-^{}\bar\chi_-^{}\nonumber \\
&\quad -2m
\big(  F_1^{} \chi_-^{} e^{\vf}+F_2^{} \bar\chi_-^{} e^{\bar\vf}
+F_3^{} \chi_-^{} e^{-\vf}+F_0^{} \bar\chi_-^{} e^{-\bar\vf}\big) \nonumber \\
&\quad  -m^2\big(\{F_1^{},F_2^{}\} e^{\vf+\bar\vf}-\{F_{2}^{},F_3^{}\}
 e^{\bar\vf-\vf}
+\{F_3^{},F_0^{}\} e^{-\vf-\bar\vf}-\{ F_0^{},F_1^{} \}
e^{\vf-\bar\vf}\big) . \label{U-sl(2|2)}
\end{align}
\end{subequations}
Here, we have used the choice \eqref{XiChoice} to fix the coefficients of $Z$.  Then, one finds that all of the equations which follow from \rf{ZCC} and \rf{boseLaxsl(2|2)} are implied by the equations of motion \rf{N=2compEOM}.

\section{Quantum affine algebras and integrable quantum field theories}
\label{CFTint}

We have seen that affine Lie (super)algebraic structures  underlie the classical integrability of the models of interest. It therefore seems natural to expect that the quantization of these
models will lead to some deformation of these structures.
In order to identify the precise form of this deformation, we are going to argue
that the algebraic structure behind the integrability
becomes visible through the algebra generated by the interaction terms in
the light-cone representation of the dynamics.  In order to explain this more
precisely, note that the light-cone representation of the classical dynamics admits a fairly direct quantization in which the interaction terms of the
equations of motion are realized as operators $\SQ_i$, $i=0,\dots,r$, on suitable Fock spaces.
The key observation to be made is that these operators
generate a representation of the nilpotent part $\CN_-$ of
some quantum affine (super)algebra. The existence of infinitely many local and non-local conserved quantities can then be deduced from this fact through purely algebraic methods \cite{FF1,FF2,FF3}. This gives highly non-trivial evidence for the claim that the quantum affine (super)algebra behind the integrability
is one whose nilpotent part is $\CN_-$.

This discussion is sharpened considerably
by the observation \cite{BLZ3,BHK} that the quantum monodromy matrices of the
corresponding massless models for {imaginary} $b$
can be \emph{directly} obtained from one of the
most basic objects associated with quantum affine (super)algebras,
the so-called
universal R-matrix, in a way to be described below. In the following section, we shall review and slightly generalize what is known about
these connections for the models of interest. Based on this discussion, we will try to formulate more precisely the proposed connection between quantum affine (super)algebras and the integrability of our models.

Relations between integrable quantum field theories 
and quantum affine algebras 
have also been found in \cite{BL1,BL2}. 
These works are concerned with the non-local 
conserved charges related to the appearance of solitonic excitations
in the infinite-volume scattering theory.
This does not seem to be {\it directly} related to the connections
discussed in our paper. One may observe, in particular, that the approach
of  \cite{BL1,BL2} was generalized to the N=2 Sine-Gordon model in \cite{KUY}, 
and it was found by these authors
that the quantum affine algebra associated with the 
non-local conserved charges is $\CU_{\tilde{q}}(\widehat{\fsl}(2))$ 
in this case, while
we will argue below that it is the quantum affine superalgebra
$\CU_q(\widehat{\fsl}(2|2))$ which is relevant in our context. 
Despite the apparent differences, it seems clear, however, that such appearances
of quantum affine algebras 
must be related on a deeper level. A better
understanding of this relation, 
in connection to integrable quantum field theories, seems highly desirable.

\subsection{Quantum affine algebras} \label{QuaAffAlg}

Let $\widehat{\alg{g}}$ be the (untwisted) affine Kac-Moody algebra associated to the simple Lie algebra $\alg{g}$.  We let $r$ denote the rank of $\alg{g}$ and assume, for simplicity, that all the real roots of $\widehat{\alg{g}}$ have the same length (this is the only case that will concern us).  The \qea{} $\QEA{q}{\widehat{\alg{g}}}$ may then be defined \cite{J,D} as the Hopf algebra generated by the elements $\wun$ (the unit), $E_i$, $F_i$, $K_i = q^{H_i}$ ($i = 0 , 1 , \ldots , r$), and $q^D$, subject to the following relations:
\begin{subequations} \label{eqnQEARels}
\begin{gather}
K_i E_j = q^{A_{ij}} E_j K_i, \qquad K_i F_j = q^{-A_{ij}} F_j K_i, \qquad E_i F_j - F_j E_i = \delta_{ij} \frac{K_i - K_i^{-1}}{q - q^{-1}}, \label{eqnQEARels1} \\
q^D E_i = q^{\delta_{i0}} E_i q^D, \qquad K_i K_j = K_j K_i, \qquad q^D K_i = K_i q^D, \qquad q^D F_i = q^{-\delta_{i0}} F_i q^D, \label{eqnQEARels2} \\
\sum_{n=0}^{1-A_{ij}} \brac{-1}^n \genfrac{[}{]}{0pt}{}{1-A_{ij}}{n}_q E_i^n E_j E_i^{1-A_{ij}-n} = \sum_{n=0}^{1-A_{ij}} \brac{-1}^n \genfrac{[}{]}{0pt}{}{1-A_{ij}}{n}_q F_i^n F_j F_i^{1-A_{ij}-n} = 0. \label{eqnQEASerre}\end{gather}
\end{subequations}
Here, $A$ is the Cartan matrix of $\widehat{\alg{g}}$ and we use the standard $q$-number notation
\begin{equation}
\genfrac{[}{]}{0pt}{}{m}{n}_q = \frac{\qnum{m}{q}!}{\qnum{n}{q}! \qnum{m-n}{q}!}, \qquad \qnum{n}{q}! = \qnum{n}{q} \qnum{n-1}{q} \cdots \qnum{1}{q}, \qquad \qnum{n}{q} = \frac{q^n - q^{-n}}{q - q^{-1}}.
\end{equation}
\eqnref{eqnQEASerre} is known as the Serre relations.  This is supplemented by a coproduct $\Delta$ given by
\begin{subequations} \label{eqnQEACoproduct}
\begin{align}
\func{\Delta}{E_i} &= E_i \otimes K_i + \wun \otimes E_i, & \func{\Delta}{K_i} &= K_i \otimes K_i, \\
\func{\Delta}{F_i} &= F_i \otimes \wun + K_i^{-1} \otimes F_i, & \func{\Delta}{q^D} &= q^D \otimes q^D.
\end{align}
\end{subequations}
There is also a counit and antipode, though their explicit forms are not important for us, except in noting that there exist Hopf
subalgebras $\CB_+$ and $\CB_-$ generated by the $E_i$, $K_i$, $q^D$ and the
$F_i$, $K_i$, $q^D$, respectively. These are the analogs of Borel subalgebras and we will refer to them as such.  The subalgebras $\CN_+$ and $\CN_-$ generated by the $E_i$ and the $F_i$, respectively, will be called the nilpotent subalgebras.  They are not Hopf subalgebras.

As in the classical case ($q = 1$) above, we will generally be interested in level $0$ representations.  Because of this, we will often denote a quantum affine algebra by $\QEA{q}{\affine{\alg{g}}_0}$, understanding that the linear combination of Cartan generators giving the level has been set to $0$.  As the level is dual to the derivation $D$ under the (extended) Killing form, it is therefore often also permissible to ignore $D$ in our computations.

\subsection{Universal R-matrices}\label{SSunivR}

The physical relevance of quantum
affine algebras stems from the existence \cite{D} of the so-called universal R-matrix $\CR$.
This is a formally invertible infinite sum of tensor products of algebra elements
\begin{equation} \label{eqnUnivRExp}
\mathcal{R}  = \sum_i a_i \otimes b_i, \qquad a_i, b_i \in \QEA{q}{\affine{\alg{g}}},
\end{equation}
which must satisfy three properties:
\begin{subequations} \label{eqnUnivRAxioms}
\begin{gather}
\mathcal{R} \func{\Delta}{x} = \func{\Delta^{\textup{op}}}{x} \mathcal{R} \qquad \text{for all $x \in \QEA{q}{\affine{\alg{g}}}$,} \label{eqnUnivRIntertwiner} \\
\func{\brac{\Delta \otimes \id}}{\mathcal{R}} = \mathcal{R}_{13} \mathcal{R}_{23}
\qquad \text{and} \qquad
\func{\brac{\id \otimes \Delta}}{\mathcal{R}} = \mathcal{R}_{13} \mathcal{R}_{12}. \label{eqnUnivROther}
\end{gather}
\end{subequations}
Here, $\func{\Delta^{\textup{op}}}{x}$ denotes the ``opposite'' coproduct of $\QEA{q}{\affine{\alg{g}}}$, formally defined as $\De^{\rm op}(x)=\si(\De(x))$,
where the permutation $\si$ acts as
\begin{equation}
\si(x\ot y) = y\ot x.
\end{equation}
We have also used the standard shorthand $\mathcal{R}_{12} = \sum_i a_i \otimes b_i \otimes \wun$, $\mathcal{R}_{13} = \sum_i a_i \otimes \wun \otimes b_i$ and $\mathcal{R}_{23} = \sum_i \wun \otimes a_i \otimes b_i$.

Quantum affine algebras have an abstract realisation in terms of a so-called quantum double \cite{D} which proves the existence of their universal R-matrices.  This realisation moreover shows that these R-matrices can be factored so as to isolate the contribution from the Cartan generators:
\begin{equation}\label{redRdef}
\CR = q^t \bar{\CR} , \qquad t = \sum_{i,j} \bigl( \affine{A}^{-1} \bigr)_{ij} \: H_i \otimes H_j.
\end{equation}
Here, $\affine{A}$ denotes the non-degenerate extension of the Cartan matrix to the entire Cartan subalgebra (including $D$).  This is achieved by identifying this matrix with that of the (appropriately normalised) standard invariant bilinear form on the Cartan subalgebra.
The so-called \emph{reduced R-matrix} $\bar{\CR}$ is
a formal linear combination of monomials of the form
$\BFE_I\ot \BFF_J:=E_{i_1}\cdots E_{i_k}\ot F_{j_1}\cdots
F_{j_{\ell}}$ ($I = \set{i_1 , \ldots , i_k}$, $J =
\set{j_1 , \ldots j_{\ell}}$).

It is worth noting \cite{KT} that $\CR$
is already uniquely defined by \eqref{eqnUnivRIntertwiner}
and \eqref{redRdef}. In order to get some idea why this is so,
let us first note that putting $x = K_i$ into
\eqref{eqnUnivRIntertwiner} shows that each term $\mathbf{E}_I \ot
\mathbf{F}_J$ in the expansion of the reduced R-matrix is constrained so
that the affine weight of $\mathbf{E}_I$ cancels that of $\mathbf{F}_J$.
Second, putting $x = F_i$ into \eqref{eqnUnivRIntertwiner} and using the relations
\begin{equation} \label{t-Constraint}
\brac{F_i\ot K_i^{-1}} q^t = q^t \brac{F_i \ot \wun}, \qquad
\brac{\wun \ot F_i} q^t = q^t \brac{K_i \ot F_i},
\end{equation}
we find that (for the algebras and superalgebras we are interested in)
\begin{equation} \label{Constraint}
\comm{\bar{\CR}}{F_i \ot \wun} =
\brac{K_i \ot F_i}  \,\bar{\CR} - \bar{\CR} \brac{K_i^{-1} \ot F_i}.
\end{equation}
This relation can be solved recursively by expanding $\bar\CR$ as
a formal series in the monomials $\mathbf{E}_I \ot
\mathbf{F}_J$.
In particular, it is easy to deduce that the expansion to first order is
\begin{equation} \label{redRlow}
\bar{\CR} = \wun \otimes \wun + \brac{q - q^{-1}} \sum_i \brac{E_i \otimes F_i} + \ldots
\end{equation}
We will use \rf{Constraint} repeatedly in \secDref{lattice}{lattice-gen} when we discuss lattice regularisations.

We note that a second solution to the defining properties \eqref{eqnUnivRAxioms} is given by\footnote{We thank A. Bytsko for pointing this out.}
\begin{equation} \label{DefR-}
\CR^- = \brac{\func{\si}{\CR}}^{-1}.
\end{equation}
This alternative universal R-matrix $\CR^-$ is then of the form
\begin{equation}\label{R-factor}
\CR^- = \bar\CR^- q^{-t},
\end{equation}
in which $\bar\CR^-$ is a formal series in monomials of the form $\mathbf{F}_I \ot \mathbf{E}_J$.  In order that the symmetry between the two universal R-matrices is emphasised, we shall also use the notation $\CR^+:=\CR$.
It easily follows from the defining properties \eqref{eqnUnivRAxioms} that $\CR^+$ and $\CR^{-}$ satisfy the abstract Yang-Baxter equations \begin{subequations} \label{YBE}
\begin{align}
\mathcal{R}_{12}^+  \mathcal{R}_{13}^+  \mathcal{R}_{23}^+  &=  \mathcal{R}_{23}^+  \mathcal{R}_{13}^+  \mathcal{R}_{12}^+ ,\label{YBE++}\\
\mathcal{R}_{12}^+  \mathcal{R}_{13}^-  \mathcal{R}_{23}^-  &=  \mathcal{R}_{23}^-  \mathcal{R}_{13}^-  \mathcal{R}_{12}^+ ,\label{YBE+-}\\
\mathcal{R}_{12}^-  \mathcal{R}_{13}^+  \mathcal{R}_{23}^+  &=  \mathcal{R}_{23}^+  \mathcal{R}_{13}^+  \mathcal{R}_{12}^- ,\label{YBE-+}\\
\mathcal{R}_{12}^-  \mathcal{R}_{13}^-  \mathcal{R}_{23}^-  &=  \mathcal{R}_{23}^-  \mathcal{R}_{13}^-  \mathcal{R}_{12}^- .\label{YBE--}
\end{align}
\end{subequations}

It is also useful to note that $\CR^+$ and $\CR^-$ may be
related by an anti-automorphism $\zeta$ given by
\begin{equation} \label{DefZeta}
\zeta(E_i) = F_i, \qquad \zeta(F_i) = E_i, \qquad \zeta(H_i) = H_i, \qquad \zeta(D) = D, \qquad \zeta(q) = q^{-1}.
\end{equation}
This action can be continued to tensor products via $\zeta(x\ot y)=\zeta(x)\ot\zeta(y)$.
In terms of $\zeta$, we can represent $\CR^-$ as
\begin{equation}\label{R-fromR+}
\CR^-=\zeta(\CR^+).
\end{equation}
Indeed, applying $\zeta$ to the defining property
\rf{eqnUnivRIntertwiner} shows that $\CR':=\zeta(\CR^+)$ likewise
satisfies \rf{eqnUnivRIntertwiner}.  Moreover, $\CR'$ is clearly of the form $\CR' = \bar\CR' q^{-t}$, with $\bar\CR'$ a formal series in monomials of the form $\mathbf{F}_i \ot \mathbf{E}_j$.
As $\CR^-$ is uniquely determined by these two properties
\rf{eqnUnivRIntertwiner} and \rf{R-factor}, we conclude that $\CR'=\CR^-$.

Applying appropriate representations of the Hopf algebras $\CB_{\pm}$ and $\QEA{q}{\affine{\alg{g}}}$ to \rf{YBE} results in more familiar forms of the Yang-Baxter equation.
In particular, we will frequently be constructing representations $\pi_{a,\lambda}$ ($\lambda \in \CC$) and $\pi_q$ so that we can apply $\pi_{a,\lambda} \otimes \pi_{a,\mu} \otimes \pi_q$ to \eqref{YBE}.  The resulting specialisation of \eqref{YBE++}, for example, then takes the form
\begin{equation} \label{eqnRLL}
\func{R_{12}}{\lambda , \mu} \func{L_{13}}{\lambda} \func{L_{23}}{\mu} = \func{L_{23}}{\mu} \func{L_{13}}{\lambda} \func{R_{12}}{\lambda , \mu},
\end{equation}
when we set
\begin{equation}
\func{R}{\lambda , \mu} = \func{\brac{\pi_{a,\lambda} \otimes \pi_{a,\mu}}}{\mathcal{R}^+} \qquad \text{and} \qquad \func{L}{\lambda} = \func{\brac{\pi_{a,\lambda} \otimes \pi_q}}{\mathcal{R}^+}.
\end{equation}
Note that this requires that $\pi_{a,\la}$ be a representation of $\QEA{q}{\affine{\alg{g}}}$, whereas $\pi_q$ need only be a representation of the Borel subalgebra $\CB_-$.

\subsection{Relation to the algebra of quantum monodromy matrices}

Let us now formulate the conjectured relation between our models and the representation theory of
quantum affine (super)algebras on a somewhat abstract level. Recall that the key objects
used to establish the classical integrability of our models were the monodromy matrices $M_a(\la)$
which can be defined for each choice of representation $\pi_{a,\la}$ of the relevant
loop algebra $\hat\fg_0$. We conjecture that the quantization of the models produces operator-valued
matrices $\SM_a(\la)$ which satisfy algebraic relations of the following general form
\begin{equation}\label{RMMMMR}
R_{ab}(\la,\mu) \bigl(\SM_a(\la)\otimes {\rm I}\bigr) \bigl({\rm I}\otimes \SM_b(\mu)\bigr)
 = \bigl({\rm I}\otimes \SM_b(\mu)\bigr) \bigl(\SM_a(\la)\otimes {\rm I}\bigr)
R_{ab}(\la,\mu)  .
\end{equation}
In order to write the relation compactly, we consider $\SM_a(\la)$ and $\SM_b(\la)$ as endomorphisms of
corresponding representation spaces $\CV_a$ and $\CV_b$, so that \rf{RMMMMR} may be read as
a relation between operator-valued endomorphisms of $\CV_a\ot\CV_b$. The entries of the matrix $R_{ab}(\la,\mu)$
in \rf{RMMMMR} are not operator-valued --- they play the role of structure constants in these algebraic relations.

The main point here is that the so-called R-matrix $R_{ab}(\la,\mu)\colon\CV_a\ot\CV_b\ra\CV_a\ot\CV_b$ is related
to the universal R-matrix $\CR$ of the affine Lie (super)algebra $\QEA{q}{\affine{\alg{g}}_0}$ deforming $\hat\fg_0$ via
\begin{equation}
R_{ab}(\lambda,\mu):=\bigl(\pi_{a,\lambda}\otimes\pi_{b,\mu}\bigr)(\CR) .
\end{equation}
In the quantum case, the representations $\pi_{a,\la}$ and $\pi_{b,\la}$ should therefore be deformations of the representations
defining the corresponding classical Lax matrices $M_a(\la)$ and $M_b(\la)$, respectively.

In order to get the quantum counterparts of the integrals of motion, it is then natural to consider
traces of the monodromy matrices, taken over the auxiliary spaces $\CV_a$:
\begin{equation}\label{maindef}
\ST_a(\lambda) := {\rm Tr}^{}_{\CV_a}(\SM_a(\lambda))  .
\end{equation}
The mutual commutativity,
\begin{equation}\label{TT}
\bigl[ \ST_a(\lambda) , \ST_b(\mu) \bigr] = 0 ,
\end{equation}
for all allowed values of $\la$ and $\mu$, and all
admissible choices of representations $\pi_{a,\la}$ and $\pi_{b,\mu}$, then
follows easily by taking the trace of \eqref{RMMMMR}
over $\CV_a\ot\CV_b$. By varying the choice of representation $\pi_{a,\la}$,
one may generate a large family $\CI$ of mutually commuting operators. We
expect that the Hamiltonians $\SH$ of our models can be constructed from
the elements of $\CI$.

Proposing the existence of operator-valued matrices $\SM_a(\la)$ which satisfy
the relations \rf{RMMMMR} may seem bold in a quantum field-theoretical
context, because of the possibility that modifications to \rf{RMMMMR} will be required by renormalization.
However, in the case of imaginary $b$, there exist \cite{BLZ1,BLZ3}
direct quantum field-theoretical
constructions of monodromy matrices $\SM_a(\la)$
satisfying \rf{RMMMMR}, as we will shortly
review. For real values of
$b$, there is strong evidence for one
of the most important consequences of the existence of
the $\SM_a(\la)$, namely the functional relations
satisfied by the eigenvalues of the transfer matrices
$\ST_a(\la)$ \cite{ByT,T,ByT2}.

\subsection{Light-cone representation for integrable quantum field theory}

A somewhat unconventional picture for integrable quantum field theory models can be obtained by taking the piecewise light-like saw-blade contour
${\mathcal C}_1$ from Section
\ref{lc} as an initial-value surface. For notational simplicity, let
us begin with the case of the Sinh-Gordon model, the generalization
to the other (bosonic) affine Toda theories being straight-forward
(we briefly discuss the $\SLA{sl}{3}$ case in \secref{QuantumSL3}).

\subsubsection{Classical dynamics in the light-cone representation}

In the light-cone picture for the classical dynamics, one takes
the values of the field $\phi$ on the two light-like segments of
${\mathcal C}_1$,
\begin{equation}
\phi^+(2u)=\phi(u,u) \quad \text{and} \quad
\phi^-(2v)=\phi(\tfrac{R}{2}-v,\tfrac{R}{2}+v), \qquad  0 \leqslant u,v \leqslant \tfrac{R}{2},
\end{equation}
as initial values
for the time-evolution from which $\phi(x,t)$
can be found for all $x$ and $t$
by solving the equations of motion.  The dynamics may still be represented in the Hamiltonian form by using the Poisson structure
\begin{equation}\label{PBlc}
\{ \phi^+(u) , \phi^+(u') \} = \frac{\pi}{2}{\rm sgn}_R(u-u') ,\qquad
\{ \phi^-(v) , \phi^-(v') \} = \frac{\pi}{2}{\rm sgn}_R(v-v')
\end{equation}
on the light-cone data $\phi^+$ and $\phi^-$ (brackets between $\phi^+$ and $\phi^-$ are zero). The Hamiltonians $H_+$ and $H_-$ which generate the time
evolution in the two light-like directions may be found by expanding the trace of the monodromy matrix $M(\la)$ around the singular points $\la=\infty$ and $\la=0$, respectively.
One finds, for example, that
\begin{equation}\label{H_-def}
H_- = \int_{0}^R\frac{dx_-}{4\pi}
\Big((\pa_-\phi^-)^2-\frac{1}{b}\pa_-^2\phi^-\Big)
+{\mu}\int_0^R{dx_+}\;2\cosh (2b\phi^+) .
\end{equation}
Using the representation \rf{M+factor}, it is easy to see
that the interaction terms in $H_-$ are directly related to
the matrix elements of $V_+(\la;m)$.
The equation of motion can now be represented in the Hamiltonian form as
\begin{equation}
\pa_-(\pa_+\phi) = \{ H_- , \pa_+\phi \} = -4\pi b\mu\sinh (2b\phi) .
\end{equation}
The same equation of motion is found by exchanging the roles
of $\phi^+$ and $\phi^-$, of course.

\subsubsection{Quantization} \label{BosonConventions}

The Poisson brackets \rf{PBlc} are those a massless free field.
The quantization is therefore standard. Let us write the expansion
of $\phi^{\pm}(x_{\pm})$ into Fourier modes in the form
\begin{equation}\label{modes}
\phi^\pm(x_{\pm}) = \sq + \frac{2\pi}{R}\spp x_{\pm} +
\phi^{\pm}_<(x_{\pm})+\phi^{\pm}_>(x_{\pm}) ,
\end{equation}
where
\begin{equation}
\phi^{\pm}_<(x_{\pm}) = \sum_{n<0} \frac{i}{n}
\sa_n^{\pm}
e^{-2\pi inx_{\pm} / R} ,\quad
\phi^{\pm}_>(x_{\pm}) =\sum_{n>0} \frac{i}{n}
\sa_n^{\pm}
e^{-2\pi inx_{\pm} / R} .
\end{equation}
The modes $\sa_n^\eps$ ($\eps = \pm$), $\sq$ and $\spp$ are required to satisfy the canonical
commutation relations
\begin{equation}\label{CCR}
\comm{\sq}{\spp} = \frac{i}{2}, \qquad
\comm{\sa_m^{\ep}}{\sa_n^{\ep'}} = \frac{1}{2} m \de_{m+n,0} \de_{\ep\ep'}.
\end{equation}
Quantum analogs of the exponential functions $e^{2\al\phi^\pm}$ are then constructed by normal ordering:
\begin{equation}\label{normord}
\normord{e^{2\al\phi^\pm(x_\pm)}} :=
\exp(2\al\phi_<^\pm(x_\pm))
e^{2\al (\sq+2\pi\spp x_\pm/R)} \exp(2\al\phi_>^\pm(x_\pm)) .
\end{equation}
The quantum Hamiltonians $\SH_+$ and $\SH_-$
corresponding to $H_+$ and $H_-$, respectively, will similarly be defined
by normal ordering $(\pa_\pm\phi^\pm)^2$ and $\cosh (2b\phi^\pm)$.

\subsubsection{Conserved quantities}

The quantum equation of motion for an
observable $\SO$ built from $\pa_+\phi^+(x_+)$ can then be represented in the form
\begin{equation}\label{q-EOM}
-i\pa_-\SO = \big[ \SH_- , \SO ] = \mu\big[ \SQ_0^++\SQ_1^+ , \SO \big] ,
\end{equation}
where the operators
\begin{equation}
\SQ_i^+ = \int_0^R dx \; \SV_i(x), \qquad
\SV_0(x) = \normord{e^{+2b\phi_+(x)}}, \quad
\SV_1(x) = \normord{e^{-2b\phi_+(x)}},
\end{equation}
are called screening charges. We see that finding conserved quantities is reduced to a purely algebraic problem:
\emph{Find all operators $\SO$} (\emph{built from $\pa_+\phi^+(x_+)$}) \emph{which commute with the screening charges $\SQ_0^+$, $\SQ_1^+$}. Note that we require the commutativity of $\SO$ with both $\SQ_0^+$ and $\SQ_1^+$ independently.  This is motivated by the fact that we could easily generalize
the right hand side of \rf{q-EOM} to $\big[ \mu\SQ_0^++\nu\SQ_1^+ , \SO \big]$
by a shift of the zero mode $\sq$.

This problem was studied in \cite{FF1,FF2,FF3}.
A key point underlying the approach used in
these references is the fact that the operators
$\SQ_i^+$, $i=0,1$, satisfy the relations 
\begin{equation} \label{eqnSGSerre}
(\SQ_i^+)^3 \SQ_j^+ - \qnum{3}{q} (\SQ_i^+)^2 \SQ_j^+ \SQ_i^+ +
\qnum{3}{q} \SQ_i^+ \SQ_j^+ (\SQ_i^+)^2 - \SQ_j^+ (\SQ_i^+)^3 = 0,
\end{equation}
with $q = e^{-\ii \pi b^2}$.  The validity of these relations
was first shown in a related context in \cite{BMP}.
It can be checked by direct calculation --- we detail
the method in \appref{appScreenAlg}.
The relations \rf{eqnSGSerre} can be identified with
the Serre-relations \rf{eqnQEASerre} of the quantum affine
algebra $\CU_q(\widehat{\fsl}(2))$. They imply that the operators
$\SQ_i^+$, $i=0,1$, generate a representation of the nilpotent
part $\CN_-$ of $\CU_q(\widehat{\fsl}(2))$.
Based on this observation, it is possible to prove
that there exist infinitely many local \cite{FF1,FF2}
and non-local \cite{FF3} conserved operators $\SO$.

These results represent a first basic link between the
integrability of the Sinh-Gordon quantum field theory and
quantum affine algebras. The main lesson that we wish to extract
from this example is that
there is a direct relation
between the algebra generated by the operators $\SQ_i^+$, describing
the perturbations in the light-cone representation, and the
integrability of the theory. The fact that the perturbing operators
$\SQ_i^+$ generate a representation of the nilpotent subalgebra
of some quantum affine algebra implies the existence of
infinitely many conserved quantities.

\subsection{Quantization of the monodromy matrices}

The connection between quantum affine algebras and integrability
can be strengthened significantly by considering the quantization
of the monodromy matrices in the massless limits.
Following \cite{BLZ1,BLZ3} we shall, in the
following, describe the quantization of the monodromy matrices
of the (m)KdV theory for $b$ imaginary together with its link to the 
representation theory of the quantum affine algebra $\CU_q(\widehat{\fsl}(2))$.

\subsubsection{Quantization of (m)KdV theory}

In the regime where $b=i\beta$, $\beta\in\BR$, it is
straight-forward to construct
the quantized counterpart $\SM_+(\la_+)$ of the
monodromy matrix $M_+(\la_+)$ as \cite{BLZ1}
\begin{equation}\label{q-monod}
\SM_+(\la_+) = e^{\pi b \spp \SH}
\CP\exp\left(\int_0^{R}dx_+\;\SW_+(x_+)\right) ,
\end{equation}
where
\begin{equation}
\SW_+(x;\lambda) = \la_+
\begin{pmatrix}
0 & \normord{e^{-2b \phi_+(x)}} \\
\normord{e^{2b\phi_+(x)}} & 0
\end{pmatrix}
.
\end{equation}
$\SM_+(\la)$ is \emph{a priori} defined as a formal power series in $\la$, whose
coefficients are represented by ordered integrals over products of
normally-ordered exponential fields. These integrals converge if $\beta^2<\frac{1}{2}$ and it can be shown \cite{BLZ3} that the summation
over powers of $\la$ is convergent in this case.


It was shown in \cite{BLZ3} that the commutation
relations satisfied by the matrix elements of $\SM_+(\la)$ can be
written as the exchange relations
\begin{equation}\label{RMMMMR'}
R(\la/\mu) \bigl(\SM_+(\la)\otimes {\rm I}\bigr) \bigl({\rm I}\otimes \SM_+(\mu)\bigr) = \bigl({\rm I}\otimes \SM_+(\mu)\bigr)
\bigl(\SM_+(\la)\otimes {\rm I}\bigr) R(\la/\mu) ,
\end{equation}
with matrix $R(\la)$ given by
\begin{equation}\label{Rsl2}
R(\la) =
\begin{pmatrix}
q^{-1}\la-q\la^{-1} & 0 & 0 & 0 \\
 0 & \la-\la^{-1} & q^{-1}-q & 0 \\
 0 & q^{-1}-q & \la-\la^{-1} & 0 \\
 0 & 0 & 0 & q^{-1}\la-q\la^{-1}
\end{pmatrix}
 .
\end{equation}
The commutation relations \rf{RMMMMR'} represent a natural
quantization of the Poisson structure \rf{FPBR}.
It follows immediately
from \rf{RMMMMR'} that the operators $\ST_+(\la):={\rm Tr}\bigl(\SM_+(\la)\bigr)$ commute
for arbitrary values of the spectral parameter:
\begin{equation}
\big[ \ST_+(\la) , \ST_+(\mu) \big] = 0 \qquad \text{for all $\la,\mu\in\BC$.}
\end{equation}
The family of operators $\ST_+(\la)$ generates the
algebra of quantum integrals of motion
in the quantized (m)KdV-theory.

The quantized counterpart $\SM_-(\la_-)$ of the
monodromy matrix $M_-(\la_-)$ can likewise be constructed as
\begin{equation}\label{q-monod-}
{\SM_-(\la_-)  =
\CP\exp\left(\int_0^{-R}dx_-\;\SW_-(x_-)\right)
e^{\pi b \spp \SH}} ,
\end{equation}
where
\begin{equation}
\SW_-(x;\lambda) =
-\la_-\begin{pmatrix}
0 & \normord{e^{+2b \phi_-(x)}} \\
\normord{e^{-2b\phi_-(x)}} & 0
\end{pmatrix}.
\end{equation}
The quantum monodromy matrix $\SM_-(\la_-)$ defines a second copy of the
quantum (m)KdV-theory which may be associated to the second chiral half of the massless free field.

\subsubsection{Representation-theoretic interpretation of the
monodromy matrices}\label{repKdV}
\label{MvsRsl2}

A beautiful relationship between the quantization of (m)KdV theory and
the representation theory of the \qea{} $\QEA{q}{\AKMA{sl}{2}}$ was found in  \cite{BLZ3}
and proven in \cite{BHK}. {It asserts the equality of $\SM_+(\la_+)$ with the
evaluation of the universal R-matrix $\CR^+$ in certain
representations $\pi_{a,\la_+}$ and $\pi_q^+$ of the Borel subalgebras 
$\CB_+$ and $\CB_-$ of the \qea{} $\QEA{q}{\AKMA{sl}{2}_0}$.}
For the representation $\pi_{a,\la_+}$, we may take the representation
defined in \rf{evrepsl2}, which may be checked
to define a representation of $\CU_q(\widehat{\fsl}(2)_0)$ for all values
of $q$. For $\pi_q^+$, we shall take
\begin{subequations} \label{piqdef}
\begin{align}
\pi_q^+(H_0) &= -2 \ii \spp / b, &
\pi_q^+(F_0) &= \tau_q^{-1} \SQ_0^+, \\
\pi_q^+(H_1) &= +2\ii \spp/b, &
\pi_q^+(F_1) &= \tau_q^{-1} \SQ_1^+,
\end{align}
\end{subequations}
where $\tau_q:=q-q^{-1}$.
It follows from \rf{eqnSGSerre} and straight-forward calculation
that \rf{piqdef} indeed defines a representation of
the Borel subalgebra $\CB_-$ of $\CU_q\bigl(\AKMA{sl}{2}_0\bigr)$.
The observation  of \cite{BLZ3} can then
be formulated as the assertion that the monodromy
matrix defined in \eqref{q-monod} is equal to
\begin{equation}\label{MfromR}
\SM_+(\la_+) = (\pi_{a,\la_+}\otimes\pi_q^+)(\CR^+) ,
\end{equation}
where $\pi_{a,\la_+}$ and $\pi_q^+$ are the representations
defined in \rf{evrepsl2} and \rf{piqdef}, respectively.
{In order to prepare for the comparison with the case of $\SM_-(\la_-)$ we have
included the proof of \rf{MfromR} (following \cite{BHK})
in Appendix \ref{MvsR-proof}.}

With very similar arguments {(see Appendix \ref{MvsR-proof})}, one may show that
\begin{equation}\label{M-fromR}
\SM_-(\la_-) = (\pi_{a,\la_-}\ot\pi_q^-)(\CR^-) ,
\end{equation}
where the representation
$\pi_q^-$ of $\CB_+$ is defined by
\begin{subequations} \label{piqdef-}
\begin{align}
\pi_q^-(H_0) &= +2i\spp/b , &
\pi_q^-(E_0) &= \tau_q^{-1} \int_0^{-R}dx_-\; \normord{e^{+2b\phi_-(x_-)}}, \\
\pi_{q}^-(H_1) &= -2i\spp/b, &
\pi_q^-(E_1) &= \tau_q^{-1} \int_0^{-R}dx_-\; \normord{e^{-2b\phi_-(x_-)}}.
\label{piqdef-b}\end{align}
\end{subequations}
It follows from \rf{YBE+-} and \rf{M-fromR} that $\SM_-(\la_-)$
satisfies relations of the form \rf{RMMMMR'} with the same
matrix $R(\la)$.

The proof of \rf{M-fromR} described in Appendix \ref{MvsR-proof}
shows that the different orientations in the integrations appearing in the
definitions \rf{q-monod} and \rf{q-monod-} of
$\SM_+(\la_+)$ and $\SM_-(\la_-)$, respectively,
are precisely accounted for by replacing $\CR^+$ in \eqref{MfromR} by $\CR^-$ in \rf{M-fromR}.
It seems quite remarkable that the two chiralities of the massless free field
are naturally related to the two universal R-matrices discussed in
Section \ref{SSunivR}. This will become even clearer in our discussion of the
lattice regularization below (\secref{ShGLatticeReps}).

\subsection{$\SLA{sl}{3}$ affine Toda theory} \label{QuantumSL3}

This story generalizes fairly easily to the affine Toda models of higher rank.
As an example, let us discuss the case of the affine Toda theory associated to
$\SLA{sl}{3}$. The integrable structure of the massless limit is related to the Boussinesq equation.

\subsubsection{Conserved quantities in the light-cone representation}

The quantization of this theory in the light-cone representation
can be performed along the same lines as
described above. We introduce chiral free fields $\phi_1^\pm$ and $\phi_2^\pm$
with mode expansions of the same form as \rf{modes}. The modes
of $\phi_i^\pm$ are required to satisfy commutation relations obtained from
\rf{modes} by the obvious replacements. Out of the $\phi_i^+$,
one may then construct the vertex operators
\begin{equation} \label{SL3VOs}
\SV_0(x) = \normord{e^{2b\phi_1^+(x)}}, \qquad
\begin{aligned}
\SV_1(x) &= \normord{e^{-b(\phi_1^+(x)+\sqrt{3}\phi_2^+(x))}}, \\
\SV_2(x) &= \normord{e^{-b(\phi_1^+(x)-\sqrt{3}\phi_2^+(x))}}.
\end{aligned}
\end{equation}
From these vertex operators, let us define the screening charges
\begin{equation}\label{SQsl3}
\SQ_i^+  =  \int_0^{R} dx\;\SV_i(x) .
\end{equation}
Using once more the technique described
in \appref{appScreenAlg}, these operators may be checked to 
satisfy the relations \cite{BMP}
\begin{equation} \label{qSerreSL3}
(\SQ_i^+)^2 \SQ_j^+ - \qnum{2}{q} \SQ_i^+ \SQ_j^+ \SQ_i^+ + \SQ_j^+(\SQ_i^+)_{}^2 = 0
\qquad \text{($i \neq j$),}
\end{equation}
again with $q = e^{-\ii \pi b^2}$. As before, it now follows from the results
of \cite{FF1,FF2,FF3} that there exist infinitely many local
and non-local conserved quantities.

\subsubsection{Quantum Boussinesq theory}\label{secQBouss}

The quantization of this theory \cite{BHK} leads to the monodromy matrix
\begin{equation}\label{q-monodsl3}
\SM_+(\la_+) = e^{\pi b(\mathbf{H},\mathbf{P})}
\CP\exp\left(\la_+\int_0^{R}dx\;(\SE_1\SV_{1}(x)+\SE_2\SV_2(x)+
\SE_0\SV_0(x))\right) ,
\end{equation}
where $(\mathbf{H},\mathbf{P})=(\spp_1+\spp_2/\sqrt{3})\SH_1+(\spp_1-\spp_2/\sqrt{3})\SH_2$, the $\SV_i$ were given in \eqnref{SL3VOs} and the $\SE_i$ in \eqnref{evrep3}. Our aim is to relate this monodromy matrix to the representation theory of a quantum affine algebra, as we did for quantum KdV theory in Section \ref{repKdV}.

We define the following representation of the Borel subalgebra $\CB_- \subset \QEA{q}{\AKMA{sl}{3}_0}$:
\begin{subequations} \label{piqdef3}
\begin{align}
\pi_q^+(H_0) &= -2i\spp_1/b, &
\pi_q^+(F_0) &= \tau_q^{-1} \SQ_0^+, \\
\pi_q^+(H_1) &= i(\spp_1+\sqrt{3}\spp_2)/b, &
\pi_q^+(F_1) &= \tau_q^{-1} \SQ_1^+, \\
\pi_q^+(H_2) &= i(\spp_1-\sqrt{3}\spp_2)/b, &
\pi_q^+(F_2) &= \tau_q^{-1} \SQ_2^+.
\end{align}
\end{subequations}
The arguments described in Section \ref{MvsRsl2} can now be used
to show that
\begin{equation}
\SM_+(\la) = (\pi_{a,\la}\otimes\pi_q^+)(\CR) ,
\end{equation}
with $\pi_{a,\la}$ and $\pi_q$ being the representations defined
in \rf{evrep3} and \rf{piqdef3}, respectively.

It follows in particular, from the abstract Yang-Baxter relation
\rf{YBE} satisfied by $\CR$, that
the operator-valued matrix $\SM_+(\la)$
satisfies Yang-Baxter type
relations of the form \rf{RMMMMR'}
with matrix $R$ given by
\begin{equation}\label{RfromR}
R(\la,\mu)  =  (\pi_{a,\la}\otimes\pi_{a,\mu})(\CR)  ,
\end{equation}
up to an irrelevant scalar factor $f(\la,\mu)$. Explicitly, this R-matrix has the form
\begin{equation} \label{sl3FundR}
\func{R}{\la,\mu} = \sum_{i,j=1}^{3} \rho_{ij}(\la,\mu) E_{ii}\ot E_{jj}
+\sum_{i,j=1}^{3}\si_{ij}(\la,\mu) E_{ij}\ot E_{ji},
\end{equation}
where $\rho_{ij}$ and $\si_{ij}$ are the $\brac{i,j}$-th
entries of the matrices
\begin{subequations} \label{Rsl3}
\begin{align}
\rho &=
\begin{pmatrix}
\la^3 q^{-1} - \mu^3 q & \la^3 - \mu^3 & \la^3 - \mu^3 \\
\la^3 - \mu^3 & \la^3 q^{-1} - \mu^3 q & \la^3 - \mu^3 \\
\la^3 - \mu^3 & \la^3 - \mu^3 & \la^3 q - \mu^3 q^{-1}
\end{pmatrix}
 ,\\
\si &= -\la \mu \brac{q-q^{-1}}
\begin{pmatrix}
0 & \mu & \la \\
\la & 0 & \mu \\
\mu & \la & 0
\end{pmatrix}
 .
\end{align}
\end{subequations}
As before, one may deduce the commutativity of the
integrals of motion of the quantized Boussinesq theory from the
Yang-Baxter type
relations  \rf{RMMMMR'}.
The modifications necessary to construct $\SM_-(\la)$ are clear.  It also satisfies the relations \eqref{RMMMMR'} with the R-matrix \eqref{sl3FundR}.

\section{Models related to quantum affine superalgebras}\label{supermodels}

Let us now discuss the modifications to the formalism of \secref{CFTint} that are necessary to treat the cases related to quantum affine superalgebras.

\subsection{Quantum affine superalgebras} \label{QuaAffSup}

As we saw in \secref{QuaAffAlg}, the defining relations \eqref{eqnQEARels} of a quantum affine algebra amount to a $q$-deformation of the presentation of the corresponding affine Kac-Moody algebra in the Chevalley basis $E_i$, $F_i$, $K_i = q^{H_i}$, $q^D$, including in particular, the Serre relations.  The definition of quantum affine superalgebras precisely mimics this deformation.  However, the analogs of the Serre relations for superalgebras are significantly more complicated than (and not nearly as well understood as) their bosonic counterparts.  Indeed, there still seems to be some controversy over the completeness of superalgebra Serre relations \cite{Z}.  One complicating factor is that the Dynkin diagram of a superalgebra need not be unique, leading to a finite number of different presentations and (potentially) a finite number of different deformations.  We refer to Yamane \cite{Y2,Y3} for these Serre relations and their $q$-deformations --- as they do not seem to admit an obvious general form, we will only give them as needed.  A second complication is that certain Lie superalgebras require two derivations.  We shall defer a discussion of this point until its consideration becomes necessary (\secref{secGrading}).

Aside from the Serre relations, the defining relations and Hopf-algebraic structure of a quantum affine superalgebra $\QEA{q}{\affine{\alg{g}}}$ (assumed for simplicity to derive from a superalgebra $\affine{\alg{g}}$ whose real roots all have the same length) are very similar to their bosonic counterparts.  Indeed, the only change at this level is that the commutator of $E_i$ and $F_j$ is replaced by
\begin{equation} \label{QESARels}
E_i F_j - \brac{-1}^{p_i p_j} F_j E_i = \delta_{ij} \frac{K_i - K_i^{-1}}{q - q^{-1}},
\end{equation}
where $A$ is a Cartan matrix of the affine superalgebra $\affine{\alg{g}}$ and $p_i = \func{p}{E_i} = \func{p}{F_i} \in \set{0,1}$ denotes the parity, even or odd (bosonic or fermionic), of the elements $E_i$ and $F_i$.  The Cartan elements $K_i$, $q^D$ are always even.

It is convenient for a compact presentation of the Serre relations to introduce the following notation.  Define the graded $q$-commutator by
\begin{equation}
[ x , y ]_q := xy-(-1)^{p(x)p(y)}q\:yx ,\qquad [ x , y ] := [ x , y ]_1.
\end{equation}
For $q=1$, this is the usual graded commutator.  From now on, $[ x , y ]$ will
denote the anticommutator if both $x$ and $y$ are fermionic.  

The parity of the generators is particularly important when considering the coproduct of a quantum affine superalgebra.  Let us first introduce the graded tensor product $\otimes_s$ which satisfies
\begin{equation}\label{gradedtensor}
(x_1\ot_s x_2)(y_1\ot_s y_2) = (-1)^{p(x_2)p(y_1)}(x_1y_1\ot_s x_2y_2) .
\end{equation}
The superalgebra coproduct is then simply \eqref{eqnQEACoproduct} with $\ot$ replaced by $\ot_s$.  With a suitable counit and antipode (which we will not need), the quantum affine superalgebra $\QEA{q}{\affine{\alg{g}}}$ becomes a Hopf superalgebra.  As before, we have Hopf subalgebras $\CB_+$ and $\CB_-$ which are generated by the $E_i$, $K_i$, $q^D$ and the $F_i$, $K_i$, $q^D$, respectively, and non-Hopf subalgebras $\CN_+$ and $\CN_-$ which are generated by the $E_i$ and $F_i$, respectively.  We will again refer to these as Borel subalgebras and nilpotent subalgebras, as appropriate.

Let us also generalize the notation $\De^{\rm op}$ to superalgebras via
\begin{equation}
\De^{\rm op}=\si\circ\De ,\qquad \si(x\ot_s y):=(-1)^{p(x)p(y)}y\ot_s x .
\end{equation}
The universal R-matrix $\CR^+$ of a quantum affine superalgebra $\QEA{q}{\affine{\alg{g}}}$ is then defined as an invertible element of the form $\sum_i a_i \otimes_s b_i$, $a_i \in \CB_+$, $b_i \in \CB_-$, that satisfies the standard universal R-matrix axioms \eqref{eqnUnivRAxioms} but with $\ot$ replaced by $\ot_s$.
The existence and uniqueness of the universal R-matrix was shown for the quantum affine superalgebras of interest to us in \cite{Y1}.   As before, this implies abstract Yang-Baxter equations identical to \eqref{YBE}.  \eqnDref{redRdef}{Constraint} are also valid for these superalgebras (with $\ot$ replaced by $\ot_s$).  \eqnref{redRlow} generalises, however, to
\begin{equation} \label{super-redRlow}
\bar{\CR}^+ = \wun \otimes_s \wun + \brac{q - q^{-1}} \sum_i \brac{-1}^{p_i} \brac{E_i \ot_s F_i} + \ldots
\end{equation}

The alternative universal R-matrix $\CR^-$ is again defined as in \eqref{DefR-}.  It may also be related to $\CR^+$ by an anti-automorphism $\zeta$ which is defined as in \eqref{DefZeta}, but with one small modification:  In order that $\zeta$ continues to define an anti-automorphism on tensor products, consistency
with \rf{gradedtensor} requires us to set
\begin{equation}\label{gradedzeta}
\zeta(x\ot_s y)=(-1)^{p(x)p(y)}\zeta(x)\ot_s \zeta(y).
\end{equation}
With this modification, $\CR^-=\zeta(\CR^+)$ as before.

\subsection{$N=2$ super Sine-Gordon model}

Our next aim will be to determine the algebraic structure
underlying the integrability of the $N=2$ Sine-Gordon model.
An interesting new feature arises
when we try to follow the path described in the previous section.
There, we observed a link between the generators
of the nilpotent parts of certain
quantum affine algebras and the interaction terms in the
light-cone representation.
In this case, we have two options to consider: According to our discussion in
Section \ref{models}, we could either take the interaction terms manifest
in the classical action \rf{dualcplxShG'} or those appearing in the
representation as a perturbed free field \rf{Ssl(2|1)}.
We will work with the second of these options.
We shall observe that these operators satisfy the Serre relations
of the affine superalgebra $\CU_q(\widehat{\fsl}(2|2))$.

Based on this observation, one may try to define quantum monodromy
matrices by evaluating the universal R-matrix of
$\CU_q(\widehat{\fsl}(2|2))$ in appropriate representations.
In order to establish the connection with the $N=2$ Sine-Gordon model,
we will then verify that the classical limit of
these monodromy matrices correctly reproduces the
the integrable structure of the massless limit of the
$N=2$ Sine-Gordon model. This turns out to
be more involved than in the previous cases.

\subsubsection{Appearance of the quantum affine superalgebra
$\CU_q(\widehat{\fsl}(2|2))$}

Following the path described in the previous section leads us to consider
four screening charges, constructed as
\begin{equation}
\SQ_i^+(x)  =  \int_0^{R}dx\;\SV_i(x) ,\qquad i=0,1,2,3,
\end{equation}
with
\begin{equation}
\begin{aligned}
\SV_0(x) &= \bar{\psi}_+(x) \normord{e^{-b(\phi_1^+(x)-i\phi_2^+(x))}}, \\
\SV_2(x) &= \bar{\psi}_+(x) \normord{e^{+b(\phi_1^+(x)-i\phi_2^+(x))}},
\end{aligned}
\qquad
\begin{aligned}
\SV_1(x) &= \psi_+(x) \normord{e^{+b(\phi_1^+(x)+i\phi_2^+(x))}}, \\
\SV_3(x) &= \psi_+(x) \normord{e^{-b(\phi_1^+(x)+i\phi_2^+(x))}}.
\end{aligned}
\end{equation}
We find that these screening charges
satisfy, in particular, the following relations:
\begin{subequations}
\begin{gather}
(\SQ_i^+)^2 = 0, \qquad \SQ_i^+ \SQ_{i+2}^+ + \SQ_{i+2}^+ \SQ_i^+ = 0, \label{QSerresl(2|2)a} \\
\SQ_{i-1,i,i+1,i}^+ - \SQ_{i+1,i,i-1,i}^+ + \qnum{2}{q} \SQ_{i,i-1,i+1,i}^+ - \SQ_{i,i+1,i,i-1}^+ + \SQ_{i,i-1,i,i+1}^+ = 0. \label{QSerresl(2|2)b}
\end{gather}
\end{subequations}
Here, $q = e^{-\ii \pi b^2}$, $i \in \ZZ_4$, and we have used the
shorthand $\SQ_{ij \cdots k}^+ = \SQ_i^+ \SQ_j^+ \cdots \SQ_k^+$.
These relations may be compared with the
Serre relations
of the quantum affine superalgebra $\CU_q(\widehat{\fsl}(2|2))$ with Dynkin diagram
\begin{center}
\parbox[c]{0.18\textwidth}{
\includegraphics[width=0.15\textwidth]{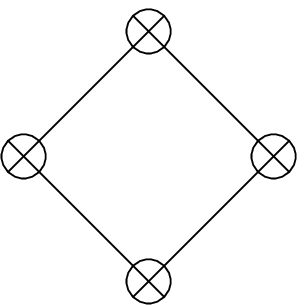}
},
\end{center}
as listed in \cite{Y3}. This list may be presented in the following manner:
\begin{subequations}\label{Serresl(2|2)}
\begin{align}
F_i^2 = 0, \qquad \comm{F_i}{F_{i+2}} &= 0 & &\text{for $i=0,1,2,3$,} \label{Serresl(2|2)a} \\
\comm{\comm{\comm{F_{i+1}}{F_i}_{q^{+1}}}{F_{i-1}}_{q^{-1}}}{F_i} &= 0 & &\text{for $i=0,2$,} \label{Serresl(2|2)b} \\
\comm{\comm{\comm{F_{i+1}}{F_i}_{q^{-1}}}{F_{i-1}}_{q^{+1}}}{F_i} &= 0 & &\text{for $i=1,3$,} \label{Serresl(2|2)c} \\
\comm{F_{i+2}}{F_i^{\brac{m}}} &= 0 & &\text{for $i=0,1,2,3$ and $m\geqslant 1$.} \label{Serresl(2|2)d}
\end{align}
\end{subequations}
Here, $F_i^{\brac{m}}$ is defined recursively for $m\geq 0$ by
\begin{equation}
F_i^{\brac{0}} = F_i, \qquad F_i^{\brac{m}} =
\begin{cases}
\comm{
\comm{
\comm{
\comm{F_i^{\brac{m-1}}}{F_{i-1}}_{q^{-1}}
}{F_{i-2}}_{q^{+1}}
}{F_{i-3}}
}{F_i} & \text{for $i = 0,2$,} \\ 
\comm{
\comm{
\comm{
\comm{F_i^{\brac{m-1}}}{F_{i-1}}_{q^{+1}}
}{F_{i-2}}_{q^{-1}}
}{F_{i-3}}
}{F_i} & \text{for $i = 1,3$.}
\end{cases}
\end{equation}
The relations \rf{Serresl(2|2)a} are easily identified with
\rf{QSerresl(2|2)a}, while relations \rf{QSerresl(2|2)a}
and \rf{QSerresl(2|2)b} ensure that the definition
$\pi_q^+(F_i):=\tau_q^{-1}\SQ_i^+$
represents the relations \rf{Serresl(2|2)b} and \rf{Serresl(2|2)c}.
We have furthermore verified that the $\SQ_i^+$ satisfy the
sixth order relation \rf{Serresl(2|2)d} with $m=1$, but
have to leave the validity of the relations \rf{Serresl(2|2)d}
for $m>1$ as conjecture.

This gives us a representation $\pi_q^+$ of the nilpotent subalgebra $\CN_-$ of $\QEA{q}{\AKMSA{sl}{2}{2}}$.  As usual, we need to extend this to a representation of the Borel subalgebra $\CB_-$.  It is easily checked that this may be accomplished by setting
\begin{subequations}\label{qrep(2|2)'}
\begin{align}
\pi_q^+(H_0) &= -i(\spp^{+}_{1}-i\spp^{+}_{2})/b, &
\pi_q^+(H_1) &= +i(\spp^{+}_{1}+i\spp^{+}_{2})/b, \\
\pi_q^+(H_2) &= +i(\spp^{+}_{1}-i\spp^{+}_{2})/b, &
\pi_q^+(H_3) &= -i(\spp^{+}_{1}+i\spp^{+}_{2})/b.
\end{align}
\end{subequations}

One should remark that the quantum affine superalgebra
$\QEA{q}{\AKMSA{sl}{2}{2}}$ contains non-trivial ideals
by which one might wish to take quotients in order to define smaller quantum affine superalgebras.  For example, $\QEA{q}{\AKMSA{psl}{2}{2}}$ may be obtained in this way.  The Serre relations of these quotients will then include those of $\QEA{q}{\AKMSA{sl}{2}{2}}$.  A more complicated example is the algebra denoted by $\QEA{q}{\bigl( A\brac{1,1}^{\brac{1}} \bigr)^{\mathcal{H}}}$ in \cite{Y2,Y3}.  This may be obtained as a quotient of a one-dimensional (non-central) extension of $\QEA{q}{\AKMSA{sl}{2}{2}}$.  Nevertheless, its Serre relations include (properly) those of $\QEA{q}{\AKMSA{sl}{2}{2}}$ \cite{Y3}.  It seems then that the Serre relations alone cannot distinguish these three quantum affine superalgebras.

However, any representation of $\QEA{q}{\AKMSA{psl}{2}{2}}$ is also a representation of $\QEA{q}{\AKMSA{sl}{2}{2}}$ in which the generators may satisfy additional relations.  Furthermore, one of the representations we want to use to construct Lax matrices is the representation defined in \rf{sl(2|2)fund-a}, which actually defines a four-dimensional representation $\pi_{a,\la}$ of $\QEA{q}{\AKMSA{sl}{2}{2}}$ for all values of $q$.  It is easy to check that this $\pi_{a,\la}$ does not descend to a representation of $\QEA{q}{\AKMSA{psl}{2}{2}}$.  For this reason, and because we have no motivation to consider the extension required to define $\QEA{q}{\bigl( A\brac{1,1}^{\brac{1}} \bigr)^{\mathcal{H}}}$, we will consider $\QEA{q}{\AKMSA{sl}{2}{2}}$ rather than any of the alternatives in what follows.

\subsubsection{Quantum monodromy matrices} \label{secGrading}

Following our previous discussions, it is natural to consider
$\SM_+(\la)=(\pi_{a,\la}\ot_s\pi_{q}^+)(\CR)$ as a candidate for the
quantum monodromy matrix describing the integrable structure
of the massless limit of the $N=2$ Sine-Gordon model.

A new feature of this quantum affine superalgebra
is that there are \emph{two} linearly independent
central elements, which we may take to be $C_0=H_0+H_2$ and $C_3=H_3+H_1$.
In order to find an explicit representation for the element
$t$ which represents the Cartan part of the universal R-matrix
we therefore now need to introduce two derivations, which will be
chosen as $D_0$ and $D_3$ with the non-trivial commutation relations
\begin{equation}
\comm{D_i}{E_j} = \delta_{ij} E_j, \qquad
\comm{D_i}{F_j} = -\delta_{ij} F_j \qquad
\text{($i = 0,3$).}
\end{equation}
We remark that $D_0$ coincides with the ``standard'' derivation $D$.
One may compute the element $t$ appearing in \eqref{redRdef} by extending the Cartan matrix (invariant bilinear form) to include these derivations or by simply requiring \rf{t-Constraint}.  The result is
\begin{equation}\label{tsl(2|2)}
t = -H_1\ot_s H_2 - H_2\ot_s H_1 + C_0\ot_s D_0 + C_3\ot_s D_3 + D_0\ot_s C_0 + D_3\ot_s C_3.
\end{equation}

As before, the representations $\pi$ that we are considering all satisfy $\pi(H_0 + H_1 + H_2 + H_3)=\pi(C_0)+\pi(C_3)=0$, so they are representations of the quantum loop algebra $\QEA{q}{\AKMSA{sl}{2}{2}_0}$. Because of this, we therefore only need to consider the combination $D':=D_0-D_3$ of the derivations. The definitions of $\pi_{a,\la}$ and $\pi_q^+$ above therefore need to be supplemented by
\begin{equation}
\pi_{a,\la}(D')=-\frac{1}{2}\brac{E_{11}+E_{22}+E_{33}-E_{44}}, \qquad
\pi_q^+(D')=\frac{1}{2}\rho_+-\frac{1}{b}\spp_2^+,
\end{equation}
where $\rho_+$ is the fermion number operator defined by
$[\rho_+,\psi_+]=\psi_+$, $[\rho_+,\bar\psi_+]=-\bar\psi_+$.

It is now easy to generalize the arguments of \cite{BHK} to the
case at hand to show that
\begin{equation}\label{Msl(2|2)}
\SM_+(\la_+) = q^{\rho_+ \SZ / 2} e^{-\pi b(\spp^+\SH+\bar\spp^+\bar\SH)/2}
\CP\exp\left(\la_+\int_{0}^R dx\;\SW_{+}(x)\right) ,
\end{equation}
with the operator-valued Lax matrix
\begin{equation}\label{qLaxsl(2|2)}
\SW_{+}(x) = \sum_{i=0}^3 \SE_i^+\SV_i(x) \qquad \text{($\SE_i^+ := \pi_q^+(E_i)$).}
\end{equation}
We use, as in \secref{secSuperLax}, the notation $\SH=\SH_2-\SH_0$,
$\bar\SH=\SH_1-\SH_3$, $\SZ=\SH_0+\SH_2$, and define
$\spp^+=\spp_1^++i\spp_2^+$, $\bar{\spp}^+=\spp_1^+-i\spp_2^+$.  The counterpart $\SM_-(\la_-)$ of this monodromy matrix may be likewise computed by slightly varying the representation $\pi_q^+$.  Explicitly, we construct a representation $\pi_q^-$ of $\CB_+$ by $\pi_q^-(E_i) = \tau_q^{-1} \SQ_i$ and defining $\pi_q^-(H_i)$ as in \eqref{qrep(2|2)'}, but with a relative sign (and exchanging all $+$ labels for $-$ labels).  With $\CR^- = \zeta(\CR^+)$, the analysis now proceeds identically.

It again follows from the Yang-Baxter relation
\rf{YBE} satisfied by $\CR$ that
the operator-valued matrix $\SM_+(\la_+)$
satisfies Yang-Baxter type
relations of the form \rf{RMMMMR'}
with matrix $R$ replaced by
$R(\la/\mu)=(\pi_{a,\la}\ot_s\pi_{a,\mu})(\CR)$.
This matrix may be calculated by analyzing the relations following from \eqref{eqnUnivRIntertwiner}. It is found to be given by
\begin{equation}
R(\la,\mu) = \sum_{i,j=1}^4\rho_{ij}(\la,\mu) E_{ii} \ot_s E_{jj} + \sum_{i,j=1}^4\si_{ij}(\la,\mu) E_{ij} \ot_s E_{ji} ,
\end{equation}
up to an inessential scalar multiple, where $\rho_{ij}$ and $\si_{ij}$ are the $\brac{i,j}$-th entries of the matrices
\begin{subequations}
\begin{align}
\rho &=
\begin{pmatrix}
\la^4 q^{-1} - \mu^4 q & \la^4 - \mu^4 & \la^4 - \mu^4 & \la^4 - \mu^4 \\
\la^4 - \mu^4 & \la^4 q^{-1} - \mu^4 q & \la^4 - \mu^4 & \la^4 - \mu^4 \\
\la^4 - \mu^4 & \la^4 - \mu^4 & \la^4 q - \mu^4 q^{-1} & \la^4 - \mu^4 \\
\la^4 - \mu^4 & \la^4 - \mu^4 & \la^4 - \mu^4 & \la^4 q - \mu^4 q^{-1}
\end{pmatrix}
, \\
\si &= -\la \mu \brac{q-q^{-1}}
\begin{pmatrix}
0 & +\la\mu & -\mu^2 & -\la^2 \\
+\la\mu & 0 & -\la^2 & -\mu^2 \\
+\la^2 & +\mu^2 & 0 & -\la\mu \\
+\mu^2 & +\la^2 & -\la\mu & 0
\end{pmatrix}
.
\end{align}
\end{subequations}
Note the relative signs in $\rho_{33}$, $\rho_{44}$ and the last two columns of $\si$.  These correlate with the (relative) fermionic nature of the third and fourth basis states in the representations $\pi_{a,\la}$.

\subsubsection{Classical limit}

We will now compare the classical limit of $\SM_+(\la_+)$ with the
monodromy matrix $M_+(\la_+)$ that would be obtained by adapting the discussion of the massless limit from Section \ref{massless} to this case.  This would
lead to the consideration of the monodromy matrix
\begin{equation}
M_+(\la_+) =
e^{-\pi b(p \SH+\bar p \bar\SH)/2}
\CP\exp\left(\int_{0}^R dx_{+}\;W_{+}(x_+)\right) ,
\end{equation}
where $p=(\vf^+(R)-\vf^+(0))/\pi b$ and $W_+(x_+)$ is given by the formula
\begin{align} \label{V+sl(2|2)}
W_+(x_+) =  \SZ\;\chi_+\bar\chi_+
&+2m \bigl( \SE_1^{} \chi_+^{} e^{2\vf^+}+\SE_2^{}
\bar\chi_+^{} e^{2\bar\vf^+}
+\SE_3^{} \chi_+^{} e^{-2\vf^+}+\SE_0^{} \bar\chi_+^{} e^{-2\bar\vf^+} \bigr) \notag \\
&-m^2\bigl( \{\SE_1^{},\SE_2^{}\} e^{2(\vf^++\bar\vf^+)}
- \{\SE_{2}^{},\SE_3^{}\} e^{2(\bar\vf^+-\vf^+)} \bigr. \notag \\
&\mspace{120mu} \bigl. + \{\SE_3^{},\SE_0^{}\} e^{-2(\vf^++\bar\vf^+)}
- \{\SE_0^{},\SE_1^{}\} e^{2(\vf^+-\bar\vf^+)} \bigr).
\end{align}
To see how $M_+(x_+)$ may be obtained from
$\SM_+(x_+)$, observe that the terms in the
second line of \rf{V+sl(2|2)} are produced in the limit $b\ra 0$
from the short-distance behavior of the higher order terms in the expansion of
\rf{Msl(2|2)}. In order to see this in more detail, let
us recall the relations \rf{N=2ident} between the respective
variables. 
The terms in $\la_+\SW_{+}(x_+)$ are easily identified
with the terms of order $m$ in the expression \rf{V+sl(2|2)} for
$W_+(x_+)$ if $\la_+$ is chosen appropriately.
When taking the limit $b\ra 0$, one encounters a subtlety
similar to that discussed in Section \ref{altacts}. To elaborate,\footnote{After submitting the first version of this paper, we were informed by A. Zeitlin that similar arguments had previously been used by him 
in \cite{Ze}.},
let us consider the term proportional to $\{\SE_1,\SE_2\}$ at order $\la^2_+$,
for example.  It is given by the integral
\begin{equation}\label{secondorder}
-(2m)^2 \int_{x_1>x_2} dx_1  dx_2 \;\chi_+(x_1) \normord{e^{2\vf^+(x_1)}}
\bar\chi_+(x_2) \normord{e^{2\bar\vf^+(x_2)}},
\end{equation}
where the minus sign is due to the fact that the $\SE_i$ anticommute
with the fermionic fields.
The contribution from the region $|x_1-x_2|<\epsilon$
may be approximated with the help of the operator product expansion\footnote{The variables $x$, $y$ appearing in \rf{lcOPE} are
related to the variables previously used in \rf{FermionOPE} by the usual map from the complex plane to the cylinder, 
that is $z=e^{ix}$, \emph{etc}...}
\begin{equation}\label{lcOPE}
\psi_+(x)\bar\psi_+(y)\sim\frac{-2}{x-y-i0}+\dots.
\end{equation}
This allows us to represent the term in \rf{secondorder} to
leading order as
\begin{equation}
(2m)^2 \int dx \int_{0}^{\epsilon}
dy \;\frac{b^2}{8} \frac{2}{y^{1+b^2}}
\normord{e^{2(\vf^+(x)+\bar\vf^+(x))}} = -\frac{m^2}{\epsilon^{b^2}}
\int dx\; \normord{e^{2(\vf^+(x)+\bar\vf^+(x))}}.
\end{equation}
We see that the result has a finite limit for $b\ra 0$.
The resulting contact terms from higher orders in the expansion
can all be taken into account by adding to the Lax connection the
term $-m^2 \{\SE_1,\SE_2\} \normord{e^{2(\vf^++\bar\vf^+)}}$.
In a similar way one finds the other terms in the
second line of \rf{V+sl(2|2)}.

In order to see where the
term containing the central element $\SZ$ comes from,
let us note that the operator product expansion \eqref{lcOPE} implies that
\begin{equation}
\{\psi_+(x),\bar\psi_+(y)\}=-4\pi i\de(x-y),
\end{equation}
which implies that the fermion number operator $\rho_+$ can be represented as
\begin{equation}\label{rhochichi}
\rho_+ = \frac{i}{4\pi}\int_0^R dx\; \normord{\psi_+(x)\bar\psi_+(x)} =
\frac{2i}{\pi b^2}\int_0^R dx\; \normord{\chi_+(x)\bar\chi_+(x)}.
\end{equation}
It follows that the term containing $\rho_+$ in \rf{Msl(2|2)} reproduces
the contribution proportional to $\SZ$ in \rf{V+sl(2|2)}. As $\SZ$ is
represented by the identity matrix, the term $\SZ\chi_+\bar\chi_+$
will give a contribution to $M_+(\la)$ that can be factored out
like the corresponding factor in \rf{Msl(2|2)}. This concludes our check
that the classical limit of $\SM_+(\la_+)$ reproduces the
monodromy matrix of the classical massless $N=2$ Sine-Gordon model.

\subsection{Fermionic $\fsl(2|1)$ affine Toda theory}

To round off the picture, we shall conclude by listing the relevant
results for the remaining case corresponding to the
fermionic $\fsl(2|1)$ affine Toda theory. Related results have been
obtained in \cite{KZ04a,KZ04b,KZ05,Ze}. The results in this 
subsection are furthermore related to
those obtained in \cite{BT} by bosonization of the fermions.

\subsubsection{Appearance of the quantum affine superalgebra
$\CU_q(\widehat{\fsl}(2|1))$}

Let us define
\begin{equation}
\SQ_i^+ = \int_0^{R}dx\;\SV_i(x) ,\qquad i=0,1,2 ,
\end{equation}
with
\begin{equation}
\SV_1(x) = \normord{e^{2b\phi_1^+(x)}}, \qquad
\begin{aligned}
\SV_0(x) &= \bar\psi_+(x)
\normord{e^{-b\phi_1^+(x)}}, \\
\SV_2(x) &= \psi_+(x)\normord{e^{-b\phi_1^+(x)}}.
\end{aligned}
\end{equation}
With the technique described in \appref{appScreenAlg}, one may then check
that the screening charges  $\SQ_i^+$, $i = 0, 1 ,2$, satisfy the following relations (with $q = e^{-\ii \pi b^2}$):
\begin{subequations} \label{QSerre2|1}
\begin{gather}
(\SQ_0^+)^2 = (\SQ_2^+)^2 = 0, \\
\SQ_i^+ (\SQ_1^+)^2 - \qnum{2}{q} \SQ_1^+ \SQ_i^+ \SQ_1^+ + (\SQ_1^+)^2 \SQ_i^+ = 0 \qquad \text{($i=0,2$),} \\
\SQ_{10202}^+ + \qnum{2}{q} \brac{\SQ_{21020}^+ + \SQ_{02120}^+ + \SQ_{02012}^+} + \SQ_{20201}^+ \mspace{200mu} \notag \\
\mspace{200mu} = \SQ_{12020}^+ + \qnum{2}{q} \brac{\SQ_{01202}^+ + \SQ_{20102}^+ + \SQ_{20210}^+} + \SQ_{02021}^+.
\end{gather}
\end{subequations}
In this last relation, we have again made use of the convenient shorthand $\SQ_{ij \cdots k}^+ = \SQ_i^+ \SQ_j^+ \cdots \SQ_k^+$.
The relations \rf{QSerre2|1} can be identified as the Serre relations of the quantum affine superalgebra $\CU_q(\widehat{\fsl}(2|1))$, given in \cite{Y2} in the form
\begin{subequations} \label{Serre2|1}
\begin{gather}
F_0^2 = F_2^2 = 0, \label{Serre2|1.} \\
\comm{\comm{F_0}{F_1}_{q^{-1}}}{F_1}_q = \comm{F_1}{\comm{F_1}{F_2}_{q^{-1}}}_q = 0, \label{Serre2|1a} \\
\comm{F_0}{\comm{F_2}{\comm{F_0}{\comm{F_2}{F_1}_{q^{-1}}}}}_q = \comm{F_2}{\comm{F_0}{\comm{F_2}{\comm{F_0}{F_1}_{q^{-1}}}}}_q. \label{Serre2|1b}
\end{gather}
\end{subequations}
It follows that setting $\pi_q^+(F_i)=\tau_q^{-1}\SQ_i^+$ defines a representation of the
nilpotent subalgebra $\CN_-$ of $\QEA{q}{\AKMSA{sl}{2}{1}}$.
We conclude that $\QEA{q}{\AKMSA{sl}{2}{1}}$
is the quantum algebraic structure
underlying the integrability of the fermionic $\fsl(2|1)$ affine Toda model.

\subsubsection{Quantum monodromy matrices}

The representation $\pi_q^+$ is extended to a representation of the Borel subalgebra $\CB_-$ by setting
\begin{equation}
\pi_q^+(H_1) = -2i\spp_1/b ,\qquad
\begin{aligned}
\pi_q^+(H_0) &= i\spp_1/b-\rho_+/2, \\
\pi_q^+(H_2) &= i\spp_1/b+\rho_+/2,
\end{aligned}
\end{equation}
where $\rho_+$ is the fermion number operator defined in the previous
subsection.
We may define, as before, $\SM_+(\la) = (\pi_{a,\la}\ot_s\pi_q^+)(\CR)$, where $\pi_{a,\la}$ is given in \eqnref{evrep(2|1)}.
This operator-valued matrix may again be shown to possess a representation as a path-ordered exponential of the form \rf{q-monod} with
\begin{equation}\label{q-monodsl(2|1)}
\SW_+(x;\la)  =  \SE_1\SV_{1}(x)+\SE_2\SV_2(x)+\SE_0\SV_0(x),
\end{equation}
where $\SE_i := \pi_{a,\la}(E_i)$.
We conclude by computing the R-matrix of $\QEA{q}{\AKMSA{sl}{2}{1}}$ in the tensor product of the representations
$\pi_{a,\lambda}$ and $\pi_{a,\mu}$.
Appealing once again to \eqnref{eqnUnivRIntertwiner}, the result is proportional to
\begin{equation}
\func{R}{\la,\mu} = \sum_{i,j=1}^{3} \rho_{ij}(\la,\mu) E_{ii}\ot_s E_{jj}
+\sum_{i,j=1}^{3}\si_{ij}(\la,\mu) E_{ij}\ot_s E_{ji},
\end{equation}
where $\rho_{ij}$ and $\si_{ij}$ are the $\brac{i,j}$-th
entries of the matrices
\begin{subequations} \label{Rsl21}
\begin{align}
\rho &=
\begin{pmatrix}
\la^3 q^{-1} - \mu^3 q & \la^3 - \mu^3 & \la^3 - \mu^3 \\
\la^3 - \mu^3 & \la^3 q^{-1} - \mu^3 q & \la^3 - \mu^3 \\
\la^3 - \mu^3 & \la^3 - \mu^3 & \la^3 q - \mu^3 q^{-1}
\end{pmatrix}
 ,\\
\si &= -\la \mu \brac{q-q^{-1}}
\begin{pmatrix}
0 & +\mu & -\la \\
+\la & 0 & -\mu \\
+\mu & +\la & 0
\end{pmatrix}
 .
\end{align}
\end{subequations}
The entry $\rho_{33}$ again reflects the (relatively) fermionic nature of the third basis state
in the evaluation representation $\pi_{a,\la}$ (note also the signs in the third column of $\si$).

\section{Lattice light-cone approach to the Sinh-Gordon model} \label{lattice}

The difficulties with real exponential interactions described in Subsection \ref{realb} have another
consequence of importance for
us. The constructions described in the previous
section do not immediately generalize.  A careful regularization
of the generating functions $\ST_a(\lambda)$ of the conserved quantities is needed and the only regularization
that is known to work at present is the lattice regularization.
In this section, we will first review the known lattice-regularization of the Sinh-Gordon
model. It will then be reformulated in a way that prepares for the generalization to the other models
of our interest. The reformulation that we will use is a lattice version of the light-cone
representation discussed previously in the classical case.
It is similar, but not equivalent to the lattice light-cone formulations introduced
in \cite{FV92,BBR}. We will discuss the precise relation between our formalism and theirs in \secref{comparison}.

\subsection{Lattice Sinh-Gordon model}\label{SSlatticeSG}

For the case of the Sinh-Gordon model, it has been known for a long time how to construct a tailor-made lattice regularization \cite{FST,IK,Sk83}.
To motivate this construction, one can introduce a minimal distance (ultraviolet cutoff) $\Delta$.  It is then natural
to formulate a regularized version of the theory in terms of averages of the
basic field variables $\phi(x,t)$ and $\Pi(x,t) := \pa_t\phi(x,t)$ over intervals of length $\Delta$.  We therefore introduce
\begin{equation}\label{discrete}
\phi_n = \frac{1}{\Delta}\int_{n\Delta}^{(n+1)\Delta}dx\;\phi(x) ,\qquad
\Pi_n = \frac{1}{4\pi}\int_{n\Delta}^{(n+1)\Delta}{dx}\;\Pi(x) .
\end{equation}
These operators will satisfy the commutation relations
\begin{equation}
\big[ \phi_n , \Pi_m \big] = \frac{i}{2} \de_{n,m} .
\end{equation}
We are looking for a matrix $L_n(\la)$ such that:
\begin{itemize}
\item[(i)] The Lax matrix $U_x(x,t;\lambda)$ is recovered in the continuum limit $\De\ra 0$ as
 \begin{equation}\label{clcontlim}
 L_n(\la) = I+\De  U_x(n\De,t;\lambda)+\CO(\De^2) .
 \end{equation}
\item[(ii)] The elements of the lattice Lax matrix $L_n(\lambda)$ satisfy the commutation
relations
\begin{equation}\label{RLLLLR-SG}
R(\la/\mu)(L_{n}^{}(\la)\ot 1)(1\ot L_{n}^{}(\mu)) = (1\ot L_{n}^{}(\mu))(L_{n}^{}(\la)\ot 1)R(\la/\mu) ,
\end{equation}
with matrix $R$ being obtained from the universal R-matrix
of $\CU_q(\widehat{\fsl}(2))$ via \rf{RfromR}.
\end{itemize}
The relations \rf{RLLLLR-SG} imply similar relations for the
elements of the monodromy matrix
\begin{equation}
\SM_a(\la) = L_N(\la)L_{N-1}(\la)\cdots L_1(\la) ,
\end{equation}
which can be seen as the most natural quantization of the Poisson bracket
relations \rf{FPBR}.

A suitable choice for $L_n(\la)$ is known \cite{FST,IK,Sk83}. It can be written as
\begin{equation}\label{LL}
L_n^{\rm\sst SG}(\la) =
\begin{pmatrix}
\su_n+m^2\Delta^2 \sv_n\su_n\sv_n &
m\Delta( \la \sv_n^{}  + \la^{-1} \sv_n^{-1}) \\
m\Delta( \la \sv_n^{-1}+ \la^{-1} \sv_n^{})
& \su_n^{-1}+m^2\Delta^2 \sv_n^{-1}\su_n^{-1}\sv_n^{-1}
\end{pmatrix}
 ,
\end{equation}
where we have used the operators
$\su_n= e^{2\pi b\Pi_n}$ and
$\sv_n= e^{-b\phi_n}$ which satisfy the relations
\begin{equation} \label{commUV}
\su_n\sv_m = q^{-\de_{nm}}\sv_m\su_n ,\qquad q = e^{-\ii \pi b^2} .
\end{equation}
It is elementary to check that this choice for $L_n(\la)$ satisfies both
requirements (i) and (ii) above. It therefore defines a suitable integrable
lattice regularization of the Sinh-Gordon model.

\subsection{KdV-theory on the lattice}

In the following, we want to explain the representation-theoretic origin of
the Lax matrix \rf{LL} on the one hand, and how all this is related
to the light-cone representation for the model on the other. In order to do this, we begin by discussing the massless limits of the model for which we have previously observed a particularly simple relation between the integrable structure and the universal R-matrix of $\QEA{q}{\AKMA{sl}{2}}$. This will turn out to have a very simple discretized version which was studied in \cite{Ge,V1,V2}.

The procedure of \secref{SG-KdV}, which gave us the integrable structure of the massless limit of the Sinh-Gordon model, can now be
also be applied to the lattice Sinh-Gordon model. Taking the limits
$m\ra 0$, $\la\ra \infty$ with $\mu_+ := \la m\De$ fixed,
or $m\ra 0$,  $\la\ra 0 $ with $\mu_- := \la/m\De$ fixed, yields the Lax matrices \cite{Ge,V1}
\begin{align}\label{Lnpm}
L_n^+(\mu_+) :=
\begin{pmatrix}
\su_n & \mu_+  \sv_n\\
\mu_+ \sv^{-1}_n & \su^{-1}_n
\end{pmatrix} ,\qquad
L_n^-(\mu_-) :=
\begin{pmatrix}
\su_n & \mu_-^{-1}  \sv_n^{-1}\\
\mu_-^{-1}  \sv_n & \su^{-1}_n
\end{pmatrix} ,
\end{align}
respectively. These matrices define interesting
quantized lattice versions of (m)KdV theory.

Remembering the discussion
in Section \ref{SG-KdV}, one would like to interpret
the degrees of freedom of the
integrable lattice model defined by the Lax matrices
$L_n^{+}(\la)$ as a discretization of the left-moving part $\phi_+(x_+)$
of the massless free field $\phi(x,t)$. This raises an
apparent problem as $L_n^{+}(\la)$ contains the same degrees
of freedom per lattice site as $L_n^{\rm\sst SG}(\la)$ did.
In order to see how this puzzle is resolved, let us consider the
family of operators
\begin{equation}
\ST^{+}(\la):=
{\rm Tr}\bigl(L_N^+(\la)L_{N-1}^+(\la)\cdots L_1^{+}(\la)\bigr) .
\end{equation}
The main observation to be made \cite{V1} is that the operators $\ST^{+}(\la)$
depend on the variables $\su_n$, $\sv_n$, $n=1,\ldots,N$, only through the
combinations
\begin{equation}
\sw_n^+ = \bigl(\su_n^{}\sv_{n}^{}\su_{n+1}^{}
\sv_{n+1}^{-1}\bigr)^{1/2} =
e^{b(\Pi_{n+1}+\Pi_n+2(\phi_{n+1}-\phi_n))/4} ,
\end{equation}
which can be seen as lattice analogs of the field variables
$e^{b(\pa_t+\pa_x)\phi(x,t)}$ (the index $n$ is of course defined \emph{modulo} $N$).
This can be verified by using the operator-valued gauge transformation
\begin{equation}\label{gaugesl2}
\tilde{L}_n^+(\mu) := g_{n+1}^{-1}
\begin{pmatrix}
\su_n^{} & \mu_+  \sv_n^{}\\
\mu_+  \sv_n^{-1} & \su_n^{-1}
\end{pmatrix} g_n^{} =
\begin{pmatrix}
q^{-1/4}\sw_n^+ & \mu_+ \sw_n^+ \\
\mu_+  q^{1/2} \brac{\sw_n^+}^{-1} & q^{-1/4} \brac{\sw_n^+}^{-1}
\end{pmatrix} ,
\end{equation}
where
\begin{equation}
g_n :=
\begin{pmatrix}
\su_{n}^{-1/2} \sv_{n}^{1/2} & 0\\
0& \sv_n^{-1/2} \su_{n}^{1/2}
\end{pmatrix} .
\end{equation}

One may also introduce the operators
\begin{equation}
{\sw}_n^- = \bigl(\su_n^{}\sv_{n}^{-1}\su_{n+1}^{}
\sv_{n+1}^{}\bigr)^{1/2} =
e^{b(\Pi_{n+1}+\Pi_n-2(\phi_{n+1}-\phi_n))/4} ,
\end{equation}
which are lattice analogs of the field variables
$e^{b(\pa_t-\pa_x)\phi(x,t)}$. We have the
following commutation relations:
\begin{equation}\label{+-comm}
\sw_n^+\sw_m^- = \sw_m^-\sw_n^+ , \qquad
\sw_n^\pm\sw_m^\pm =
\begin{cases}
q^{\pm \brac{n-m} / 2} \sw_m^\pm\sw_n^\pm & \text{if $\abs{n-m} = 1$,} \\
\sw_m^\pm\sw_n^\pm & \text{otherwise.}
\end{cases}
\end{equation}
It follows that all operators $\sw_n^-$, $n=1,\dots,N$,
commute with $\ST^{+}(\la)$.
Similarly, one may see that
\begin{equation}
\ST^{+}(\la):=
{\rm Tr}\bigl(L_N^+(\la)L_{N-1}^+(\la)\cdots L_1^{+}(\la)\bigr) ,
\end{equation}
commutes with the operators $\sw_n^+$, $n=1,\dots,N$.
This shows how the Lax matrices $L_n^\pm(\la)$ describe two
decoupled integrable structures for the lattice discretization of
a free field corresponding to two decoupled lattice KdV theories
associated with left- and right-movers, respectively.

\subsection{Representation-theoretic origin of the massless Lax matrix} \label{ShGLatticeReps}

Our first concern is to discuss how the Lax matrices $L_n^{\pm}(\la)$
are embedded into the general representation-theoretic
scheme described in the previous sections. This is rather simple.
Let us consider $L_n^+(\la)$. It is easy to check that
\begin{subequations} \label{q-osc}
\begin{align}
\sk_{0,n} &:= \pi_{q,n}^+(K_0) = \su_n^{-2}, &
\sf_{0,n} &:= \pi_{q,n}^+(F_0) = \tau_q^{-1} \su_n^{+1}\sv_n^{-1}, \\
\sk_{1,n} &:= \pi_{q,n}^+(K_1) = \su_n^{+2}, &
\sf_{1,n} &:= \pi_{q,n}^+(F_1) = \tau_q^{-1} \su_n^{-1}\sv_n^{+1},
\end{align}
\end{subequations}
with $\tau_q = q-q^{-1}$, defines a representation $\pi_{q,n}^+$ of the Borel subalgebra $\CB_-$ of $\QEA{q}{\AKMA{sl}{2}_0}$ (the Serre relations follow trivially from the fact that $\sf_0$ and $\sf_1$ commute).
We mention that this representation is a close relative of those referred to as $q$-oscillator representations in \cite{BLZ3}.

We are going to show that there exists a function $f(\la)$ such that the following equality holds:
\begin{equation}\label{L=pipiR}
L_n^{+}(\la) =f(\la) \bigl(\pi_{a,\lambda}\otimes\pi_{q,n}^+\bigr)(\CR) .
\end{equation}
Here, the representations $\pi_{a,\la}$ and $\pi_{q,n}^+$ are defined in
\rf{evrepsl2} and \rf{q-osc},
respectively. Indeed, thanks to the simplicity of the
representations $\pi_{a,\la}$ and $\pi_{q,n}^+$, we will only need
to use generic properties of $\CR$ to establish \rf{L=pipiR}.

It is useful to keep in mind the factorization \rf{redRdef} of the
universal R-matrix $\CR$
into a Cartan part $q^{t} = q^{\brac{H_1 \ot H_1}/2}$ and the reduced R-matrix $\bar\CR$.
First note that in the representation $\pi_{a,\la}$,
the non-trivial monomials in the
operators $\SE_i:=\pi_{a,\la}(E_i)$ are of the form
\begin{equation}\label{monom}
(\SE_0\SE_1)^n = \SE_0\SE_1, \qquad (\SE_1\SE_0)^n = \SE_1\SE_0, \qquad (\SE_0\SE_1)^n\SE_0 = \SE_0, \qquad (\SE_1\SE_0)^n\SE_1 = \SE_1,
\end{equation}
which represent a basis in the space of two-by-two matrices.
Next, recall that for each term $\BFE_I\ot \BFF_J$ appearing in the expansion of the reduced R-matrix $\bar\CR$,
the affine weight (with respect to the $K_i$) of the monomial $\BFE_I$ must cancel that of $\BFF_J$.  As the monomial basis elements $\pi_{a,\la}(\BFE_I)$
have weights taking values in $\{1,q^{2},q^{-2}\}$ and
as $\sf_{0,n}\sf_{1,n}=\sf_{1,n}\sf_{0,n}$ is a multiple of the identity, the corresponding monomials $\pi_{q,n}^+(\BFF_J)$ may be taken from the set
$\{1, \sf_{0,n}, \sf_{1,n}\}$.  It follows from these
observations that the reduced R-matrix has the form
\begin{equation}\label{R-LKdV}
\bigl(\pi_{a,\lambda}\otimes\pi_{q,n}^+\bigr)(\bar\CR) =
\begin{pmatrix}
a(\la) & b(\la) \sf_{1,n} \\
c(\la) \sf_{0,n} & d(\la)
\end{pmatrix} ,
\end{equation}
where $a(\la)$, $b(\la)$, $c(\la)$ and $d(\la)$ are proportional
to the identity operator in $\pi_{q,n}^+$ (they may also possess an implicit $q$-dependence).  We can compute $a$, $b$, $c$ and $d$ by evaluating \eqref{Constraint} in the representation $\pi_{a,\lambda}\otimes\pi_{q,n}^+$, both for $F_0$ and $F_1$.  This yields constraints whose solutions are
\begin{equation}
a(\la)=d(\la) \qquad \text{and} \qquad b(\la)=c(\la)=\la(q-q^{-1}) a(\la).
\end{equation}
Our claim \rf{L=pipiR} now follows easily upon premultiplying by the Cartan part
\begin{equation}
\bigl(\pi_{a,\lambda}\otimes\pi_{q,n}^+\bigr)(q^t) = q^{\brac{\SH_1 \ot \sh_{1,n}} / 2} =
\begin{pmatrix}
\sk_{1,n}^{1/2} & 0 \\
0 & \sk_{1,n}^{-1/2}
\end{pmatrix}
.
\end{equation}

In order to understand the representation-theoretic
origin of $L_n^-(\la)$, we introduce
the representation $\pi_{q,n}^-$ of $\CB_+$ which is defined by
\begin{subequations} \label{q-osc'}
\begin{align}
\pi_{q,n}^-(K_0) &= \su_n^{+2}, &
\pi_{q,n}^-(E_0) &= \tau_{q}^{-1}\sv_n^{-1}\su_n^{+1}, \\
\pi_{q,n}^-(K_1) &= \su_n^{-2} &
\pi_{q,n}^-(E_1) &= \tau_{q}^{-1}\sv_n^{+1}\su_n^{-1}.
\end{align}
\end{subequations}
Repeating the above analysis now, we obtain
\begin{equation}
L_n^-(\la) = g(\la)(\pi_{a,\la}\ot\pi_{q,n}^-)(\CR^-) ,
\end{equation}
where $g(\la)$ is some scalar function.
It now follows from the abstract Yang-Baxter equation \rf{YBE+-} that $L_n^+(\la)$ and $L_n^-(\la)$ both
satisfy an RLL-relation of the form
\rf{RLLLLR-SG} with the \emph{same} R-matrix $R(\la)$
as that which appears in the relation satisfied
by $L_n^{\rm\sst SG}(\la)$.

\subsection{Recombining left-and right-movers}

We have seen that the two simple Lax matrices $L_n^{\pm}(\la)$ for the lattice
(m)KdV theory can be obtained from the Lax matrix $L_n^{\rm\sst SG}(\la)$ of the lattice
Sinh-Gordon model by a limiting procedure.
It is easy to see that by taking classical continuum limits of
$L_n^+(\la)$ and $L_n^{-}(\la)$,
similar to the limit taken in \rf{clcontlim},
one recovers the classical light-cone Lax matrices
$U_+(\la)$ and $U_-(\la)$
defined in \rf{LaxSG+-}, respectively.

Recall the representation of the monodromy matrix in terms of the saw-blade contour ${\mathcal C}_N$ of \eqnref{Tmodmod}.  This naturally suggests an alternative approach to the
discretization of the model: Use averages of the fields $\phi_n$, $\Pi_n$ over the light-like segments ${\mathcal C}_k^\pm$ as basic variables. Out of these, construct the quadruple of operators
$\su_n^\ep=e^{2\pi b\spp_n}$, $\sv_n^\ep=e^{-b\phi_n}$, $\ep=\pm$, with commutation relations
$\su_n^\ep\sv_m^{\ep'}=q^{-\de_{nm}\de_{\ep\ep'}}\sv_m^{\ep'}\su_n^\ep$.
If we now redefine
\begin{align}\label{Lnpm'}
L_n^+(\mu_+) :=
\begin{pmatrix}
\su_n^+ & \mu_+  \sv_n^+\\
\mu_+ (\sv^{+}_n)^{-1} & (\su^{+}_n)^{-1}
\end{pmatrix} ,\qquad
L_n^-(\mu_-) :=
\begin{pmatrix}
\su_n^- & \mu_-^{-1}  (\sv_n^-)^{-1}\\
\mu_-^{-1}  \sv_n^- & (\su^{-}_n)^{-1}
\end{pmatrix}
,
\end{align}
then a natural discrete version of the saw-blade representation
\rf{Tmodmod} for $M(\la)$ may be constructed as
\begin{equation}\label{Lfactor}
\CM(\la) =  \CL_N(\la)\cdots\CL_1(\la) ,\qquad
\CL_n(\la):=L^{-}_{n}(\mu_-)
{L}_{n}^+(\mu_+) ,
\end{equation}
where $\mu_+=\De m\la$ and $\mu_-=\la/m\De$.
It follows from the RLL-type relations \rf{RLLLLR-SG},
satisfied by $L_n^+(\mu_+)$ and $L_n^-(\mu_-)$,
that the monodromy matrix $\CM(\la)$
satisfies RLL-type relations with the
same R-matrix as $L_n^+(\mu_+)$, $L_n^-(\mu_-)$, and hence $L_n^{\rm\sst SG}(\la)$.

What may be confusing is the apparent doubling of the
number of degrees of freedom assigned to a lattice
site with label $n$. We are are therefore going to show that the lattice discretization
defined by \rf{Lfactor} with Lax matrices \rf{Lnpm'} is physically
equivalent to the one introduced in Section \ref{SSlatticeSG}.

\subsubsection{An ultralocal representation}\label{ultraloc}

To this purpose, it is useful to note that $L_n^+(\mu_+)$ and $L_n^-(\mu_-)$ can be factorized as
\begin{subequations}
\begin{align}
L_n^+(\mu_+) &=
\begin{pmatrix}
\su_n^+ & 0 \\
0 & (\su_n^+)^{-1}
\end{pmatrix}
\begin{pmatrix}
1 & \mu_+  \sw_n^+ \\
\mu_+ q  (\sw_n^+)^{-1} & 1
\end{pmatrix}
 , \\
L_n^-(\mu_-) &=
\begin{pmatrix}
1 & {\mu}_-^{-1} q^{-1}  \brac{\sw_n^-}^{-1} \\
{\mu}_-^{-1}  \sw_n^- & 1
\end{pmatrix}
\begin{pmatrix}
\su_n^- & 0 \\
0 & (\su_n^-)^{-1}
\end{pmatrix}
,
\end{align}
\end{subequations}
where $\sw_n^+=(\su_n^+)^{-1}\sv_n^+$ and $\sw_n^-=\sv_n^-(\su_n^-)^{-1}$.
It follows that the Lax matrix
\begin{equation} \label{FactorSL2Lax}
\CL_{n}(\lambda) =
\begin{pmatrix}
1 & {\mu}_-^{-1} q^{-1}  (\sw_{n}^-)^{-1} \\
{\mu}_-^{-1} \sw_{n}^- & 1
\end{pmatrix}
\begin{pmatrix}
\su_{n}^-\su_{n}^+ & 0 \\
0 & \brac{\su_{n}^-\su_{n}^+}^{-1}
\end{pmatrix}
\begin{pmatrix}
1 & \mu_+  \sw_{n}^+ \\
\mu_+ q  (\sw_{n}^+)^{-1} & 1
\end{pmatrix}
\end{equation}
only depends upon the operators $\sw_{n}^+$, $\sw_{n}^-$ and $\SU_n := \su_{n}^-\su_{n}^+$.  Note that $\sw_{n}^+$ and $\sw_{n}^-$ commute as
they act on different tensor factors. The combination
$\sw_{n}^{-}(\sw_{n}^+)^{-1}$ also commutes with $\SU_n$ and is therefore
central in the algebra generated by $\sw_{n}^+$, $\sw_{n}^-$ and $\SU_n$.
It follows that we may consider a representation in which
$\sw_{n}^{-}(\sw_{n}^+)^{-1}$
is represented by a scalar multiple of the identity.
Taking this multiple to be $q^{-1}$ and defining $\SV_n := \brac{\SU_n \sw_n^+}^{-1}$ then gives
\begin{equation}
\CL_{n}(\lambda)=
\begin{pmatrix}
\SU_n+\mu_+{\mu}_-^{-1}
\SV_n\SU_n\SV_n & \mu_+ \SV_n^{-1}+{\mu}_-^{-1} \SV_n \\
{\mu}_-^{-1} \SV_n^{-1}+\mu_+ \SV_n & \SU_n^{-1}+\mu_+{\mu}_-^{-1} \SV^{-1}_n \SU_n^{-1}\SV^{-1}_n
\end{pmatrix}
.
\end{equation}
This Lax matrix is equivalent to that defined in \rf{LL}
when we take $\mu_+$ and $\mu_-$ as in \eqref{Lfactor}.

\subsubsection{A non-ultralocal representation} \label{non-ultraloc}

Another way to identify the variables that the monodromy matrix $\CM(\la)$
depends upon is to use a gauge transformation
similar to that used in \rf{gaugesl2}.
Specifically, with
\begin{subequations}
\begin{align}
g_n^+ &=
\begin{pmatrix}
\brac{\su_n^+}^{-1/2} \brac{\sv_n^+}^{1/2} & 0 \\
0 & \brac{\sv_n^+}^{-1/2} \brac{\su_n^+}^{1/2}
\end{pmatrix}
, \\
g_n^- &=
\begin{pmatrix}
\brac{\su_n^-}^{-1/2} \brac{\sv_n^-}^{-1/2} & 0 \\
0 & \brac{\sv_n^-}^{1/2} \brac{\su_n^-}^{1/2}
\end{pmatrix}
,
\end{align}
\end{subequations}
we can write $\CM(\la)$ in the form
\begin{equation}\label{monosl2}
\CM(\la) = g_{1}^+\tilde{L}_{N}^-(\mu_-)
\tilde{L}_N^+(\mu_+)\cdots\tilde{L}_{1}^-(\mu_-)
\tilde{L}_1^+(\mu_+) (g_{1}^+)^{-1}.
\end{equation}
Here,
\begin{subequations}
\begin{align}
\tilde{L}_{n}^+(\mu_+) &= \brac{g_n^-}^{-1} L_n^+(\mu_+)  g_n^+ =
\begin{pmatrix}
\mst_n^+ & \mu_+ q^{1/4} \mst_{n}^+ \\
\mu_+ q^{1/4} \brac{\mst_{n}^+}^{-1} & q^{-1/2} \brac{\mst_{n}^+}^{-1}
\end{pmatrix}
, \\
\tilde{L}_{n}^-({\mu}_-) &= \brac{g_{n+1}^+}^{-1}  L_n^-(\bar{\mu}_-)
 g_n^- =
\begin{pmatrix}
\mst_n^- & {\mu}_-^{-1} q^{-1/4} \mst_{n}^- \\
{\mu}_-^{-1} q^{-1/4} \brac{\mst_{n}^-}^{-1} & q^{1/2} \brac{\mst_{n}^-}^{-1}
\end{pmatrix}
,
\end{align}
\end{subequations}
and the $\mst_n^{\pm}$ are given by
\begin{equation}
\mst_n^+ = (\sv_n^-)^{1/2} (\su_n^-)^{1/2} (\su_n^+)^{1/2} (\sv_n^+)^{1/2}, \qquad \mst_n^- = (\sv_{n+1}^+)^{-1/2} (\su_{n+1}^+)^{1/2} (\su_n^-)^{1/2} (\sv_n^-)^{-1/2}.
\end{equation}
In this form, it is manifest that
$\CM(\la)$ depends on the correct number of local degrees
of freedom. The price to pay is that we now have non-vanishing commutation
relations between the operators associated to neighboring sites
(non-ultralocality):
\begin{equation}
\mst_{n+1}^+\mst_n^-  = q^{1/2} \mst_n^-\mst_{n+1}^+  ,
\qquad \mst_{n}^-\mst_n^+  = q^{-1/2} \mst_n^+\mst_{n}^- .
\end{equation}
We mention that the variables
$(\mst_n^{\pm})^2$ have the virtue that they make the form of the (discrete) time evolution equations particularly nice \cite{FV94}.

\subsection{Comparison to other approaches}\label{comparison}

The construction \rf{Lfactor} of ${\CL}_n(\la)$
is inspired by similar constructions in
\cite{FV92,BBR}, but differs in detail.
In \cite{FV92,BBR}, the authors proposed
a Lax matrix $\tilde{\CL}_n(\la)$ which, in our notation, would be obtained by
replacing the matrix $L_n^-(\mu_-)$ in \rf{Lnpm'} by
\begin{equation}
\tilde{L}_n^-(\mu_-) :=
\begin{pmatrix}
\su_n^- & \mu_-^{-1}  \sv_n^-\\
\mu_-^{-1}  (\sv_n^-)^{-1} & (\su^{-}_n)^{-1}
\end{pmatrix}  .
\end{equation}
Reducing to the physical degrees of freedom as described
in Section \ref{ultraloc}, one would obtain a Lax
matrix $\tilde{L}_n(\la)$ that is equivalent to the Lax matrix $L_n^{\rm\sst XXZ}(\la)$ defining
a non-compact version of the XXZ-model \cite{ByT}. This Lax matrix  $L_n^{\rm\sst XXZ}(\la)$ is
related to $L_n^{\rm\sst SG}(\la)$ by multiplication with
$\si_1$ and a simple equivalence transformation in quantum
space (see \cite{ByT} for details). This relationship implies the physical equivalence
of the two approaches when the number $N$ of lattice
sites is \emph{even}, while the lattice models are physically inequivalent in the case of \emph{odd} $N$ (see \cite[Appendix D]{NT}
for a detailed discussion of this point in
the closely related case of the lattice Sine-Gordon model).
It is of course quite possible that the inequivalence of the
two approaches for odd $N$ disappears in the continuum limit.

A detailed study of the spectrum of these models and of their
continuum limits has so far been carried out only
for the lattice Sinh-Gordon model defined by
the Lax matrix $L_n^{\rm\sst SG}(\la)$ on lattices with odd
$N$ \cite{ByT,T}. This is due
to the fact observed in \cite{ByT} that this case is
the most convenient one for the analysis of the spectrum of
the respective lattice models.
The results obtained in \cite{ByT,T,ByT2} demonstrate that
our approach is indeed suitable for defining the Sinh-Gordon
continuum quantum field theory by taking the continuum limit
of the lattice Sinh-Gordon model discussed in this paper.

For us, the main advantage of the Lax matrix ${\CL}_n(\la)$ defined in
\rf{Lfactor} will be that it will turn out to
have a very natural generalization to the other models, as
we are now going to explain.


\section{Generalization to the other models}\label{lattice-gen}

In tailor-made lattice regularizations, we want to
preserve as much of the structure of the quantum field theories as possible.
This will include the algebraic relations \rf{RMMMMR} that the
elements of a quantum monodromy matrix are supposed to satisfy.
The discussion of the lattice Sinh-Gordon model suggests a natural
way to realize this feature automatically, as we will now discuss.

\subsection{The general scheme}\label{scheme}

In the case of a lattice model with $N$ sites, one has
$\CH=\CH_1\ot\CH_2\ot\dots\ot\CH_N$.
We will construct the monodromy matrix $\CM_a(\la)$ of the lattice model
as a product of local Lax matrices
\begin{equation}\label{SMdef}
\CM(\la) = \CL_{N}(\la)\CL_{N-1}(\la)\cdots
\CL_{1}(\la)  ,
\end{equation}
which are themselves constructed
from the universal R-matrix in the following way:
\begin{equation}\label{L+-def}
\CL_{n}^{}(\la) = L_{n}^-(\mu_-) L_{n}^+(\mu_+) ,\qquad L_{n}^\pm(\mu_{\pm}):=(\pi_{a,\mu_{\pm}}\ot\pi_{q,n}^\pm)(\CR^\pm).
\end{equation}
Here, $\mu_+ = \la m \De$, $\mu_- = \la / m \De$, and the $\pi_{q,n}^\pm$ are representations of the Borel subalgebras $\CB_\mp$ on $\CH_n^\pm$ such that $\CH_n=\CH_n^+\ot\CH_n^-$.
It follows from \rf{YBE++} and \rf{YBE+-} that both $L_{n}^-(\mu_-)$ and
$L_{n}^+(\mu_+)$ satisfy
\begin{equation}\label{RLLLLR}
R(\la,\mu) \bigl(L_{n}^\pm(\la)\otimes {\rm I}\bigr)
\bigl({\rm I}\otimes L_{n}^\pm(\mu)\bigr)
 = \bigl({\rm I}\otimes L_{n}^\pm(\mu)\bigr) \bigl(L_{n}^\pm(\la)\otimes {\rm I}\bigr)R(\la,\mu)  ,
\end{equation}
with the same matrix
\begin{equation}
R(\lambda,\mu):=\bigl(\pi_{a,\lambda}\otimes\pi_{a,\mu}\bigr)(\CR) .
\end{equation}
The monodromy matrix constructed in \rf{SMdef} therefore satisfies
\rf{RMMMMR}, as desired.

When applying this construction to the remaining models, we therefore need
to:
\begin{itemize}
\item[(i)] Find representations $\pi_{q,n}^+$ and $\pi_{q,n}^{-}$ of
the relevant Borel subalgebras $\CB_-$ and $\CB_+$, respectively,
such that the Lax matrices $L_{n}^+$ and $L_{n}^-$ defined in \rf{L+-def}
reproduce correctly the corresponding
classical Lax matrices in the classical continuum limit.
\item[(ii)] Make sure that the physical degrees of freedom
of the lattice model, defined initially with an
auxiliary doubling of the lattice
degrees of freedom, are indeed in one-to-one correspondence
with discretized versions of the field variables.
\end{itemize}
We are now going to apply this strategy to the remaining models of interest.

\subsection{The Boussinesq model on the lattice} \label{BoussinesqLat}

We begin by applying the general scheme described in Section \ref{scheme} to the $\fsl(3)$ affine Toda theory. Let us begin by explaining how to find the Lax matrix $L_{n}^+(\la)$ associated to the
left-moving degrees of freedom in the massless limit.
It was previously argued that the relevant
algebraic structure is the quantum affine algebra $\CU_q(\widehat{\fsl}(3))$.
The main task is then to find suitable representations $\pi_{a,\mu}$ and $\pi_{q,n}^+$ with which to construct the Lax matrix $L_{n}^+(\mu)$
as $(\pi_{a,\mu}\ot\pi_{q,n}^+)(\CR)$.

To begin with, we shall consider the case in which $\pi_{a,\mu}$ is the
representation defined in \rf{evrep3}.
In order to motivate our choice for $\pi_{q,n}^+$, it will be useful to make some observations on the generic structure of Lax matrices representing a universal R-matrix $\CR$.
First, recall the factorization \eqref{redRdef} of $\CR$ into a
part $q^t$ containing only Cartan generators and a reduced R-matrix $\bar\CR$, the latter being a formal sum of monomials in the generators $E_i\ot \wun$ and $\wun\ot F_i$.
The factor $t = \tfrac{1}{3} \sum_i H_i \ot H_i$ yields a diagonal matrix under $\pi_{a,\mu} \ot \pi_{q,n}^+$.  With $\SH_i:=\pi_{a,\mu}(H_i)$ and $\sh_{i,n}:=\pi_{q,n}^+(H_i)$, we may write
\begin{equation}\label{Cartan3}
\func{\brac{\pi_{a,\mu}\ot\pi_{q,n}^+}}{q^t} = q^{\sum_i \SH_i \ot \sh_{i,n} / 3} ={\rm diag}(\su_{1,n},\su_{2,n},\su_{0,n}),
\end{equation}
where
\begin{equation}
\su_{0,n} = \sk_{1,n}^{-1/3} \sk_{2,n}^{-2/3} \quad
\su_{1,n} = \sk_{1,n}^{2/3} \sk_{2,n}^{1/3}, \quad
\su_{2,n} = \sk_{1,n}^{-1/3} \sk_{2,n}^{1/3}, \qquad
\text{($\sk_{i,n} = q^{\sh_{i,n}}$).}
\end{equation}
In order to calculate the factor
$(\pi_{a,\mu}\ot\pi_{q,n}^+)(\bar\CR)$, we will again use the intertwining property \eqref{eqnUnivRIntertwiner}
of $\CR$ in the form of \eqnref{Constraint}.
But as our choice of $\pi_{a,\mu}$ is such that $\SE_i:=\pi_{a,\mu}(E_i)$
is proportional to $\mu$, the first order expansion \eqref{redRlow} already gives the representative of $\bar{\CR}$ as
\begin{equation}\label{barLexp}
\bar{L}^+_{n}(\mu):=(\pi_{a,\la}\ot\pi_{q,n}^+)(\bar\CR) = \id+\mu (q-q^{-1})
\begin{pmatrix}
0 & \sf_{1,n} & 0 \\
0  &  0 & \sf_{2,n} \\
\sf_{0,n} & 0 & 0
\end{pmatrix}+\CO(\mu^2) ,
\end{equation}
with $\sf_{i,n}:=\pi_{q,n}^+(F_i)$.

This should be compared with the form of the classical Lax matrix \rf{sl3Lax}
to which $L_{n}^+(\mu)$ should reduce in a classical continuum limit
analogous to \rf{clcontlim}. The comparison suggests that
the operators $\su_{i,n}$ should be
constructed from exponential functions of the averages $\spp_{i,n}$ of $\pi_{i}(x)=\pa_t\phi_i(x,0)$ over light-like
segments ${\mathcal C}_n^+$, while the $\sf_{i,n}$ should be proportional to operators
$\sv_{i,n}$ which represent discrete versions of the exponential functions one finds in the
off-diagonal elements of \rf{sl3Lax}. A more detailed comparison suggests that we take
\begin{subequations} \label{uvpq}
\begin{align}
\su_{0,n} &= e^{-\pi b(\spp_{1,n}-\spp_{2,n}/\sqrt{3})}, &
\sv_{0,n} &= e^{+2b\sq_{1,n}} \\
\su_{1,n} &= e^{+\pi b(\spp_{1,n}+\spp_{2,n}/\sqrt{3})}, &
\sv_{1,n} &= e^{-b(\sq_{1,n}+\sqrt{3}\sq_{2,n})}, \\
\su_{2,n} &= e^{-2\pi b\spp_{2,n}/\sqrt{3}}, &
\sv_{2,n} &= e^{-b(\sq_{1,n}-\sqrt{3}\sq_{2,n})}.
\end{align}
\end{subequations}
We note that we do not have to take $\sf_{i,n}$ strictly equal to $\sv_{i,n}$. It is possible to
multiply $\sv_{i,n}$ by combinations of the $\su_{i,n}$ which would disappear in the continuum limit
since $\spp_{i,n}=\CO(\De)$. From the point of view of the representation theory, such a modification will not change the affine weight of the $\sf_{i,n}$, but is, in this case, necessary for satisfying the Serre-relations of $\QEA{q}{\AKMA{sl}{3}}$. It is easy to check that defining
\begin{equation} \label{qrep(3')}
\sf_{i,n} := \tau_q^{-1} \su_{i,n}^{-1} \sv_{i,n}^{}
\end{equation}
allows us to achieve all the requirements above. Indeed, it follows that
\begin{equation} \label{FF=qFF}
\sf_{i,n} \sf_{i+1,n} = q \; \sf_{i+1,n} \sf_{i,n},
\end{equation}
in which the first indices take values in $\ZZ_3$.  The Serre relations are now trivial to check.

Terms of higher order in the expansion \rf{barLexp} can be straight-forwardly calculated by evaluating \eqnref{Constraint}
in the representation $\pi_{a,\mu}\ot\pi_{q,n}^+$. It is useful to organize the calculation
as an expansion in powers of $\mu$. In the case at hand, we easily find that the terms proportional to $\mu^2$ vanish due to the relations \rf{FF=qFF}. In this way, remembering to multiply by \eqref{Cartan3}, we arrive at the Lax matrix
\begin{equation}\label{LBoussi}
L^+_{n}(\mu) = \func{\ell}{\mu}
\begin{pmatrix}
\su_{1,n} & \mu \sv_{1,n} & 0 \\
0  &  \su_{2,n} & \mu \sv_{2,n} \\
\mu \sv_{0,n} & 0 & \su_{0,n}
\end{pmatrix}
,
\end{equation}
where $\func{\ell}{\mu}$ is an unimportant scalar function.

It is also interesting to repeat this computation using the representation $\pi'_{a,\la}$ given in \eqnref{evrep3'} (but the \emph{same} $\pi_{q,n}^+$).  The resulting Lax matrix may be expressed in the form
\begin{equation}\label{LBoussi'}
L^+_{n}{}'(\mu) =
\begin{pmatrix}
\su_{0,n}^{-1} & \mu \su_{1,n} \sv_{2,n} & \mu^2 \sv_{0,n}^{-1} \\
-\mu^2 \sv_{2,n}^{-1} &  \su_{2,n}^{-1} & \mu \su_{0,n} \sv_{1,n} \\
-\mu \su_{2,n} \sv_{0,n} & -\mu^2 \sv_{1,n}^{-1} & \su_{1,n}^{-1}
\end{pmatrix}
.
\end{equation}
We observe additional off-diagonal terms in this case.  Note that these are perfectly consistent with the expected classical continuum limit, as $\mu^2=\CO(\De^2)$ is then of sub-leading order.

\subsection{The $\SLA{sl}{3}$ affine Toda theory on the lattice} \label{LatticeSL3}

Inspired by the example of the Sinh-Gordon model, we will now
look for a monodromy matrix $\CM(\la)$ for the lattice $\SLA{sl}{3}$ affine Toda
theory of the form \eqref{SMdef}.
We have already determined the local Lax matrix $L_{n}^+$.  To determine $L_n^-$, we must repeat the analysis of \secref{BoussinesqLat} with the representation $\pi_{q,n}^+$ of $\CB_-$ replaced by a representation $\pi_{q,n}^-$ of $\CB_+$.  It is easy to see that sending $H_i$ to $-H_i$ and $F_i$ to $E_i$ achieves this, giving $\sk_{i,n}^- := \pi_{q,n}^-(K_i)$ and $\se_{i,n}^- := \pi_{q,n}^-(E_i)$ as
\begin{subequations}
\begin{gather}
\su_{0,n}^- = (\sk_{1,n}^-)^{1/3} (\sk_{2,n}^-)^{2/3}, \quad
\su_{1,n}^- = (\sk_{1,n}^-)^{-2/3} (\sk_{2,n}^-)^{-1/3}, \quad
\su_{2,n}^- = (\sk_{1,n}^-)^{1/3} (\sk_{2,n}^-)^{-1/3}, \\
\se_{i,n}^- = \tau_q^{-1} \sv_{i,n}^- (\su_{i,n}^-)^{-1}.
\end{gather}
\end{subequations}
We mention that we have commuted the operators in the expression for the $\se_{i,n}$, dropping the $q$-factor thereby obtained, for computational convenience.  Affixing similar labels to the operators in $L_n^+$, we now have two local Lax matrices:
\begin{subequations}
\begin{align}
L_n^+(\mu_+) &=
\begin{pmatrix}
\su_{1,n}^+ & \mu_+ \sv_{1,n}^+ & 0 \\
0  &  \su_{2,n}^+ & \mu_+ \sv_{2,n}^+ \\
\mu_+ \sv_{0,n}^+ & 0 & \su_{0,n}^+
\end{pmatrix}
, \label{sl3Ln+} \\
L_n^-(\mu_-) &=
\begin{pmatrix} \su_{1,n}^- & 0 & \mu_-^{-1} \sv_{0,n}^- \\
\mu_-^{-1} \sv_{1,n}^-  &  \su_{2,n}^- & 0 \\
0 & \mu_-^{-1} \sv_{2,n}^- & \su_{0,n}^- \label{sl3Ln-}
\end{pmatrix}
.
\end{align}
\end{subequations}
To be clear, the operators $\su_{i,n}^\ep$ and $\sv_{i,n}^\ep$ are constructed as in \rf{uvpq}, but with the substitutions
$\spp_i\ra\spp_{i,n}^\ep$ and $\sq_i\ra\sq_{i,n}^\ep$ for $i=1,2$,
$\ep=\pm$, $n=1,\dots,N$ (the local position and momentum modes are now taken to satisfy $\comm{\spp_{i,n}^{\ep}}{\sq_{j,m}^{\ep'}} = \tfrac{1}{2i}\de_{nm}\de_{ij}\de_{\ep\ep'}$).  We remark that one can check \eqref{sl3Ln-} by applying the anti-automorphism $\zeta$ to \eqref{sl3Ln+}, while simultaneously considering the slight differences between $\pi_{q,n}^+$ and $\pi_{q,n}^-$.

The key observation to make now is that $\CM(\la)$ actually depends upon only $4N+2$
algebraically independent combinations of the $8N$ variables
$\spp_{i,n}^{\ep}$ and $\sq_{i,n}^{\ep}$.
This is an easy consequence of the following observations:

First, the operators which appear in the matrix elements of the product
$\CL_{n}(\la)=L_n^-(\mu_-) L_n^+(\mu_+)$ can all be expressed in terms
of the six operators
$\sx_{i,n}^{}=\sk_{i,n}^-\sk_{i,n}^+$, $\sy_{i,n}^{}=\se_{i,n}^{-}
\sf_{i,n}^{+}$ and
$\sz_{i,n}=\se_{i,n}^{-}(\sf_{i,n}^{+})^{-1}$,
where $i=1,2$. The operators $\sx_{i,n}$ and $\sy_{i,n}$ commute
with the $\sz_{j,n}$, but $\sz_{1,n}$ does not commute
with $\sz_{2,n}$. Using this observation, one can show directly
that the algebra $\mathfrak{A}_n$ generated
by the $\sx_{i,n}$, $\sy_{i,n}$ and $\sz_{i,n}$ has no
non-trivial central elements. This is an important difference as compared
with the case previously discussed in Section \ref{ultraloc}.

Note, on the other hand, that
the algebra generated by the matrix elements
of the individual factors $L^{-}_{n}(\mu_-)$ and
${L}_{n}^+(\mu_+)$ contains a non-commutative
subalgebra $\mathfrak{B}_n$ which is
generated by
\begin{equation}
\eta_1^{} = \sf_{2,n}^{+}(\sk_{1,n}^{+}\sf_{0,n}^{+})^{-1}
\bigl(\se_{2,n}^{-}(\sk_{1,n}^{-}\se_{0,n}^{-})^{-1}\bigr)^{-1}, \quad
\eta_2^{} = \sf_{1,n}^{+}\sk_{2,n}^{+}(\sf_{0,n}^{+})^{-1}
\bigl(\se_{1,n}^{-}\sk_{2,n}^{-}(\se_{0,n}^{-})^{-1}\bigr)^{-1}.
\end{equation}
It can be checked
that $\mathfrak{B}_n$ commutes with the algebra $\mathfrak{A}_n$.
We conclude that the
monodromy matrix  does not depend on any function of the
elements of $\mathfrak{B}_n$. This means
that $\CM(\la)$ depends only on $6N$ combinations
formed out of the $8N$
operators $\spp_{i,n}^{\ep}$ and $\sq_{i,n}^{\ep}$.

We may repeat this argument for the products
$L^{+}_{n+1}(\mu_+) L_{n}^-(\mu_-)$, $i=1,\dots,N-1$,
of Lax matrices associated to $N-1$ neighboring sites. It allows us to
find another $2(N-1)$ combinations of the basic variables
that the monodromy
matrix $\CM(\la)$ does not depend upon.
We conclude that $\CM(\la)$ depends on only
$4N+2$ independent variables.

Another way to explicitly identify a minimal set of operators
from which all elements of $\CM(\la)$
can be constructed goes as follows:
Insert the identity in the form $(\sg_n^\ep)^{-1} \sg_n^\ep$ to the right of each factor $L_n^\ep(\la)$
in \rf{SMdef}. We will choose the $\sg_n^+$ and $\sg_n^-$ to be the respective diagonal matrices with elements
\begin{equation}
\begin{matrix}
\normord{(\su_{1,n}^+)^{-\frac{2}{3}}(\sv_{1,n}^+)^{+\frac{2}{3}}(\su_{2,n}^+)^{-\frac{1}{3}}(\sv_{2,n}^+)^{+\frac{1}{3}}}, \\
\normord{(\su_{1,n}^+)^{+\frac{1}{3}}(\sv_{1,n}^+)^{-\frac{1}{3}}(\su_{2,n}^+)^{-\frac{1}{3}}(\sv_{2,n}^+)^{+\frac{1}{3}}}, \\
\normord{(\su_{1,n}^+)^{+\frac{1}{3}}(\sv_{1,n}^+)^{-\frac{1}{3}}(\su_{2,n}^+)^{+\frac{2}{3}}(\sv_{2,n}^+)^{-\frac{2}{3}}}
\end{matrix}
\quad \text{and} \quad
\begin{matrix}
\normord{(\su_{1,n}^-)^{-\frac{1}{3}}(\sv_{1,n}^-)^{-\frac{2}{3}}(\su_{2,n}^-)^{+\frac{1}{3}}(\sv_{2,n}^-)^{-\frac{1}{3}}}, \\
\normord{(\su_{1,n}^-)^{-\frac{1}{3}}(\sv_{1,n}^-)^{+\frac{1}{3}}(\su_{2,n}^-)^{-\frac{2}{3}}(\sv_{2,n}^-)^{-\frac{1}{3}}}, \\
\normord{(\su_{1,n}^-)^{+\frac{2}{3}}(\sv_{1,n}^-)^{+\frac{1}{3}}(\su_{2,n}^-)^{+\frac{1}{3}}(\sv_{2,n}^-)^{+\frac{2}{3}}}.
\end{matrix}
\end{equation}
This induces a gauge transformation $L_n^\ep(\la)\ra\tilde{L}_n^\ep(\mu)$
of the form
\begin{subequations}
\begin{align}
\tilde{L}_n^+(\mu_+) &=
\begin{pmatrix} q^{-1/3}\mst_{1,n}^+ & q^{+1/3} \mu_+
\mst_{1,n}^+ & 0 \\
0  &  q^{-1/3} \mst_{2,n}^+ & q^{+1/3} \mu_+ \mst_{2,n}^+ \\
q^{+1/3} \mu_+ \mst_{0,n}^+ & 0 & q^{-1/3} \mst_{0,n}^+
\end{pmatrix}
, \\
\tilde{L}_n^-(\mu_-) &=
\begin{pmatrix}
q^{+1/3}\mst_{1,n}^- & 0 & q^{-1/3} \mu_-^{-1} \mst_{0,n}^- \\
q^{-1/3} \mu_-^{-1} \mst_{1,n}^- & q^{+1/3} \mst_{2,n}^- & 0 \\
0 & q^{-1/3} \mu_-^{-1} \mst_{2,n}^- & q^{+1/3} \mst_{0,n}^-
\end{pmatrix}
,
\end{align}
\end{subequations}
where $\mst_{i,n}^+ := \brac{\sg_n^-}_{ii}^{-1} \su_{i,n}^+ \brac{\sg_n^+}_{ii}$, $\mst_{i,n}^- := \brac{\sg_n^+}_{ii}^{-1} \su_{i,n}^- \brac{\sg_n^-}_{ii}$.
The monodromy matrix $\CM(\la)$ is then represented as
\begin{equation}\label{monosl3'}
\CM(\la) = \sg_{1}^+\tilde{L}_{N}^-(\mu_-)\tilde{L}_N^+(\mu_+)
\;{\dots}\;\tilde{L}_{1}^-(\mu_-)\tilde{L}_1^+(\mu_+)
 (\sg_{1}^+)^{-1} .
\end{equation}
In this form, it is manifest that $\CM(\la)$ depends only upon the
$4N$ variables $\mst_{i,n}^\ep$, $i=1,2$, $n=1,\dots,N$, $\ep=\pm$, together with the two $\brac{\sg_1^+}_{ii}$, $i=1,2$.
As in \secref{non-ultraloc}, the price to pay for making manifest the correct number of local degrees of freedom is the presence of non-ultralocal commutation
relations:  We cannot guarantee that $\mst_{i,n}^\ep$ and $\mst_{j,m}^{\ep'}$ will commute with each other unless $\abs{n-m}>1$.

\subsection{Fermionic $\fsl(2|1)$ affine Toda theory on the lattice} \label{LatticeSL(2|1)}

To discretize the fermionic fields $\psi_\pm(x)$, $\bar\psi_{\pm}(x)$
in a way that is compatible with our previous fermion
conventions, we introduce a set of operators
$\psi_n^{\ep}$, $\bar\psi_n^{\ep}$, $\ep=\pm$, satisfying the algebra
\begin{equation}
\acomm{\psi_n^{\ep}}{\psi_m^{\ep'}} = 0, \qquad
\acomm{\psi_n^{\ep}}{\bar\psi_m^{\ep'}} = -\ep i\delta_{nm} \delta_{\ep\ep'}, \qquad
\acomm{\bar\psi_n^{\ep}}{\bar\psi_m^{\ep'}} = 0.
\end{equation}
Defining $\rho_n^{\ep} := \ep i\comm{\psi_n^{\ep}}{\bar\psi_n^{\ep}}$,
we then have
\begin{equation}
\comm{\rho_n^{\ep}}{\psi_m^{\ep'}} = \delta_{nm} \delta_{\ep\ep'} \psi_n^{\ep}, \qquad
\comm{\rho_n^{\ep}}{\bar\psi_m^{\ep'}} = -\delta_{nm} \delta_{\ep\ep'} \bar\psi_n^{\ep}.
\end{equation}
Finally, let $\spp^{\ep}_{n}$, $\sq^{\ep}_{n}$ be operators which satisfy
\begin{equation}
\comm{\spp^{\ep}_{i,n}}{\spp^{\ep'}_{j,m}} = 0, \qquad
\comm{\spp^{\ep}_{i,n}}{\sq^{\ep'}_{j,m}} = \frac{1}{2i}\de_{ij}\de_{nm}\de_{\ep\ep'}, \qquad
\comm{\sq^{\ep}_{i,n}}{\sq^{\ep'}_{j,m}} = 0.
\end{equation}
The operators $\psi_n^\ep$, $\bar\psi_n^\ep$ will represent
the discretized fermionic fields $\psi^\ep(x)$, $\bar\psi^\ep(x)$, while $\sq_n^\ep$, $\spp_n^\ep$
will represent $\phi_1^\ep$ and its conjugate momentum, respectively, at the lattice site $n$.
Out of these operators, let us construct the following representation of the Borel subalgebra $\CB_-$ of
$\CU_q\bigl(\widehat{\fsl}(2|1)\bigr)$:
\begin{subequations} \label{qrep(2|1)}
\begin{align}
\pi_{q,n}^+(H_0) &= i\spp_n^+/b-\rho_n^+/2, &
\sf_{0,n}^+:=\pi_{q,n}^+(F_0) &= -\tau_q^{-1}e^{-{b}\sq_n^+} \bar\psi_n^+  q^{-\rho^+_n/2},\\
\pi_{q,n}^+(H_1) &= -2i\spp_n^+/b, &
\sf_{1,n}^+:=\pi_{q,n}^+(F_1) &= +\tau_q^{-1}e^{+2b\sq_n^+} q^{-\rho_n^+}, \\
\pi_{q,n}^+(H_2) &= i\spp_n^+/b+\rho_n^+/2, &
\sf_{2,n}^+:=\pi_{q,n}^+(F_2) &= -\tau_q^{-1}e^{-{b}\sq_n^+} \psi_n^+  q^{-\rho^+_n/2}. \label{qrep(2|1)+c}
\end{align}
\end{subequations}
As usual, $\tau_q = q - q^{-1}$.  The signs in the above expressions for the $\sf_{i,n}^+$ have been chosen to ensure consistency with the classical Lax matrix \eqref{U+sl(2|1)}.
The Serre relations \eqref{Serre2|1} follow from the observation that
\begin{equation}
\comm{\sf_{0,n}^+}{\sf_{1,n}^+}_{q^{-1}} = \comm{\sf_{1,n}^+}{\sf_{2,n}^+}_{q^{-1}} = 0,
\end{equation}
along with some manipulation of the left hand side of \eqref{Serre2|1b}.  We note for later use that
\begin{equation} \label{2|1acomm}
\comm{\sf_{2,n}^+}{\sf_{0,n}^+}_q = -iq^{1/2} \tau_q^{-2} e^{-2b \sq_n^+} q^{-\rho_n^+},
\end{equation}
recalling that this is a $q$-anticommutator by the conventions of \secref{QuaAffSup}.

The corresponding Lax matrix is again defined as $L_n^+(\mu_+)=(\pi_{a,\mu_+}\ot_s\pi_{q,n}^+)(\CR^+)$ with $\pi_{a,\mu_+}$ as in \eqref{evrep(2|1)}.
We will sketch the derivation (up to the usual irrelevant scalar multiplier) of
\begin{equation}\label{Laxsl(2|1)}
L_n^+(\mu_+) = \ell_n^+ \bar L_n^+(\mu_+) ,
\end{equation}
where
\begin{equation}
\ell_n^+ = q^{\rho_n^+/2} \brac{e^{-\pi b\spp_n^+} E_{11} + e^{\pi b\spp_n^+} E_{22} + q^{\rho_n^+/2} E_{33}}
\end{equation}
and
\begin{align} \label{Lbarsl(2|1)}
\bar L_n^+(\mu_+) = \id
&+ \mu_+ \brac{E_{12} e^{2b\sq_n^+} q^{-\rho_n^+/2}
+ E_{23} e^{-b\sq_n^+} \bar\psi_n^+
+ E_{31} e^{-b\sq_n^+} \psi_n^+} q^{-\rho^+_n/2} \notag \\
&+i\mu_+^2 q^{1/2} \tau_q^{-1} E_{21} e^{-2b\sq_n^+} q^{-\rho_n^+}.
\end{align}
Here as before, $E_{ij}$ denotes the $3 \times 3$ matrix with $1$ in position $\brac{i,j}$ and $0$ elsewhere.
For clarity, we will defer this analysis to the end of the section.

Similar Lax matrices (with the roles of quantum and auxiliary
spaces exchanged) have been presented without proof in \cite{BT}.

A similar analysis computes $L_n^-(\mu_-) = \bar L_n^-(\mu_-) \ell_n^-$ from the representation
\begin{subequations} \label{qrep(2|1)-}
\begin{align}
\pi_{q,n}^-(H_0) &= -i\spp_n^-/b + \rho_n^-/2, &
\se_{0,n}^- := \pi_{q,n}^-(E_0) &= +\tau_q^{-1} e^{-b\sq_n^-} q^{-\rho^-_n/2} \bar\psi_n^-, \\
\pi_{q,n}^-(H_1) &= 2i\spp_n^-/b, &
\se_{1,n}^- := \pi_{q,n}^-(E_1) &= +\tau_q^{-1} e^{2b\sq_n^-} q^{-\rho_n^-}, \\
\pi_{q,n}^-(H_2) &= -i\spp_n^-/b - \rho_n^-/2, &
\se_{2,n}^- := \pi_{q,n}^-(E_2) &= -\tau_q^{-1} e^{-b\sq_n^-} q^{-\rho^-_n/2} \psi_n^-. \label{qrep(2|1)-c}
\end{align}
\end{subequations}
The signs in the $\se_{i,n}^-$ have been chosen for consistency with the classical Lax matrix \eqref{U-sl(2|1)}.  One can check that these signs do not affect the validity of the Serre relations \eqref{Serre2|1}.
It is easy to see now that $\ell_n^-$ may be obtained from $\ell_n^+$ by merely changing the $+$ labels to $-$ labels:
\begin{equation}
\ell_n^- = q^{\rho_n^-/2} \brac{e^{-\pi b\spp_n^-} E_{11} + e^{\pi b\spp_n^-} E_{22} + q^{\rho_n^-/2} E_{33}}.
\end{equation}
The story is somewhat more subtle for $\bar L_n^-$ because of the signs associated with certain fermions (for example in $\pi_{a,\mu_{\pm}}$).  The result is
\begin{align} \label{Lbarsl(2|1)-}
\bar L_n^-(\mu_-) = \id
&- \mu_-^{-1} q^{-\rho^-_n/2} \brac{E_{21} e^{2b\sq_n^-} q^{-\rho_n^-/2}
+ E_{32} e^{-b\sq_n^-} \bar\psi_n^-
+ E_{13} e^{-b\sq_n^-} \psi_n^-} \notag \\
&{+i}\mu_-^{-2} q^{-1/2} \tau_q^{-1} E_{12} e^{-2b\sq_n^-} q^{-\rho_n^-}.
\end{align}
As before, this can be checked using the anti-automorphism $\zeta$, remembering that its action on graded tensor products is given in \eqref{gradedzeta}.
The full Lax matrix is finally constructed as $\CL_n(\la)=L_n^-(\mu_-)L_n^+(\mu_+)$, as before. By repeating the discussion in \secDref{ultraloc}{LatticeSL3}, it is easy to check that the resulting lattice model has the correct number of degrees of freedom per site.

It is interesting to observe that the continuum limit $\De\ra 0$
would suppress the terms in the second line of \rf{Lbarsl(2|1)}:
These terms would be of order $\CO(\De^2)$ in the limit, since
$\mu_+=\CO(\De)$. In this way, one recovers \rf{q-monodsl(2|1)}.
One may, however, combine the
limit $\De\ra 0$ with the classical limit $b\ra 0$
in such a way that $\tau_q^{-1}\mu_+^2=\CO(\De)$.  Assuming that
$\De=b^2$ and $\mu_+=\De\la_+$,
it is easy to see that this combination of the classical and the
continuum limits allows us to recover the classical Lax matrix
\rf{U+sl(2|1)}. What we observe here is directly analogous
to the phenomenon discussed in Section \ref{pertFFsl(2|1)} ---
the term in the second line of \rf{Lbarsl(2|1)}
corresponds to the contact term produced in the classical limit.

The expression for $L_n^+$ can be derived as follows:  First, note that $\ell_n^+ = (\pi_{a,\mu_+}\ot_s\pi_{q,n}^+)(q^t)$ is obtained by substituting $t = H_0 \ot_s H_2 + H_2 \ot_s H_0$.  To evaluate $\bar L_n^+(\mu_+) = (\pi_{a,\mu_+} \ot_s \pi_{q,n}^+)(\bar\CR^+)$, we consider \eqref{Constraint} in this representation.  Substituting
\begin{equation}
\bar{L}_n^+ = \sum_{a,b} E_{ab} \ot_s \BL_{a,b}, \quad
\pi_{a,\mu_+}(F_i) = \brac{-1}^{\de_{i,2}} \mu_+^{-1} E_{-i,-i-1}, \quad
\pi_{a,\mu_+}(K_i) = \sum_c q^{m_{i,c}} E_{cc}
\end{equation}
and extracting $E_{ab}$ from each term, we arrive at
\begin{multline}\label{sl(2|1)recrel}
\brac{-1}^{p_i \brac{1 - \de_{a,3}}} \de_{b,-i-1} \BL_{a,-i} -
\brac{-1}^{\de_{i,2}} \de_{a,-i} \BL_{-i-1,b} = \\
\mu_+ \sqbrac{\brac{-1}^{p_i \brac{\de_{a,3} + \de_{b,3}}} q^{m_{i,a}} \sf_{i,n}^+ \BL_{a,b} - q^{-m_{i,b}} \BL_{a,b} \sf_{i,n}^+}.
\end{multline}
Here, the indices $a,b,i$ are taken in $\ZZ_3$, though we conventionally take $a,b \in \set{1,2,3}$, $i \in \set{0,1,2}$.  The $m_{i,a}$ are the diagonal entries of the matrices $\pi_{a,\mu_+}(H_i)$, so $m_0 = \brac{0,1,1}$, $m_1 = \brac{1,-1,0}$, $m_2 = \brac{-1,0,-1}$.  This represents $27$ equations in $9$ unknowns (though they are far from being independent) and can be used to recursively calculate the coefficients of the expansion $\BL_{a,b} = \sum_{k=0}^{\infty} \BL_{a,b}^{(k)} \mu_+^k$ in powers of $\mu_+$.

We commence the recursion by using the expansion \eqref{super-redRlow} of $\bar\CR$.  This gives $\BL_{i,j}^{(0)} = \de_{ij}$ and $\BL_{i,j}^{(1)} = \brac{-1}^{p_{-j}} \tau_q \sf_{-j,n}^+ \de_{i,j-1}$.  More explicitly, the non-zero $\BL_{i,j}^{(1)}$ are
\begin{equation}\label{SL(2|1)Rec1}
\BL_{2,3}^{(1)} = -\tau_q \sf_{0,n}^+, \qquad
\BL_{1,2}^{(1)} = +\tau_q \sf_{1,n}^+, \qquad
\BL_{3,1}^{(1)} = -\tau_q \sf_{2,n}^+.
\end{equation}
Substituting these results into the second order recursion relations and noting that weight considerations and the properties of $\pi_{a,\mu_+}$ force $\BL_{i,j}^{(2)} \propto \de_{a,b+1}$, we obtain
\begin{equation} \label{SL(2|1)Rec2}
\BL_{1,3}^{(2)} = -\tau_q \comm{\sf_{0,n}^+}{\sf_{1,n}^+}_{q^{-1}}, \qquad
\BL_{3,2}^{(2)} = -\tau_q \comm{\sf_{1,n}^+}{\sf_{2,n}^+}_{q^{-1}}, \qquad
\BL_{2,1}^{(2)} = -\tau_q \comm{\sf_{2,n}^+}{\sf_{0,n}^+}_{q}.
\end{equation}
At this point, we can significantly simplify our calculations by using the properties of the representation $\pi_{q,n}^+$.  Indeed, the coefficient of $\rho_n^+$ in $\sf_{1,n}^+$ was chosen so as to simplify \eqref{SL(2|1)Rec2} as much as possible.  Because of this, $\BL_{1,3}^{(2)}$ and $\BL_{3,2}^{(2)}$ actually \emph{vanish} and \eqref{2|1acomm} gives
\begin{equation}
\BL_{2,1}^{(2)} = +iq^{1/2} \tau_q^{-1} e^{-2b \sq_n^+} q^{-\rho_n^+}.
\end{equation}
The third order recursion now gives $\BL_{1,1}^{(3)} = \BL_{2,2}^{(3)} = \BL_{3,3}^{(3)}$.  Moreover, the fourth order equations with $a = b = -i+1$ show that $\BL_{a,a}^{(3)}$ commutes with each $\sf_{i,n}^+$.  $\BL_{a,a}^{(3)}$ likewise commutes with each Cartan representative (it has no affine weight), hence we may set it to a scalar multiple of the identity:  $\BL_{a,a}^{(3)} = l^{(3)}\id$.  The above analysis immediately generalises, resulting in $\BL_{a,b}^{(3k+r)} = l^{(3k)} \BL_{a,b}^{(r)}$, where $r=0,1,2$.  The formula \eqref{Lbarsl(2|1)} for $\BL_+(\mu_+)$ follows easily from these considerations (after dropping the tensor product symbols). 

\subsection{The $N=2$ super Sine-Gordon model on the lattice}\label{N=2lattice}

Let $\psi_n^{\ep}$, $\bar\psi_n^{\ep}$, $\rho_n^{\ep}$ (with $\ep=\pm$) be matrices as in \secref{LatticeSL(2|1)}.  These again represent the discretisation in the fermionic sector.  As the super Sine-Gordon model has bosonic fields $\phi_i^{\ep}$ with $i=1,2$, we let $\spp^{\ep}_{i,n}$, $\sq^{\ep}_{i,n}$ be operators which satisfy
\begin{equation}
\comm{\spp^{\ep}_{i,n}}{\sq^{\ep'}_{j,m}} = \frac{1}{2i} \de_{ij} \de_{nm} \de_{\ep\ep'}.
\end{equation}
From these, we introduce the following operators:
\begin{subequations} \label{qrep(2|2)}
\begin{align}
\sh_{0,n}^+ &= -i (\spp^+_{1,n} - i \spp^+_{2,n})/b, &
\sf_{0,n}^+ &= -\tau_q^{-1}
e^{-b (\sq^+_{1,n} - i \sq^+_{2,n})} \bar\psi_n^+ q^{-\rho_n^+/2}, \\
\sh_{1,n}^+ &= +i (\spp^+_{1,n} + i \spp^+_{2,n})/b, &
\sf_{1,n}^+ &= -\tau_q^{-1}
e^{+b (\sq^+_{1,n} + i \sq^+_{2,n})} \psi_n^+ q^{-\rho_n^+/2}, \\
\sh_{2,n}^+ &= +i (\spp^+_{1,n} - i \spp^+_{2,n})/b, &
\sf_{2,n}^+ &= -\tau_q^{-1}
e^{+b (\sq^+_{1,n} - i \sq^+_{2,n})} \bar\psi_n^+ q^{-\rho_n^+/2}, \\
\sh_{3,n}^+ &= -i (\spp^+_{1,n} + i \spp^+_{2,n})/b, &
\sf_{3,n}^+ &= -\tau_q^{-1}
e^{-b (\sq^+_{1,n} + i \sq^+_{2,n})} \psi_n^{+} q^{-\rho_n^+/2}.
\end{align}
\end{subequations}
This we will supplement with
${\sd'_n}^+ = \tfrac{1}{2} \rho_n^+- \spp^+_{2,n} / b$.
It is not hard to check that setting $\pi_{q,n}^+(H_i) := \sh_{i,n}^+$,
$\pi_{q,n}^+(F_i) := \sf_{i,n}^+$ and $\pi_{q,n}^+(D') := {\sd'_n}^+$
defines a representation of the Borel subalgebra $\CB_-$ of
$\CU_q\bigl(\widehat{\fsl}(2|2)\bigr)$.  The Serre relations \eqref{Serresl(2|2)a} are obvious and the rest follow immediately from the observation that the coefficient of $\rho_n^+$ in the $\sf_{i,n}^+$ has been tuned to guarantee that
\begin{subequations} \label{(2|2)TripComm}
\begin{align}
\comm{\comm{\sf_{i+1,n}^+}{\sf_{i,n}^+}_{q^{+1}}}{\sf_{i-1,n}^+}_{q^{-1}} &= 0 & \text{for $i=0,2$,} \\
\comm{\comm{\sf_{i+1,n}^+}{\sf_{i,n}^+}_{q^{-1}}}{\sf_{i-1,n}^+}_{q^{+1}} &= 0 & \text{for $i=1,3$.}
\end{align}
\end{subequations}
As before, we shall define the Lax matrix by $L_n^+(\mu_+) = (\pi_{a,\mu_+} \ot_s \pi_{q,n}^+)(\CR^+)$, where $\pi_{a,\mu_+}$ was given in \eqnref{sl(2|2)fund-a}.
It again factors as
\begin{equation}\label{Laxsl(2|2)}
L_n^+(\mu_+) = \ell_n^+ \bar L_n^+(\mu_+),
\end{equation}
where (up to the usual irrelevant scalar multiple)
\begin{equation}
\ell_n^+ = q^{\rho_n^+/2} \brac{e^{\pi b\spp^+_{1,n}} E_{11}
+ e^{-\pi b\spp^+_{1,n}} E_{22} + e^{-i\pi b\spp^+_{2,n}} E_{33}
+ e^{i\pi b\spp^+_{2,n}} E_{44}}
\end{equation}
and
\begin{align} \label{Lbarsl(2|2)}
\bar{L}_n^+(\mu_+) &= \id +
\mu_+ \brac{
E_{13} \psi_n^+ e^{b\sq^+_n} + E_{32} \bar\psi_n^+ e^{b\bar\sq^+_n} +
E_{24} \psi_n^+ e^{-b\sq^+_n} + E_{41} \bar\psi_n^+ e^{-b\bar\sq^+_n}
} q^{-\rho_{n}^+/2} \\
&-i \mu_+^2 \tau_q^{-1} q^{-\rho_n^+} \sqbrac{
q^{-1/2} \brac{E_{12} e^{2b\sq^+_{1,n}} + E_{21} e^{-2b\sq^+_{1,n}}} -
q^{1/2} \brac{E_{43} e^{2ib\sq^+_{2,n}} + E_{34} e^{-2ib\sq^+_{2,n}}}}.
\notag \end{align}
Here, we have used the shorthand $\sq^+_n = \sq^+_{1,n} + i \sq^+_{2,n}$, $\bar\sq^+_n = \sq^+_{1,n} - i \sq^+_{2,n}$.

To compute $L_n^-(\mu_-) = \BL_n^-(\mu_-) \ell_n^-$, we define a representation of $\CB_+$ by $\pi_{q,n}^-(E_i) := \se_{i,n}^-$, $\pi_{q,n}^-(H_i) := \sh_{i,n}^-$ and $\pi_{q,n}^-(D') := {\sd'_n}^-$, where
\begin{subequations} \label{qrep(2|2)-}
\begin{align}
\sh_{0,n}^- &= +i (\spp^-_{1,n} - i \spp^-_{2,n})/b, &
\se_{0,n}^- &= +\tau_q^{-1}
e^{-b (\sq^-_{1,n} - i \sq^-_{2,n})} q^{-\rho_n^-/2} \bar\psi_n^-, \\
\sh_{1,n}^- &= -i (\spp^-_{1,n} + i \spp^-_{2,n})/b, &
\se_{1,n}^- &= -\tau_q^{-1}
e^{+b (\sq^-_{1,n} + i \sq^-_{2,n})} q^{-\rho_n^-/2} \psi_n^-, \\
\sh_{2,n}^- &= -i (\spp^-_{1,n} - i \spp^-_{2,n})/b, &
\se_{2,n}^- &= +\tau_q^{-1}
e^{+b (\sq^-_{1,n} - i \sq^-_{2,n})} q^{-\rho_n^-/2} \bar\psi_n^-, \\
\sh_{3,n}^- &= +i (\spp^-_{1,n} + i \spp^-_{2,n})/b, &
\se_{3,n}^- &= -\tau_q^{-1}
e^{-b (\sq^-_{1,n} + i \sq^-_{2,n})} q^{-\rho_n^-/2} \psi_n^+
\end{align}
\end{subequations}
and ${\sd'_n}^- = -\tfrac{1}{2} \rho_n^- + \spp^-_{2,n} / b$.  We then set $L_n^-(\mu_-) = (\pi_{a,\mu_-} \ot_s \pi_{q,n}^-)(\CR^-)$ as usual.  Explicitly, we obtain
\begin{equation}
\ell_n^- = q^{\rho_n^-/2} \brac{e^{\pi b\spp^-_{1,n}} E_{11}
+ e^{-\pi b\spp^-_{1,n}} E_{22} + e^{-i\pi b\spp^-_{2,n}} E_{33}
+ e^{i\pi b\spp^-_{2,n}} E_{44}}
\end{equation}
and
\begin{align} \label{Lbarsl(2|2)-}
\bar{L}_n^-(\mu_-) &= \id -
\mu_-^{-1} q^{-\rho_{n}^-/2} \brac{
E_{31} \psi_n^- e^{b\sq^-_n} + E_{23} \bar\psi_n^- e^{b\bar\sq^-_n} +
E_{42} \psi_n^- e^{-b\sq^-_n}+ E_{14} \bar\psi_n^- e^{-b\bar\sq^-_n}
}  \\
&{-i} \mu_-^{-2} \tau_q^{-1} q^{-\rho_n^-} \sqbrac{
q^{1/2} \brac{E_{21} e^{2b\sq^-_{1,n}} + E_{12} e^{-2b\sq^-_{1,n}}} -
q^{-1/2} \brac{E_{34} e^{2ib\sq^-_{2,n}} + E_{43} e^{-2ib\sq^-_{2,n}}}}.
\notag\end{align}
The full Lax matrix is again constructed as $\CL_n(\mu_-)=L_n^-(\mu_-)L_n^+(\mu_+)$ and one may check that the resulting lattice model has the correct number of degrees of freedom per site.  Taking the classical continuum limit in the manner discussed in \secref{LatticeSL(2|1)},
we recover the classical Lax matrices.

The calculations leading to these results are very similar to those of the previous section.  In particular, the computation of $\BL_n^+ = \sum_{a,b} E_{ab} \ot_s \BL_{a,b}$ is again based on converting \eqref{Constraint} into recursion relations for the coefficients of $\BL_{a,b} = \sum_k \BL_{a,b}^{(k)} \mu_+^k$.
This time, $\BL_{a,b}^{(0)} = \delta_{ab}$ yields
\begin{subequations}
\begin{align}
\BL_{4,1}^{(1)} &= -\tau_q \sf_{0,n}^+, &
\BL_{2,1}^{(2)} &= +\tau_q \comm{\sf_{0,n}^+}{\sf_{3,n}^+}_{q^{-1}}, &
\BL_{3,1}^{(3)} &= \tau_q \comm{\comm{\sf_{0,n}^+}{\sf_{3,n}^+}_{q^{-1}}}{\sf_{2,n}^+}_{q^{+1}}, \\
\BL_{1,3}^{(1)} &= -\tau_q \sf_{1,n}^+, &
\BL_{4,3}^{(2)} &= -\tau_q \comm{\sf_{1,n}^+}{\sf_{0,n}^+}_{q^{+1}}, &
\BL_{2,3}^{(3)} &= \tau_q \comm{\comm{\sf_{1,n}^+}{\sf_{0,n}^+}_{q^{+1}}}{\sf_{3,n}^+}_{q^{-1}}, \\
\BL_{3,2}^{(1)} &= -\tau_q \sf_{2,n}^+, &
\BL_{1,2}^{(2)} &= +\tau_q \comm{\sf_{2,n}^+}{\sf_{1,n}^+}_{q^{-1}}, &
\BL_{4,2}^{(3)} &= \tau_q \comm{\comm{\sf_{2,n}^+}{\sf_{1,n}^+}_{q^{-1}}}{\sf_{0,n}^+}_{q^{+1}}, \\
\BL_{2,4}^{(1)} &= -\tau_q \sf_{3,n}^+, &
\BL_{3,4}^{(2)} &= -\tau_q \comm{\sf_{3,n}^+}{\sf_{2,n}^+}_{q^{+1}}, &
\BL_{1,4}^{(3)} &= \tau_q \comm{\comm{\sf_{3,n}^+}{\sf_{2,n}^+}_{q^{+1}}}{\sf_{1,n}^+}_{q^{-1}}.
\end{align}
\end{subequations}
By \eqnref{(2|2)TripComm}, the third order coefficients vanish and the rest of the derivation proceeds in an identical fashion to that of \secref{LatticeSL(2|1)}.



\section{Outlook}

These examples illustrate
our proposed scheme for the construction
of integrable lattice regularizations. We expect that this scheme can be applied to large classes
of integrable quantum field theories. The key ingredients are the light-cone
representation and the realization that the lattice Lax matrices $L^+_n(\la)$ and $L^-_n(\la)$, which describe parallel transport in the light-cone directions, can be obtained from the universal R-matrices $\CR^+$ and $\CR^-$
of certain quantum affine (super)algebras by evaluating them in suitable representations.

What we have described here should of course be seen as the very first
step towards the solution of the models in question. However, the
relations we have discussed with the representation theory of quantum affine (super)algebras will determine the next steps to a large extent.
The reader may in particular note that we have not yet defined a discrete
analog of the dynamical evolution law. However, within the framework
of the quantum inverse scattering method, there are standard recipes
for defining lattice Hamiltonians from the so-called
fundamental R-matrix $\SR(\la)$ which can be obtained
from the universal R-matrix by choosing the same representation
in auxiliary and quantum spaces.
A variant of this construction turns out to work for the class of 
lattice models discussed in our paper.
An object replacing the fundamental R-matrix can be
obtained from the universal R-matrix by choosing 
a certain infinite-dimensional representation in auxiliary
space instead of the finite-dimensional representations 
$\pi_{a,\la}$ used in this paper. The monodromy matrices
defined from these analogs of the 
fundamental R-matrices turn out to be related to
the Baxter Q-operators. They may furthermore be used to
construct natural lattice Hamiltonians and discrete time-evolution operators.
For the case of the lattice Sinh-Gordon model,
we recover the generator of the discrete time evolution 
of \cite{FV94} in this way, which was obtained from the
Q-operator of the lattice Sinh-Gordon model in \cite{ByT2}.
We shall defer a proper discussion of these topics
to a forthcoming publication.

\section*{Acknowledgements}

DR thanks Hiroyuki Yamane and Ruibin Zhang for helpful correspondence/discussions.  He was partially supported by the Marie Curie Excellence Grant MEXT-CT-2006-042695 and partially by an Australian Research Council Discovery Project DP0193910.  JT thanks the EU for financial support via the Marie Curie Excellence Grant MEXT-CT-2006-042695.


\appendix

\section*{Appendices}

\section{The Algebra of Screening Operators} \label{appScreenAlg}

The aim of this appendix is to briefly describe how to identify the quantum symmetry algebra generated by the various \emph{screening charges} which we have constructed for our sigma models.  This follows from a variant of the standard treatment for screening charges in the free field description of \cfts{} (see \cite[Ch.~11.4]{GR-AS} for example) and a simple algorithm described in \cite[App.~A]{BLZ3}.  We outline the method here as it is fundamental to our constructions.

Let $\SC{i}{x}$ denote a vertex operator for some collection of free bosons.  The standard derivation of the quantum symmetry generated by a given set of screening operators
\begin{equation}\label{screenings}
\SQ_i = \int \dd x \; \SC{i}{x}
\end{equation}
results in an action of operators $\sf_i$, $\sk_i$ on the vector space of screened vertex operators.  If $V$ is such a screened vertex operator, one identifies $\sf_i$ with left-multiplication of $V$ by $\SQ_i$ and $\sk_i$ with multiplication by the \emph{braiding factor} of $\SC{i}{w}$ and $V$.  The natural generalisation of this action to tensor products of screened vertex operators gives coproduct formulae:
\begin{equation} \label{eqnCoProdFK}
\func{\Delta}{\sf_i} = \sf_i \otimes \wun + \sk_i^{-1} \otimes \sf_i, \qquad \func{\Delta}{\sk_i} = \sk_i \otimes \sk_i.
\end{equation}
With the conventions of \secref{BosonConventions}, the braiding factors for the $\sk_i$-action may be determined from the formula for a single boson:
\begin{equation} \label{eqnBF2}
\normord{e^{\alpha \func{\phi}{x}}} \normord{e^{\beta \func{\phi}{y}}} = e^{-i \pi \alpha \beta / 2} \normord{e^{\beta \func{\phi}{y}}} \normord{e^{\alpha \func{\phi}{x}}} \qquad \text{($x > y$).}
\end{equation}
Elementary computation then gives
\begin{equation} \label{eqnKF=FK}
\sk_i \sf_j = \omega_{i,j}^{-1} \sf_j \sk_i,
\end{equation}
where $\omega_{i,j}$ is the factor obtained from braiding $\SC{i}{z}$ with $\SC{j}{w}$.

If we can identify the braiding factors as
\begin{equation} \label{eqnIdentification}
\omega_{i,j} = q^{A_{ij}},
\end{equation}
where $A$ is the Cartan matrix of some Lie algebra $\alg{g}$, then \eqref{eqnKF=FK} suggests that the $\sf_i$ and $\sk_i$ define a representation of the Borel subalgebra $\mathcal{B}_-$ of $\QEA{q}{\alg{g}}$.  To prove this, it only remains to check the appropriate Serre relations.  Before discussing methods for this, let us first remark that we have also found instances in which the braiding factors have the form
\begin{equation} \label{eqnIdentification2}
\omega_{i,j} = \brac{-1}^{p_i p_j} q^{A_{ij}},
\end{equation}
in which $A$ is the Cartan matrix of a Lie superalgebra $\alg{g}$ and $p_i \in \set{0,1}$ denotes the parity of the corresponding simple root.  This signals that we should replace $\otimes$ by the graded tensor product $\otimes_s$ of \secref{QuaAffSup} in \eqnref{eqnCoProdFK}.  Repeating the above derivation now corrects the braiding factors by a sign.  The upshot is that \eqnref{eqnIdentification2} reverts to \eqnref{eqnIdentification}, consistent with a representation of the Borel subalgebra $\mathcal{B}_-$ of the superalgebra $\QEA{q}{\alg{g}}$.

To verify the Serre relations in either case, we rewrite all products of $n$ \gsos{} $\SQ_{i_1} , \ldots , \SQ_{i_n}$ in terms of a fixed basis and then search for linear relations between them.
We may then choose the basis elements for the vector space of products spanned by the $\SQ_{\func{\sigma}{i_1}} \cdots \SQ_{\func{\sigma}{i_n}}$ ($\sigma$ a permutation) to be defined by
\begin{equation}
\SJ_{j_1, j_2, \dotsc , j_n} = \underset{x_1 > x_2 > \dotsb > x_n}{\int \int \dotsi \int} dx_1 dx_2 \cdots dx_n \; \VOp{j_1}{x_1} \VOp{j_2}{x_2} \dotsm \VOp{j_n}{x_n} .
\end{equation}
That these elements really do constitute a basis is a simple consequence of the braiding relations \eqref{eqnBF2}.

As always, an example best illustrates the method.  When $n=2$ and $\omega_{i,j} = q^{A_{ij}}$ with $A = \left(
\begin{smallmatrix}
+2 & -2 \\
-2 & +2
\end{smallmatrix}
\right)$, the Cartan matrix of $\AKMA{sl}{2}$, we can express the product of $\SQ_1$ and $\SQ_2$ in terms of $\SJ_{1,2}$ and $\SJ_{2,1}$ as follows:
\begin{align}
\SQ_1 \SQ_2
&= \underset{x_1 > x_2}{\iint} dx_1 dx_2 \; \VOp{1}{x_1} \VOp{2}{x_2} +
\underset{x_1 < x_2}{\iint} dx_1 dx_2 \; \VOp{1}{x_1} \VOp{2}{x_2} \notag \\
&= \SJ_{1,2} + \underset{x_2 < x_1}{\iint} dx_1 dx_2 \; \VOp{1}{x_2} \VOp{2}{x_1}
= \SJ_{1,2} + q^2 \underset{x_1 > x_2}{\iint} dx_1 dx_2 \; \VOp{2}{x_1} \VOp{1}{x_2} \notag \\
&= \SJ_{1,2} + q^2 \SJ_{2,1}.
\end{align}
The third equality uses the braiding relations \eqref{eqnBF2}.  Similarly, we can derive that $\SQ_2 \SQ_1 = \SJ_{2,1} + q^2 \SJ_{1,2}$.  Basic linear algebra therefore allows us to conclude that for generic $q$, $\SQ_1 \SQ_2$ and $\SQ_2 \SQ_1$ are not linearly related.\footnote{We use the term ``generic'' to mean that $q$ should not be a root of unity.  In this case, we only require $q^4 \neq 1$, but other similar computations end up excluding other roots of unity.}  This calculation therefore finds no Serre relations involving these products of \gsos{}.

Of course, we can search for Serre relations involving other products and for higher $n$.  The number of basis elements can be as large as $n!$, so this quickly becomes tedious.  However, each calculation reduces to an exercise involving only permutations and linear algebra, hence is easy to implement in a computer algebra package.  With \textsc{Maple}, we were able to quickly find all Serre relations with $n \leqslant 7$ for the quantum symmetries of the models considered here, and determine which were algebraically independent.  We have no proof that the relations found are exhaustive (they should not be in at least one case), but they suffice to identify the quantum symmetry as a quantum affine (super)algebra.

\section{Quantum monodromy matrices from universal R-matrix}\label{MvsR-proof}

In this appendix, we present a proof of the assertions \eqref{MfromR} 
and \rf{M-fromR} following the arguments in \cite{BHK}.  This assertion exhibits the monodromy matrix of the quantum Sinh-Gordon model (with imaginary $b$) as the universal R-matrix of $\QEA{q}{\AKMA{sl}{2}}$ in a suitably chosen representation $\pi_{a,\la} \ot \pi_q^+$.  We refer to \secref{repKdV} for further context.

To begin, it will be useful to consider
\begin{equation}
{\CP}^+_{a,\la} := (\pi_{a,\la}\ot {\rm id})(\bar{\CR}^+) ,
\end{equation}
which may be considered as a kind of
universal monodromy matrix. $\CP_{a,\la}^+$ can then be expressed as a
formal series $\CP_{a,\la}^+(F_i)$
of matrices whose entries are monomials formed out of the $F_i$.
Rewriting the basic property $(\id\ot\Delta)(\CR^+)=\CR_{13}^+\CR_{12}^+$ in terms of $\bar{\CR}^+$ and applying $\pi_{a,\la} \otimes \id \otimes \id$
leads to the non-trivial identity
\begin{equation}\label{factorid}
{\CP}^+_{a,\la}\bigl(X_{i,1}+X_{i,2}\bigr) = {\CP}^+_{a,\la}(X_{i,2})
{\CP}^+_{a,\la}(X_{i,1}) ,
\end{equation}
where $X_{i,1}$, $X_{i,2}$ are the generators $X_{i,1} := F_i\otimes \wun$ and $X_{i,2}:= K_i^{-1}\otimes F_i$.
As the identity \rf{factorid} holds in the sense of formal power series,
it implies that
\begin{equation}\label{factorid'}
{\CP}^+_{a,\la}\bigl(\SX_{i,1}+\SX_{i,2}\bigr) = {\CP}^+_{a,\la}(\SX_{i,2})
{\CP}^+_{a,\la}(\SX_{i,1})
\end{equation}
will hold for any set of operators $\SX_{i,1}$,
$\SX_{i,2}$ that satisfy the same relations
as $X_{i,1}$, $X_{i,2}$, namely the Serre relations
\eqref{eqnSGSerre} and
\begin{equation}\label{SXexch}
\SX_{i,2}\SX_{j,1} = q^{A_{ij}}\SX_{j,1}\SX_{i,2} .
\end{equation}
The main idea is to compare the factorization \rf{factorid'}
with the factorization of the path-ordered exponential
appearing in the definition \rf{q-monod}:
\begin{equation}\label{Pexpfactor}
\CP\exp\left(\int_0^{R}dx\;\SW^+(x)\right) =
\CP\exp\left(\int_y^{R}dx\;\SW^+(x)\right)
\CP\exp\left(\int_0^{y}dx\;\SW^+(x)\right) .
\end{equation}
In order to do this, let us consider the ``partial'' screening charges
\begin{equation}
\SX_{i,1} = \frac{1}{q-q^{-1}}\int_0^ydx \;\SV_i(x), \qquad
\SX_{i,2} = \frac{1}{q-q^{-1}}\int_y^Rdx \;\SV_i(x),
\end{equation}
which appear in the expansion of the factors on the
right hand side of \rf{Pexpfactor}.  It follows
easily from the braid relations \eqref{eqnBF2}
that the operators $\SX_{i,1}$,
$\SX_{i,2}$ satisfy the relations \rf{SXexch}.
The Serre relations \eqref{eqnQEASerre}
are verified by means of the technique described in \appref{appScreenAlg}.

Considering the limit $y\ra 0$, where {$\SX_{i,1}\sim\CO(y)$},
and using \rf{redRlow}, we observe that
\begin{equation}\label{CPlow}
{\CP}^+_{a,\la}(\SX_{i,1}) =
\id
+\int_0^ydx\;\SW^+(x;\la)+\CO(y^2).
\end{equation}
As the identities \rf{Pexpfactor} and \rf{CPlow} together uniquely characterize the path-ordered exponential, this allows us to conclude that
\begin{equation}
\CP_{a,\la}^+ =
\CP\exp\left(\int_0^R dx\;\SW^+(x;\lambda)\right) ,
\end{equation}
from which \rf{MfromR} follows easily.

We may similarly consider ${\CP}^-_{a,\la} := (\pi_{a,\la}\ot {\rm id})(\bar{\CR}^-)$.
Rewriting $(\id\ot\Delta)(\CR^-)=\CR_{13}^-\CR_{12}^-$ now leads to the identity
\begin{equation}
{\CP}^-_{a,\la}\bigl(Y_{i,1}+Y_{i,2}\bigr) = {\CP}^-_{a,\la}(Y_{i,2})
{\CP}^-_{a,\la}(Y_{i,1}) ,
\end{equation}
where $Y_{i,1}$, $Y_{i,2}$ are the generators
$Y_{i,1} := E_i\otimes K_i$ and
$Y_{i,2}:= \wun\otimes E_i$.
As before, it follows that
{\begin{equation}
{\CP}^-_{a,\la}\bigl(\SY_{i,1}+\SY_{i,2}\bigr) = {\CP}^-_{a,\la}(\SY_{i,2})
{\CP}^-_{a,\la}(\SY_{i,1})
\end{equation}}
will hold for any set of operators $\SY_{i,1}$,
$\SY_{i,2}$ that satisfy the relations
\begin{equation}\label{SYexch}
\SY_{i,2}\SY_{j,1} = q^{-A_{ij}}\SY_{j,1}\SY_{i,2} .
\end{equation}
We note that the difference in the signs of the exponent in the
braiding phases appearing in \rf{SXexch} and \rf{SYexch} is precisely
accounted for by the different orientations of the integration
contours that appear in the definitions of $\SM_+(\la_+)$
and $\SM_-(\la_-)$, respectively.


\end{document}